\documentclass[aps,prb,reprint,amsmath,amssymb,superscriptaddress]{revtex4-2}
\usepackage{graphicx}
\usepackage{svg}
\usepackage{dcolumn}
\usepackage{bm}
\usepackage{soul}
\usepackage{color}
\usepackage{array}
\usepackage{booktabs}
\usepackage{multirow, makecell}
\usepackage{physics}
\usepackage[colorlinks=true, allcolors=blue]{hyperref}

\begin{document}

\title{First-principles electron-phonon interactions with self-consistent Hubbard interaction: an application to transparent conductive oxides}

\author{Wooil Yang}
\affiliation{Korea Institute for Advanced Study, Seoul 02455, Korea}
\affiliation{Oden Institute for Computational Engineering and Sciences, The University of Texas at Austin, Austin, Texas 78712, USA}

\author{Sabyasachi Tiwari}
\affiliation{Oden Institute for Computational Engineering and Sciences, The University of Texas at Austin, Austin, Texas 78712, USA}
\affiliation{Department of Physics, The University of Texas at Austin, Austin, Texas 78712, USA}

\author{Feliciano Giustino}
\email{Email: fgiustino@oden.utexas.edu}
\affiliation{Oden Institute for Computational Engineering and Sciences, The University of Texas at Austin, Austin, Texas 78712, USA}
\affiliation{Department of Physics, The University of Texas at Austin, Austin, Texas 78712, USA}

\author{Young-Woo Son}
\email{Email: hand@kias.re.kr}
\affiliation{Korea Institute for Advanced Study, Seoul 02455, Korea}

\date{\today}

\begin{abstract}
The {\it ab initio} computational method known as Hubbard-corrected density functional theory (DFT+$U$) captures well ground electronic structures of a set of solids that are poorly described by standard DFT alone.  
Since lattice dynamical properties are closely linked to electronic structures, the Hubbard-corrected density functional perturbation theory (DFPT+$U$) can calculate them at the same level of accuracy. 
To investigate the effects of $U$ on electron-phonon (el-ph) interactions, we implemented DFPT+$U$ with a Hartree-Fock-based pseudohybrid functional formalism to determine $U$ self-consistently and applied our method to compute optical and transport properties of transparent conductive oxides of CdO and ZnO.
For CdO, we find that opening a band gap due to $U$ restores the long-range Fr\"ohlich interaction and that its calculated mobility and absorption spectrum are in excellent agreement with experiments. 
For ZnO where a band gap already appears at the DFT level, DFPT+$U$ brings the results into much closer alignment with experiment, thus demonstrating improved accuracy of our method in dealing with el-ph interactions in these technologically important materials. 
\end{abstract}

\maketitle

\section{Introduction\label{sec:intro}}
Carrier mobility and optical absorption are fundamental characteristics in metals and semiconductors and strongly affected by electron-phonon (el-ph) interactions~\cite{Grimvall1969phys,Giustino2017RMP}. Beyond their scientific interest, these properties play an important role in a wide range of technologies, including energy efficient transistors, solar cells, photodetectors, and transparent electrodes to name a few~\cite{Ginley2010transparent,Ellmer2012np,Schmidt2015chem,Huo2018Advmat}. In quantum simulations, both mobility and optical absorption can be calculated \textit{ab initio} using density functional theory (DFT)~\cite{HK,KS,Ihm1988RPP} and density functional perturbation theory (DFPT)~\cite{Baroni1987prl,Giannozzi1991prb,Gonze1995pra,Gonze1996pra,Baroni2001RMP}. 
These methods have been successfully applied to understand and predict the diverse physical properties of various materials~\cite{Jones2015RMP,Kurt2016Science,Baroni2001RMP,Giustino2017RMP}. 

Notwithstanding the successful applications of DFT and DFPT, the exchange-correlation functionals commonly used in standard DFT calculations, such as the local density approximation (LDA)~\cite{LDApz} and the generalized gradient approximation (GGA)~\cite{PBE}, are known to suffer from self-interaction errors (SIEs)~\cite{Cohen2008Science}. These errors typically lead to an underestimation of the electronic band gap and, in severe cases, result in qualitatively inaccurate ground state properties~\cite{Cohen2008Science,Anisimov1997JPC}. 
To overcome SIE, one widely employed method is DFT corrected with the Hubbard functional (DFT+$U$), inspired by the Hubbard model~\cite{Anisimov1997JPC,Dudarev1998PRB,Himmetoglu2014IJQC, Anisimov1991PRB,Anisimov1993PRB,Liechtenstein1995PRB}. Although SIE generally manifests in localized $d$ and $f$ orbitals, the DFT+$U$ has been extended to cases involving $p$ orbitals successfully~\cite{Korotin2012JPCM,Agapito2015PRX}.

Traditionally, DFT+$U$ employs an empirical $U$, chosen to reproduce structural or electronic properties matching experiments~\cite{Dudarev1998PRB,Anisimov1997JPC,Himmetoglu2014IJQC}. To avoid such empiricism, various methods have been proposed to compute $U$ in a fully self-consistent manner~\cite{Cococcioni2005PRB,Agapito2015PRX,Kulik2006prl,Timrov2018prb,Miyake2008prb,Miyake2009prb,Aichhorn2009prb,Mosey2007prb,Mosey2008jcp}. Among them, the pseudohybrid functional approach developed by Agapito, Curtarolo, and Buongiorno Nardelli (ACBN0) provides a way to evaluate $U$ based on Hartree-Fock (HF) approximations at a moderate computational cost~\cite{Agapito2015PRX}. Furthermore, it has been extended to include inter-site Hubbard interactions~\cite{Campo2010JPCM} self-consistently~\cite{Lee2020PRR}. This method has demonstrated reliable accuracy for electronic and lattice dynamical properties of various materials~\cite{Lee2020PRR,Rubio2020PRB,Yang2021PRB,Yang2022JPC,Yang2024prb,Jang2023PRL} and of those with spin-orbit couplings~\cite{Rubio2020PRB,Yang2024prb}.

Alongside progress in DFT+$U$, a recent development of Hubbard-corrected DFPT (DFPT+$U$) allows for the calculation of lattice dynamics from the DFT+$U$ ground state, which is essential to obtain the accurate el-ph coupling in the late transition-metal monoxides (TMOs)~\cite{Floris2011prb,Floris2020prb}. 
This combination successfully opens the band gap and achieve dynamical stability in TMOs such as CoO~\cite{Floris2020prb,Zhou2021prl}. 
Another notable approach is the linear-response extension of $GW$ approximation~\cite{Hedin1965pr,Hybertsen1986prb,Shishkin2007prl}, known as $GW$ perturbation theory~\cite{Zhenglu2019prl}.
Although it is quite accurate for systems having correlation-enhanced el-ph interactions~\cite{Zhenglu2019prl,Zhenglu2021prl,Zhenglu2024prl}, it still requires significant resources compared to DFPT~\cite{Zhanglu2024CPC}, highlighting the need for cost-effective yet accurate {\it ab initio} methods to calculate el-ph interactions in a wide range of materials.

In this work, we investigate the effects of self-consistent on-site Hubbard interactions on carrier mobility and optical absorption of semiconductors without empirical parameters. To obtain transport and optical properties of semiconductors without SIEs, we combine the ACBN0 functional~\cite{Agapito2015PRX,Lee2020PRR} with the {\it ab initio} Boltzmann transport 
equation ($ai$BTE)~\cite{Restrepo2009apl,Fiorentini2016prb,Ponce2016EPW,ponce2020rpp}.
Our method has a merit in that multiple Hubbard parameters are evaluated simultaneously and self-consistently within DFT routines~\cite{Agapito2015PRX, Lee2020PRR} and that those are integrated into an open-source software for {\it ab initio} el-ph interactions~\cite{Lee2023npj}.
To demonstrate applicability of our method, 
we compute the drift and Hall mobility as well as phonon-assisted optical absorption of transparent conductive oxides (TCOs) such as zinc oxide (ZnO) and cadmium oxide (CdO) and obtain good agreement with experiments.
 We anticipate that the efficiency and accuracy of our method without empirical parameters could play important roles not only in the proper understanding of the transport characteristics of semiconductors but also in the construction of reliable databases for data-driven materials sciences.

This paper is organized as follows: in Sec.~\ref{sec:DFPTU}, we provide a brief review of ACBN0 functional, DFPT+$U$ and {\it ai}BTE along with other related formalisms. 
The computational parameters used in the calculations are presented in Sec.~\ref{sec:details}. In Sec.~\ref{sec:comp_rlt}, we show the results for CdO and ZnO. Finally, we conclude in Sec.~\ref{sec:con} with discussions on future developments in the present methods.   

\section{Formalisms \label{sec:DFPTU}}

In DFT+$U$, the total energy is expressed as,
\begin{equation}
E_\textrm{tot}=E_\textrm{DFT}+E_\textrm{Hub},
\label{Eq:tot}
\end{equation}
where $E_\textrm{DFT}$ can be any local or semilocal density functional and  $E_\textrm{Hub}$ is the Hubbard energy. The latter term is expressed in a rotationally invariant form with $U_\textrm{eff}\equiv U-J$ excluding double 
countings~\cite{Dudarev1998PRB,Himmetoglu2014IJQC},
\begin{eqnarray}
E_\textrm{Hub} &=&\frac{1}{2}\sum_I \sum_{i,j} U_{\textrm{eff}}^I (\delta_{ij}-n^{I\sigma}_{ij})n^{I\sigma}_{ji},
\label{Eq:Hub}
\end{eqnarray}
where the occupation matrix is written as
\begin{eqnarray}
n^{I\sigma}_{ij}&\equiv& n^{I,n,l,\sigma}_{ij}\nonumber \\
&=& \sum_{m{\bf k}}f^0_{m\bf k}
\langle \psi^{\sigma}_{m{\bf k}} |\phi^{I,n,l}_i \rangle \langle \phi^{I,n,l}_j |\psi^{\sigma}_{m{\bf k}}\rangle \nonumber\\
&\equiv& \sum_{m{\bf k}}f^0_{m\bf k}
\langle \psi^{\sigma}_{m{\bf k}} |\phi^{I}_i \rangle \langle \phi^{I}_j |\psi^{\sigma}_{m{\bf k}}\rangle.
\label{Eq:occp}
\end{eqnarray}
Here $f^0_{m\bf k}$ is the Fermi-Dirac function of 
the Bloch state $|\psi^{\sigma}_{m\bf k}\rangle$ 
of the $m$-th band at a momentum ${\bf k}$.
For the $I$-th atom, the principal ($n$), azimuthal ($l$), angular ($i$) and spin ($\sigma$) quantum numbers  
are assigned and the first two numbers are implicitly written, that will be used hereafter. We use the L\"{o}wdin orthonormalized atomic orbital (LOAO)~\cite{Lowdin1950jcp} for a projector $|\phi^{I}_i \rangle$.

To obtain the self-consistent Hubbard parameters, we follow the ACBN0 functional~\cite{Agapito2015PRX},
\begin{align}
    E_{\rm ACBN0}=\frac{1}{2}\sum_{ijkl,\sigma\sigma'}P^{I\sigma}_{ij}P^{I\sigma'}_{kl}\left[(ij|kl) -\delta_{\sigma\sigma'}(ik|jl)
    \right],
    \label{Eq:ACBN0}
\end{align}
where the renormalized occupation matrices are $P^{I\sigma}_{ij}= \sum_{m{\bf k}}
f_{n\bf k} N^{I\sigma}_{{m{\bf k}}}
\langle \psi_{m{\bf k}}^\sigma |\phi^I_i \rangle \langle \phi^I_j |\psi_{m{\bf k}}^\sigma\rangle$, the renormalized occupation numbers are $N^{I\sigma}_{m{\bf k}}= \sum_{\{I\}}\sum_{i} \langle \psi_{m{\bf k}}^\sigma |\phi^{I}_i \rangle \langle \phi^{I}_i |\psi_{m{\bf k}}^\sigma\rangle$ and the bare Coulomb interaction of $(ij|kl)$ between electrons on atomic orbitals of $I$-th atom are 
$
(ik|jl) \equiv\int d{\bf r}_1 d{\bf r}_2 
\frac{\phi^{I*}_i ({\bf r}_1)\phi^I_k({\bf r}_1)\phi^{I*}_j ({\bf r}_2)\phi^I_l ({\bf r}_2)}{|{\bf r}_1-{\bf r}_2|}
$.
Here $\{I\}$ denotes all the orbitals sharing the same quantum numbers $n$ and $l$ of $I$-th atom such that $N^{I\sigma}_{{m{\bf k}}}$ can be viewed as the L\"{o}wdin charge of atom $I$ decomposed for each Bloch state. The first and second terms in Eq.~\eqref{Eq:ACBN0} correspond to Hartree and Fock terms, respectively. By comparing Eq.~\eqref{Eq:ACBN0} with Dudarev's formulation of Hubbard interaction~\cite{Dudarev1998PRB}, we obtain functionals for $U^I$ and $J^I$, respectively~\cite{Agapito2015PRX}:
\begin{align}
    U^I &=\frac{\sum_{ijkl,\sigma\sigma'}P^{I\sigma}_{ij}P^{I\sigma'}_{kl}(ij|kl)}{\sum_{i\neq j,\sigma}  n^{I\sigma}_{ii}n^{I\sigma}_{jj}
+\sum_{ij,\sigma}n^{I\sigma}_{ii}n^{I-\sigma}_{jj}}, \label{ACBN0_U}\\
    J^I &=\frac{\sum_{ijkl,\sigma}P^{I\sigma}_{ij}P^{I\sigma}_{kl}(ik|jl)}{\sum_{i\neq j,\sigma}n^{I\sigma}_{ii}n^{I\sigma}_{jj}}\label{ACBN0_J}.
\end{align}
We note that, compared to the Anisimov's formulation of DFT+$U$~\cite{Anisimov1997JPC}, the prefactor $P^{I\sigma}_{ij}P^{I\sigma'}_{kl}$ of bare Coulomb interactions in Eq.~\eqref{Eq:ACBN0} can be interpreted as the position-dependent mixing parameters of the HF energy.

From Eqs.~\eqref{Eq:tot},~\eqref{Eq:Hub} and~\eqref{Eq:occp}, the generalized Kohn-Sham (KS) equation can be obtained as
\begin{eqnarray}
\left[-\frac{1}{2}\nabla ^2+\widehat{V}_{\rm KS}^\sigma+\widehat{V}_{\rm Hub}^\sigma\right]|
\psi_{n\bf{k}}^\sigma\rangle=\epsilon_{n\bf{k}}^\sigma|\psi_{n\bf{k}}^\sigma\rangle, 
\label{Eq:KS_Hamiltonian}
\end{eqnarray}
where $\widehat{V}_{\rm KS}^\sigma$ is the KS potential, $\widehat{V}_{\rm Hub}^\sigma$ the Hubbard potential, and $\epsilon_{n\bf{k}}^\sigma$ the KS eigenvalues. We use atomic unit, $m_e = \hbar = k_B =1$ where $m_e$ is an electron mass, $\hbar$ Planck constant and $k_B$ Boltzmann constant.
The Hubbard potential is written as
\begin{eqnarray}
\widehat{V}_{\rm Hub}^\sigma = \sum_{I,i,j} U_{\textrm{eff}}^I 
\left(\frac{\delta_{ij}}{2}-n^{I\sigma}_{ij}\right)|\phi^{I}_j \rangle \langle \phi^{I}_i |,
\label{Eq:v_hub}
\end{eqnarray}
where $U_{\textrm{eff}}^I\equiv U^I - J^I$ obtained from Eqs.~\eqref{ACBN0_U} and~\eqref{ACBN0_J}.

Having discussed self-consistent DFT+$U$, we now present calculation methods to evaluate drift mobility~\cite{Ponce2021prr} and phonon-assisted optical absorption~\cite{Tiwari2024prb}.
The key ingredient for them is the matrix element of the el-ph interaction $g_{mn\nu}({\bf{k}},{\bf{q}})$, which represents the probability amplitude for an electron to scatter from an initial Bloch state  $|\psi_{n\bf{k}}^\sigma\rangle$ to a final state $|\psi_{m\bf{k}+\bf{q}}^\sigma\rangle$ by absorbing or emitting a phonon with branch index $\nu$ and momentum ${\bf q}$~\cite{Giustino2007prb}. 

A simplified matrix element can be written as ~\cite{Zhou2021prl},
\begin{equation}
   g_{mn\nu}({\bf{k}},{\bf{q}}) \equiv g^{\rm KS}_{mn\nu}({\bf{k}},{\bf{q}}) + g^{\rm Hub}_{mn\nu}({\bf{k}},{\bf{q}})
   \label{Eq:elphmat}
\end{equation}
where each term can be expressed as,
\begin{eqnarray}
   g^{\rm KS}_{mn\nu}({\bf{k}},{\bf{q}}) &=&
   \langle\psi_{m\bf{k}+\bf{q}}|
   \Delta_{{\bf q}\nu}\widehat{V}^{\rm KS}
   |\psi_{n\bf{k}}\rangle, 
   \label{Eq:elphmat_KS}\\
   g^{\rm Hub}_{mn\nu}({\bf{k}},{\bf{q}}) &=&
   \langle\psi_{m\bf{k}+\bf{q}}|
   \Delta_{{\bf q}\nu}\widehat{V}^{\rm Hub}
   |\psi_{n\bf{k}}\rangle. 
   \label{Eq:elphmat_Hub}
\end{eqnarray}
Here,  
$\Delta_{{\bf q}\nu}\widehat{V}^{\rm KS}$ and $\Delta_{{\bf q}\nu}\widehat{V}^{\rm Hub}$ are the perturbations to the KS and Hubbard potentials by a collective ionic displacement corresponding to a phonon. 
The $g^{\rm KS}_{mn\nu}$ from KS potentials in Eq.~\eqref{Eq:elphmat_KS} can be obtained from DFPT~\cite{Giustino2017RMP,Baroni2001RMP}. Similarly, assuming $\Delta_{\bf q\nu}U^I_{\rm eff}=0$,  
evaluating Eq.~\eqref{Eq:elphmat_Hub} for the Hubbard potential 
requires computing $\Delta_{\bf q\nu}n^I_{ij\sigma}$ and $|\Delta_{\bf q\nu}\phi^I_{j}\rangle$ in addition to the usual $|\Delta_{\bf q\nu}\psi_{n{\bf k}}\rangle$. 
All of these can be obtained from DFPT+$U$~\cite{Floris2011prb,Floris2020prb,Zhou2021prl}. 

Once all conventional and Hubbard responses have been evaluated, one can construct the dynamical matrix to compute phonon dispersions that include the effects of self-consistent $U$~\cite{Floris2011prb,Floris2020prb}.
To compute the force constants originated from the Hubbard potential, the second order derivatives of the projectors, $|\partial_{\bf q\nu}\partial_{\bf q'\nu'}\phi^I_i\rangle$, needs to be evaluated~\cite{Floris2020prb}. 
Since we use the LOAO $|\phi^I_i\rangle$~\cite{Lowdin1950jcp} as projectors for $U$, additional computational procedures for a part of the dynamical matrix are required to handle the first and second derivative of $1/\sqrt{\mathcal O}$. Here $\mathcal O$ is the overlap matrix converting nonorthormalized orbitals ($\bar{\phi}^J_j$) to the LOAO projectors such as
$|\phi^I_i \rangle=\sum_{J,j}(1/{\sqrt{\mathcal O}})^{JI}_{ji}|\bar{\phi}^J_j\rangle
$~\cite{Timrov2020prb,Wu2006jpca}.
Though the first derivative of ${\mathcal O}^{-1/2}$ with respect to phonon displacements
can in principle be solved using the Lyapunov (or Sylvester) operator approach~\cite{Timrov2020prb,Wu2006jpca}, 
we omit the derivative terms of ${\mathcal O}^{-1/2}$ to avoid additional complexity. We find that the errors from this omission are minimal as will be discussed later. We also note that the remaining non-analytic part of the dynamical matrix can be computed using DFPT+$U$~\cite{Floris2011prb, Floris2020prb} with self-consistent $U$ values.

Since the computationally intensive parts of Eq.~\eqref{Eq:elphmat_Hub} for the Hubbard potential 
are related to phonons, we follow similar procedures for the ionic part of the KS potentials~\cite{Giustino2007prb,Ponce2016EPW} to reduce computational burdens by confining its calculations to inside and boundary of the irreducible wedge of Brillouin zone (BZ) for phonon momentum of $\bf q$.
Following the convention defined previously~\cite{Maradudin1968rmp}, a symmetry operation $\{\mathcal{S}|\bf{v}\}$ on a real space point $\bf{r}$ is $\{\mathcal{S}|\bf{v}\}\bf{r}=\mathcal{S}\bf{r}+\bf{v}$, with $\mathcal{S}$ the rotation part and $\bf{v}$ the possible fractional translation. Since the Hubbard energy in Eq.~\eqref{Eq:Hub} is expressed in a rotationally invariant form, the Hubbard potential in Eq.~\eqref{Eq:v_hub} is invariant under $\mathcal{S}$ such that 
$
\Delta_{\mathcal{S}{\bf{q}}\nu}\widehat{V}^{\rm Hub}({\bf{r}})
=\Delta_{{\bf{q}}\nu}\widehat{V}^{\rm Hub}
(\{\mathcal{S}|{\bf{v}}\}^{-1}{\bf{r}})$.
With this, when the momentum transfer is $\mathcal{S}{\bf{q}}$,
the el-ph matrix elements for $U$ can be written as

\noindent
\begin{align}
&g^{\rm Hub}_{mn\nu}({\bf{k}},\mathcal{S}{\bf{q}})= \nonumber\\
&
\langle\psi_{m{\bf{k}}+\mathcal{S}{\bf{q}}}(\{\mathcal{S}|{\bf{v}}\}{\bf{r}})|\Delta_{{\bf q}\nu}\widehat{V}^{\rm Hub}({\bf{r}})|\psi_{n\bf{k}}(\{\mathcal{S}|{\bf{v}}\}{\bf{r}})\rangle,
\label{Eq:nosHub}
\end{align}
where we use a phonon frequency $\omega_{\mathcal{S}{\bf{q}}\nu}=\omega_{{\bf{q}}\nu}$ and 
apply a change of spatial variable, $\{\mathcal{S}|{\bf{v}}\}^{-1}{\bf{r}}\rightarrow {\bf r}$.
We may consider a further simplification such as 
$
\psi_{m{\bf{k}}+\mathcal{S}{\bf{q}}}(\{\mathcal{S}|{\bf{v}}\}{\bf{r}})=\psi_{m\mathcal{S}^{-1}{\bf{k}}+{\bf{q}}}({\bf{r}})
$ but this does not hold in electronic structure codes unless the gauge-fixing procedure has been performed. 
Furthermore, to consider systems without inversion symmetry, we could take advantage of time-reversal symmetry to reduce the required $\bf q$-points on the coarse grid~\cite{Ponce2016EPW}.

Following the similar procedure introduced to the ionic part for KS potential, we utilize the completeness of the plane-wave (PW) basis $\sum_{\bf G}|{\bf k}+{\bf G}\rangle\langle{\bf k}+{\bf G}|=1$ and use the relationship of 
$\langle {\bf k}+{\mathcal S}{\bf q}+{\bf G}|{\bf r}\rangle
=
\langle {\mathcal S}^{-1}{\bf k}+{\bf q}+{\bf G}|{\bf r}\rangle$, then we can rewrite Eq.~\eqref{Eq:nosHub} 
in time reversed version, which was implemented in the code~\cite{Ponce2016EPW,Lee2023npj}:

\begin{align}
g^{\rm Hub}_{mn\nu}({\bf{k}},&-\mathcal{S}{\bf{q}}) 
=
\sum_{\bf{GG'}} 
\langle \psi_{m{\bf{k}}-\mathcal{S}{\bf{q}}}(\{\mathcal{S}|{\bf{v}}\}{\bf{r}})|{\bf{k}-\mathcal{S}\bf{q}+\bf{G}} \rangle \nonumber\\
&\times\langle \mathcal{S}^{-1}{\bf{k}}-{\bf{q}}+{\bf{G}}|\Delta_{-{\bf q}\nu}\widehat{V}^{\rm Hub}({\bf{r}})|\mathcal{S}^{-1}{\bf{k}}+{\bf{G'}}\rangle \notag\\ 
& \times 
\langle {\bf{k}}+{\bf{G'}}|\psi_{n{\bf{k}}}(\{\mathcal{S}|{\bf{v}}\}{\bf{r}})\rangle. \label{Eq:tHub}
\end{align}
Then, one can obtain $g^{\rm Hub}_{mn\nu}({\bf k},{\bf q})$ only using the coarse mesh for phonon momentum $\bf q$ within the irreducible BZ.

Having prepared the key quantities, we can compute phonon-limited drift mobility  ($\mu^d_{\alpha\beta}$) along the $\alpha$-direction based on the linear response of the current density with respect to the electric field of ${\bf E}$ along the $\beta$-direction ($E_\beta$)~\cite{ponce2020rpp,Ponce2021prr}
with and without self-consistent $U$. The $\mu^d_{\alpha\beta}$ can be written as
\begin{equation}
    \mu^d_{\alpha\beta}=\left|
    \frac{1}{n_{\rm c} V_{\rm uc}}
    \sum_n
    \int \frac{d^3 k}{\Omega_{\rm BZ}}
    v^\alpha_{n\bf k}\partial_{E_\beta} f_{n\bf k}
    \right|,
\end{equation}
where $f_{n\bf k}$ is the carrier distribution function, $V_{\rm uc}$ and $\Omega_{\rm BZ}$ volume of unitcell and BZ, 
$v^\alpha_{n\bf k}\equiv\partial\epsilon_{n\bf k}/\partial k_\alpha$
and $\partial_{E_\beta} f_{n\bf k}\equiv\partial f_{n\bf k}/\partial E_\beta |_{{\bf E} =0}$.
Here, $n_{\rm c}$ is charge carrier density defined as 
$n_{\rm c}=\frac{1}{V_{\rm uc}}
\sum_{n}\int \frac{d^3 k}{\Omega_{\rm BZ}}
[f^0_{n\bf k}(\mu, T)-f^0_{n\bf k}(\epsilon_F,0)]$ where 
$\mu$ chemical potential, $T$ temperature and $\epsilon_F$ the Fermi level. The sum of the band index ($n$) is performed for all conduction bands or valence bands depending on the carriers (electrons or holes). 
To obtain  $\partial_{E_\beta} f_{n\bf k}$, we need to solve {\it ai}BTE given as follows~\cite{ponce2020rpp},
\begin{align}
    ev^\beta_{n\bf k} \frac{\partial f^0_{n\bf k}}{\partial \epsilon_{n\bf k}}=
    \sum_m \int \frac{d^3 q }{\Omega_{\rm BZ}}\left[
\frac{\partial _{E_\beta}f_{n{\bf k}}}{\tau^{\rm ph}_{n{\bf k}\rightarrow m{\bf k}+{\bf q}}}
    -
\frac{\partial _{E_\beta}f_{m{\bf k}+{\bf q}}}{\tau^{\rm ph}_{m{\bf k}+{\bf q}\rightarrow n{\bf k} }}
    \right]
    \label{Eq:aiBTE}.
\end{align}
Here, the partial scattering rate from the KS state $|\psi_{n\bf k}\rangle$ to 
$|\psi_{m{\bf k}+{\bf q}}\rangle$ with el-ph scattering can be written as 
\begin{align}
    &\frac{1}{\tau^{\rm ph}_{n{\bf k}\rightarrow m{\bf k}+{\bf q}}}
    =2\pi\sum_{\nu} |g_{mn\nu}({\bf k},{\bf q})|^2\nonumber \\
   &~~~~~~~~~~\times \big[(n_{\bf q \nu}+1-f^0_{m{\bf k}+{\bf q}})
   \delta(\epsilon_{n\bf k}-\epsilon_{m{\bf k}+{\bf q}}-\omega_{\bf q\nu})\nonumber \\
   &~~~~~~~~~~ +(n_{\bf q \nu}+f^0_{m{\bf k}+{\bf q}})
    \delta(\epsilon_{n\bf k}-\epsilon_{m{\bf k}+{\bf q}}+\omega_{\bf q\nu})
    \big],
    \label{Eq:tau}
\end{align}
where $n_{\bf q\nu}$ is the Bose-Einstein distribution evaluated at $\omega_{\bf q \nu}$. 
By swapping $n\bf k$ and $m{\bf k}+{\bf q}$ in Eq.~\eqref{Eq:tau}, one can calculate $1/\tau^{\rm ph}_{m{\bf k}+{\bf q}\rightarrow n{\bf k}}$ in Eq.~\eqref{Eq:aiBTE}.
We note that to obtain the mobility with $U$, the full el-ph matrix element in Eq.~\eqref{Eq:elphmat} is plugged into Eq.~\eqref{Eq:tau} with eigenvalues of generalized KS equation in Eq.~\eqref{Eq:KS_Hamiltonian} and phonon frequencies from DFPT+$U$ while the conventional mobility can be obtained with Eq.~\eqref{Eq:elphmat_KS} and $\omega_{\bf q \nu}$ from DFPT. We also note that to include charge impurity scattering, the partial scattering rate in Eq.~\eqref{Eq:aiBTE} is replaced with $1/\tau^{\rm ph}_{n{\bf k}\rightarrow m{\bf k}+{\bf q}}+1/\tau^{\rm imp}_{n{\bf k}\rightarrow m{\bf k}+{\bf q}}$ as implemented in EPW code~\cite{Leveillee2023prb,Lee2023npj}. As discussed before~\cite{Leveillee2023prb},
the charge impurity scattering rate is proportional to $N_{\rm imp}Z^2_{\rm imp}$ where $N_{\rm imp}$ is the number of impurities per crystal unit cell and $Z_{\rm imp}$ the charge of each impurity so that, by adjusting those parameters, the {\it ab initio} drift mobility including effects from both Hubbard repulsion of $U$ and appropriate ionized impurity concentration can be computed.

For the phonon-limited Hall mobility of $\mu^{H}_{\alpha\beta\gamma}$, we assume a very weak magnetic field ($\bf B$) along the $\gamma$-direction such that its effects on electronic and vibrational properties are neglected~\cite{Ponce2021prr,Lee2023npj}. With this assumption, the additional term of $e({\bf v}_{n\bf k}\times{\bf B})\frac{\partial}{\partial\bf k}\partial_{E_\beta}f_{n\bf k}$ adds to the right side of Eq.~\eqref{Eq:aiBTE} and then we can solve {\it ai}BTE for $\partial_{E_\beta}f_{n\bf k}({\bf B})$ with the same scattering rates in Eq.~\eqref{Eq:tau} in the limit of ${\bf B}\rightarrow 0$~\cite{Ponce2021prr,Lee2023npj}. Thus, for $\mu^{H}_{\alpha\beta\gamma}$, one also can compare the cases with and without $U$.

Finally, the optical absorption $\alpha(\omega)$ is computed as
\begin{equation}
\alpha(\omega) = \frac{\omega\varepsilon_{2}(\omega)}{cn(\omega)},
\end{equation}
where $\omega$ is the photon frequency, $\varepsilon_{2}(\omega)$ the imaginary part of the dielectric function, $c$ the speed of light, and $n(\omega)$ the real part of the refractive index~\cite{Tiwari2024prb,GiustinoBook}. Since the $n(\omega)$ can be calculated from  $n^2(\omega)=[\sqrt{\varepsilon_1^2(\omega)+\varepsilon^2_2(\omega)}+\varepsilon_1(\omega)]/2$
where $\varepsilon_1(\omega)$ is the real part of the dielectric function which can be obtained from a Kramers-Kronig transformation of $\varepsilon_2(\omega)$~\cite{GiustinoBook}, all we need to evaluate is $\varepsilon_2(\omega)$. When a photon can promote direct and phonon-assisted indirect absorption simultaneously, the standard theory of phonon-assisted absorption becomes ill-defined due to the resonance condition, and the oscillator strength diverges~\cite{Lee2023npj,Tiwari2024prb}. A recently developed many-body perturbation theory succeeded in resolving such cases by diagonalizing the electron-phonon Hamiltonian in the quasidegenerate subspace of electron-hole-phonon excitations~\cite{Tiwari2024prb} so that both direct and indirect absorption can be considered simultaneously. By choosing either Eq.~\eqref{Eq:elphmat} or Eq.~\eqref{Eq:elphmat_KS} for el-ph matrix elements in $\varepsilon_2(\omega)$, one can compare direct and phonon-assisted indirect optical absorption with and without $U$ appropriately.

\section{Computational details\label{sec:details}}

First-principles calculations based on DFT were performed using the Quantum ESPRESSO (QE) suite~\cite{Giannozzi2009JPC,Giannozzi2017JPC}. We employed LDA parameterized by Perdew and Zunger (PZ)~\cite{LDApz} for ZnO and GGA by Perdew-Burke-Ernzerhof (PBE) ~\cite{PBE} for CdO. The pseudopotentials are optimized norm-conserving Vanderbilt (ONCV) types, with parameters obtained from the Pseudo Dojo database~\cite{Setten2018CPC,hamann2013prb}. The kinetic energy cutoff for a plane-wave basis set was used at 110 Ry for ZnO and 135 Ry for CdO. The structural optimization was obtained with the criteria for force less than $10^{-5}$ Ry/Bohr, and pressure less than 0.05 kbar. We employed our modified in-house QE package to obtain self-consistent on-site Hubbard parameter $U$ on the oxygen 2$p$ (O-2$p$) states~\cite{Lee2020PRR} using LOAO for projectors as was done in previous studies~\cite{Lee2020PRR, Yang2021PRB}. The converged $U$ values for ZnO and CdO were 2.90 eV and 5.13 eV, respectively. For ZnO, the phonon calculation using the lattice constant optimized with PBE shows a deviation from the experimental results, especially for the optical phonon modes (Appendix~\ref{app:ZnO_PBE}). After comparing the results by varying the calculation conditions, we determined that the optimized lattice constant with LDA is relatively suitable. The optimized lattice parameters of ZnO structure were $\textit{a}=\textit{b}=3.197$~\AA~and $\textit{c}=5.134$~\AA. 
We then applied the Hubbard correction to the 2$p$ states of the oxygen atom for the coarse mesh calculations needed for Wannier interpolation. For CdO, we performed the cell optimization with DFT+$U$ after obtaining the Hubbard parameter $U$ on oxygen 2$p$ states.
To obtain dielectric constants, Born effective charges and phonon frequencies using DFPT~\cite{Baroni2001RMP,Gonze1997PRB}, the BZ is sampled on $9\times9\times9$ and $10\times10\times10$ $q$-point grid for ZnO and CdO, respectively. The same mesh size is used for the coarse $k$-point grid.

The el-ph matrix elements are computed using the EPW code~\cite{Giustino2007prb,Lee2023npj,Ponce2016EPW} with the WANNIER90 in library mode~\cite{Pizzi2020jpcm}. 
To calculate mobility and optical absorption,
the KS Hamiltonian, the dynamical matrix, and the el-ph matrix elements are interpolated using Wannier-Fourier interpolation~\cite{Giustino2007prb,Mostofi2014cpc} onto fine momentum grids.
Specifically, for ZnO, a $144\times144\times144$ $k$- and $q$-point grid is used for electron mobility, and a $90\times90\times90$ $k$- and $q$-point grid is used for hole mobility, while for CdO, a $140\times140\times140$ $k$- and $q$-point grid and an $80\times80\times80$ $k$- and $q$-point grid are employed for electron and hole mobility, respectively. Since the low-energy phonon modes play a crucial role in el-ph scattering in piezoelectric materials~\cite{Lihm2024prb}, we set the acoustic phonon energy cutoff to 0.01 meV for the ZnO calculations.

To check the quality of Wannier interpolations for $g_{mn\nu}$ without having to deal with arbitrary phases, we compare the total deformation potentials for the $\nu$-th phonon mode~\cite{Ponce2021prr} with direct DFPT or DFPT+$U$ results as a function of phonon momentum $\bf q$. The potential is written as 
\begin{equation}
    D^\nu (\Gamma,{\bf q})=\frac{1}{N_w}
    \left [
    2\rho V_{\rm uc} \omega_{\bf q \nu}\sum_{nm}\left|g_{mn\nu}(\Gamma,{\bf q})\right|^2
    \right]^{1/2},
    \label{Eq:DeformPot}
\end{equation}
where $N_w$ is a number of states within Wannier manifold, $\rho$ mass density of the crystal and $\bf k$ set to $\Gamma$.
Here we note that one can check the effects of $U$ easily through comparison between $D^\nu(\Gamma,\bf q)$ including $g^{\rm KS}_{mn\nu}$ in Eq.~\eqref{Eq:elphmat_KS} 
together with $\omega_{\bf q \nu}$ from DFPT and one including additional Hubbard contribution to el-ph matrix element in Eq.~\eqref{Eq:elphmat_Hub} with $\omega_{\bf q\nu}$ from DFPT+$U$.

\section{Results\label{sec:comp_rlt}}

The TCOs possess the simultaneous characteristics of transmitting most of the visible light spectrum and exhibiting high carrier mobility~\cite{Ginley2010transparent}. Due to these properties, they are widely used in optoelectronic industries such as panel displays and in photovoltaic applications such as solar cells, attracting ongoing significant attention~\cite{Ellmer2012np}. Among TCOs, we will consider prototypical examples, ZnO and CdO. CdO has a low resistivity that is only about an order of magnitude larger than the widely used indium tin oxide films in TCO layers~\cite{Badeker1907adp,Minami2005sst}. ZnO has a wide direct band gap of approximately $3.3\sim 3.4$ eV~\cite{Mang1995ssc,Reynolds1999prb,Dong2004prb}. 
It benefits from relatively easy methods for high-quality crystal growth and is environmentally harmless. These advantages have led to extensive research on its practical applications using various dopants~\cite{Minami2002thin,Sato1993thin,Minami1984jjap,Kim2013jmcc,Sun2018acsam,Song2015angchem,Cao2017acsam,Heiba2019jms,Turkyilmaz2017jppa,Joshi2016jac,Lung2017jac,Alexandrov2020scirep}.

\begin{table}[b]
\caption{Values of Hubbard parameters $U_d$ (Cd/Zn $d$ orbitals) and $U_p$ (O $p$ orbitals), and minimal band gap $E_g$, in eV. Experimental gaps are reported in Table~\ref{tabular:CdO_str} for CdO and Table~\ref{tabular:ZnO_str} for ZnO.}
\label{tabular:Hub_para}
\centering
\begin{ruledtabular}
\begin{tabular}{lccc}
                     & $U_d$  & $U_p$ & $E_g$  \\ \hline
\multirow{3}{*}{CdO} & 13.29 & -    & -0.07 \\
                     & 12.36 & 8.65 & 2.67  \\
                     & -     & 5.13 & 0.91  \\ \hline
\multirow{3}{*}{ZnO} & 15.80 & -    & 1.76  \\
                     & 14.71 & 7.86 & 4.20  \\
                     & -     & 2.90 & 1.56 
\end{tabular}
\end{ruledtabular}
\end{table}

Before discussing detailed results, here we briefly summarize our calculation results. 
In both CdO and ZnO, we found that calculations with on-site Hubbard correction to both $d$ and $p$ orbitals overestimate the band gap, as shown in Table~\ref{tabular:Hub_para}. For CdO, the correction for the $d$ orbital alone cannot open the band gap. Instead, the self-consistent $U$ on oxygen $p$ orbitals opens the band gap, and the resulting indirect and direct band gap values agree with the experiments. This correction restores the long-range Fr\"{o}hlich el-ph interaction also. 
In the case of ZnO, while DFT-LDA/GGA yields a band gap, and $U$ on the $d$ orbital increases it, $U$ on $p$ orbital also contributes to increase the gap and modifies the phonon dispersion, agreeing with experimental measurement.
To treat both materials in a similar manner and with minimal additional interactions, we will focus solely on the calculation results that include Hubbard corrections only to the oxygen $p$-orbitals.  
 Then, we performed phonon-assisted optical absorption calculations based on many-body quasidegenerate perturbation theory~\cite{Tiwari2024prb} for both ZnO and CdO. We also include ionized-impurity scattering in the calculation of carrier drift mobility, depending on the carrier concentration~\cite{Lu2022prm,Leveillee2023prb}
and obtain a good agreement with the experiments.

\begin{figure}[b]
	\begin{center}
		\includegraphics[width=0.9\columnwidth]{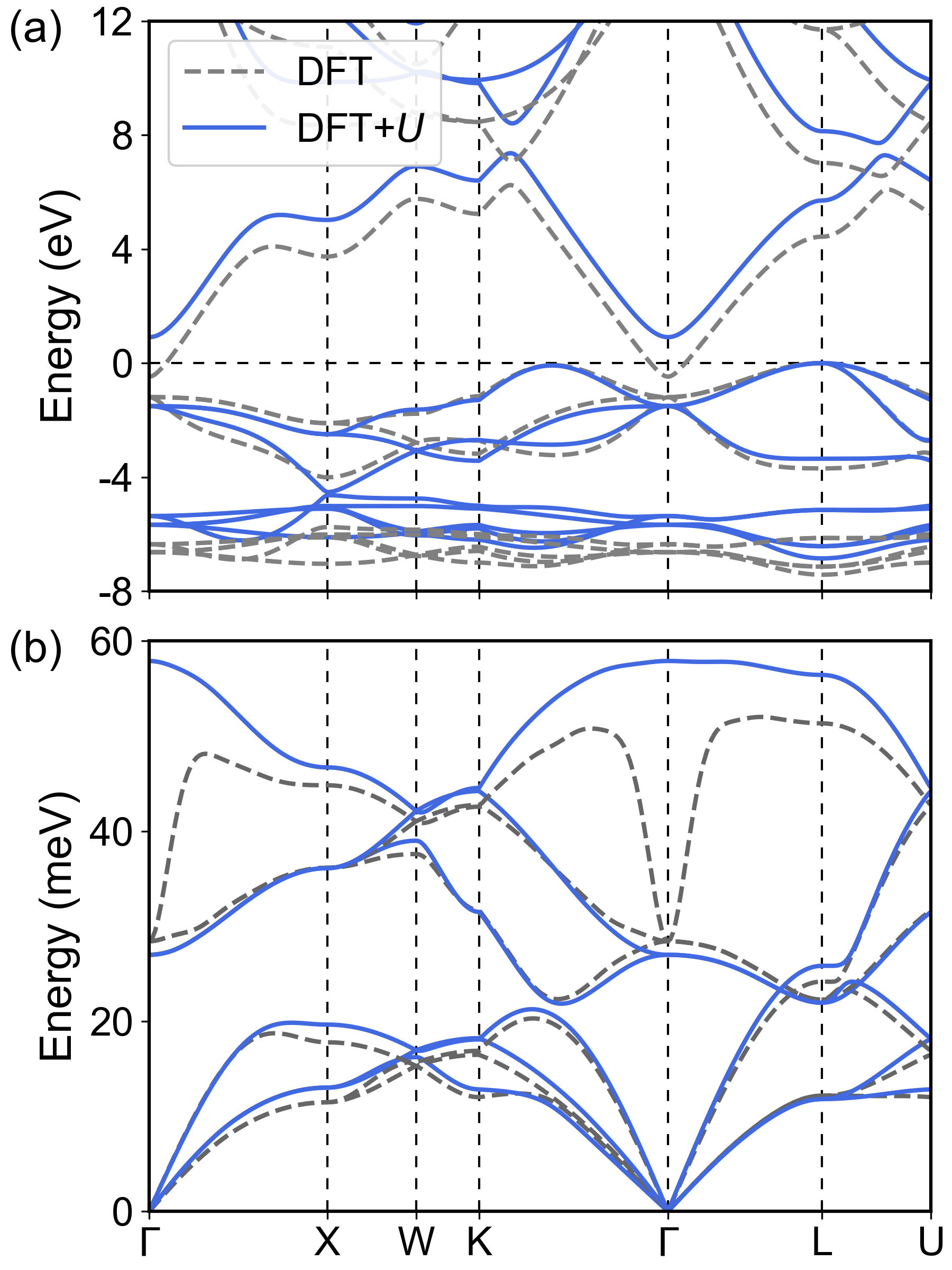}
	\end{center}
	\caption{(a) DFT (dotted grey) and DFT+$U$ (blue) band structures of CdO plotted along the high-symmetry lines of the Brillouin zone, with the valence band maximum set to zero. (b) Phonon band structures of CdO obtained with DFPT (dotted grey and DFPT+$U$ (blue), plotted along the high-symmetry lines of the Brillouin zone.
	}
	\label{fig:CdO_el_ph}
\end{figure}

\subsection{CdO}

Given that the effects of the $U$ corrections are pronounced in CdO, we chose it as the first test case for our method. Here, we used the rock salt structure for CdO. 
The approximate functionals, such as LDA and GGA, due to SIEs~\cite{Cococcioni2005PRB, Cohen2008Science}, yield a metallic ground state for CdO, as illustrated in Fig.~\ref{fig:CdO_el_ph}(a). In contrast, DFT+$U$ calculations applying a Hubbard correction to O-2$p$ state result in an insulating ground state. The calculated structural parameters closely match those obtained using other correction methods and experimental values~\cite{Burbano2011jacs,Dixit2012jpcm,Zuniga2004jcg,Demchenko2010prb,Vasheghani2013apl,King2009prb}, as presented in Table~\ref{tabular:CdO_str}. Moreover, except for the position of the 4$d$ state, both the band gap size and valence band width agree with experiments very well.

\begin{table}[t]
\caption{Lattice parameter of CdO ($a$) in {\AA} and direct ($E^{dir}_g$) and indirect band gap ($E^{ind}_g$). The valence band width ($\Delta_{\rm VB}$) and location of the Cd 4$d$ states with respect to the valence band maximum ($\epsilon_{4d}$). All energies are in eV.}
\label{tabular:CdO_str}
\centering
\begin{ruledtabular}
\begin{tabular}{lccccc}
      & $a$     & $E^{dir}_g$    & $E^{ind}_g$           & $\Delta_{\rm VB}$         & $\epsilon_{4d}$   \\ \hline

DFT   & 4.767     & 0.72    & -0.48          & 4.00          & -6.52          \\
DFT+$U$ & 4.718 & 2.42 & 0.91 & 4.52       & -5.55       \\
HSE06 \footnote{\label{HSE}Reference~\cite{Burbano2011jacs}} & 4.720 & 2.18 & 0.89       & 4.45       & -7.4        \\
$GW$ \footnote{\label{GW}Reference~\cite{Dixit2012jpcm}}    & 4.650 & 2.88 & 1.68       & $\sim$4.75 & $\sim$-9 \\ 
Expt   & 4.695\footnote{\label{exp1}Reference~\cite{Zuniga2004jcg}} & 2.18\footnote{\label{exp2}Reference~\cite{Vasheghani2013apl}}, 2.4\footnote{\label{exp3}Reference~\cite{Demchenko2010prb}}& 0.9\footref{exp3} & 4.8\footnote{\label{exp4}Reference~\cite{King2009prb}}  & -8.80\footref{exp4}  
\end{tabular}
\end{ruledtabular}
\end{table}

Using a newly implemented DFPT+$U$, we compared the phonon band structures from DFT+$U$ calculations with those from standard DFT calculations, as shown in Fig.~\ref{fig:CdO_el_ph}(b). This comparison reveals that, due to the polar semiconductor nature of CdO, the longitudinal optical (LO) and transverse optical (TO) phonon modes split near the $\Gamma$ point as a result of long-range nature of the Coulomb interaction~\cite{Gonze1997PRB}. This splitting significantly alters the dispersion of the LO phonon modes.
As we mentioned earlier, although we do not include the second derivative of the LOAO projectors to compute the dynamical matrix with $U$, the resulting phonon frequencies shown in Figs.~\ref{fig:CdO_el_ph} and~\ref{fig:ZnO_el_ph} (b) agree well with the experimental results~\cite{Hewat1970ssc,Thoma1974ssc,Serrano2010prb}, implying that the errors from such an approximation are not so significant.

\begin{figure}[b]
	\begin{center}
		\includegraphics[width=0.95\columnwidth]{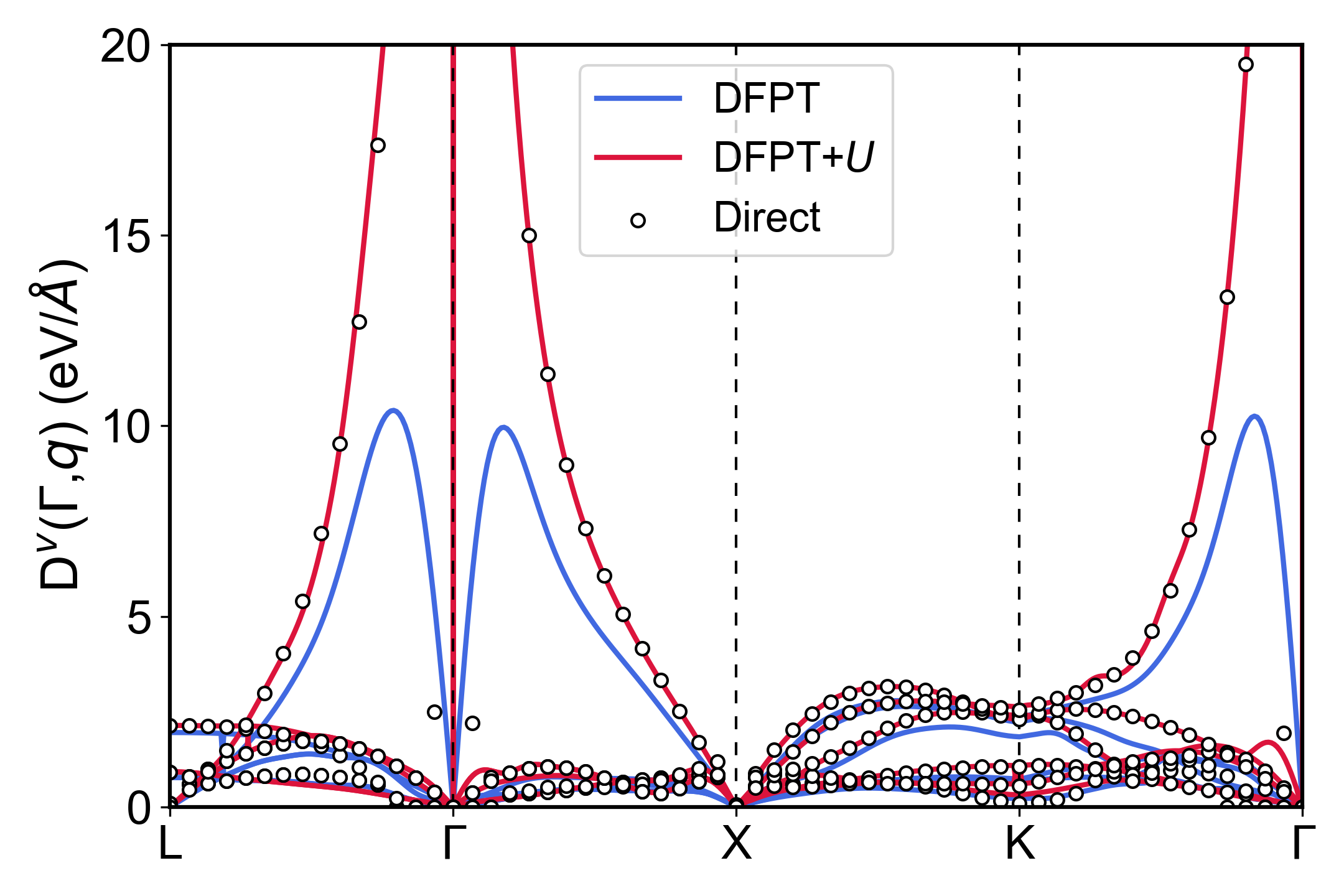}
	\end{center}
        \caption{\label{fig:CdO_elphmat} Comparison of the deformation potential of the valence band manifold for CdO, obtained from the direct DFPT+$U$ calculations (dots) and the Wannier interpolations based on DFT (blue) and DFT+$U$ calculations (red). The \textbf{k} point is set to the $\Gamma$ point. }
\end{figure}

The effect of the band gap opening from $U$ in CdO is also clearly evident in the el-ph matrix elements as shown in Fig.~\ref{fig:CdO_elphmat}. By computing the deformation potential of $D^\nu(\Gamma,q)$ in Eq.~\eqref{Eq:DeformPot} from these matrix elements focusing on the linear variation of the potential~\cite{Ponce2021prr}, we observe that in the DFPT+$U$ results, the LO phonon mode near the $\Gamma$ point diverges due to the Fröhlich interaction. In contrast, in the DFPT results without $U$, it converges to a finite value. Furthermore, we find that the interpolated Hubbard corrections through Eq.~\eqref{Eq:nosHub} agree well with direct computations using DFPT+$U$.

To confirm the optical transparency and simultaneous conductivity of CdO, we first calculated the carrier mobility. As shown in Table~\ref{tabular:CdO_mobility}, the mobility of electrons is more than ten times higher than that of holes. The conduction band near the Fermi energy is predominantly derived from Cd-5$s$ orbitals, which form relatively delocalized and spherical states. These states facilitate efficient electron transport, leading to high electron mobility. In contrast, the valence band near the Fermi energy is dominated by O-2$p$ orbitals. Since these orbitals are more localized and directional, they overlap less effectively, producing less dispersive valence bands. This leads to heavier hole masses and consequently lower mobility, consistent with the data presented in Table~\ref{tabular:CdO_mobility}. As a result, there is the significant difference in mobility between carrier types in CdO.

\begin{table}[t]
\caption{Theoretical phonon-limited drift ($\mu^{d}$), Hall ($\mu^{H}$) mobility (cm$^2$V$^{-1}$s$^{-1}$) and conductivity effective mass ($m^*$, in units of $m_e$) at 300 K of electrons and holes in CdO computed with the DFT+$U$ framework. For comparison, experimental electron mobility ($\mu^{Exp}$) values are taken from Ref.~\cite{Vasheghani2013apl,Adhikari2023tsf}.}
\label{tabular:CdO_mobility}
\centering
\begin{ruledtabular}
\begin{tabular}{ccccc} 
Carrier type & $\mu^d$  & $\mu^H$ & $m^*$ & $\mu^{Exp}$  \\
\hline
Electron    & 275.1  &  308.9 & 0.245 & 71.5-351.2 \\
Hole        & 16.6 &  14.3 & 0.914 & - \\
\end{tabular}
\end{ruledtabular}
\end{table}

\begin{figure}[b]
	\begin{center}
		\includegraphics[width=1\columnwidth]{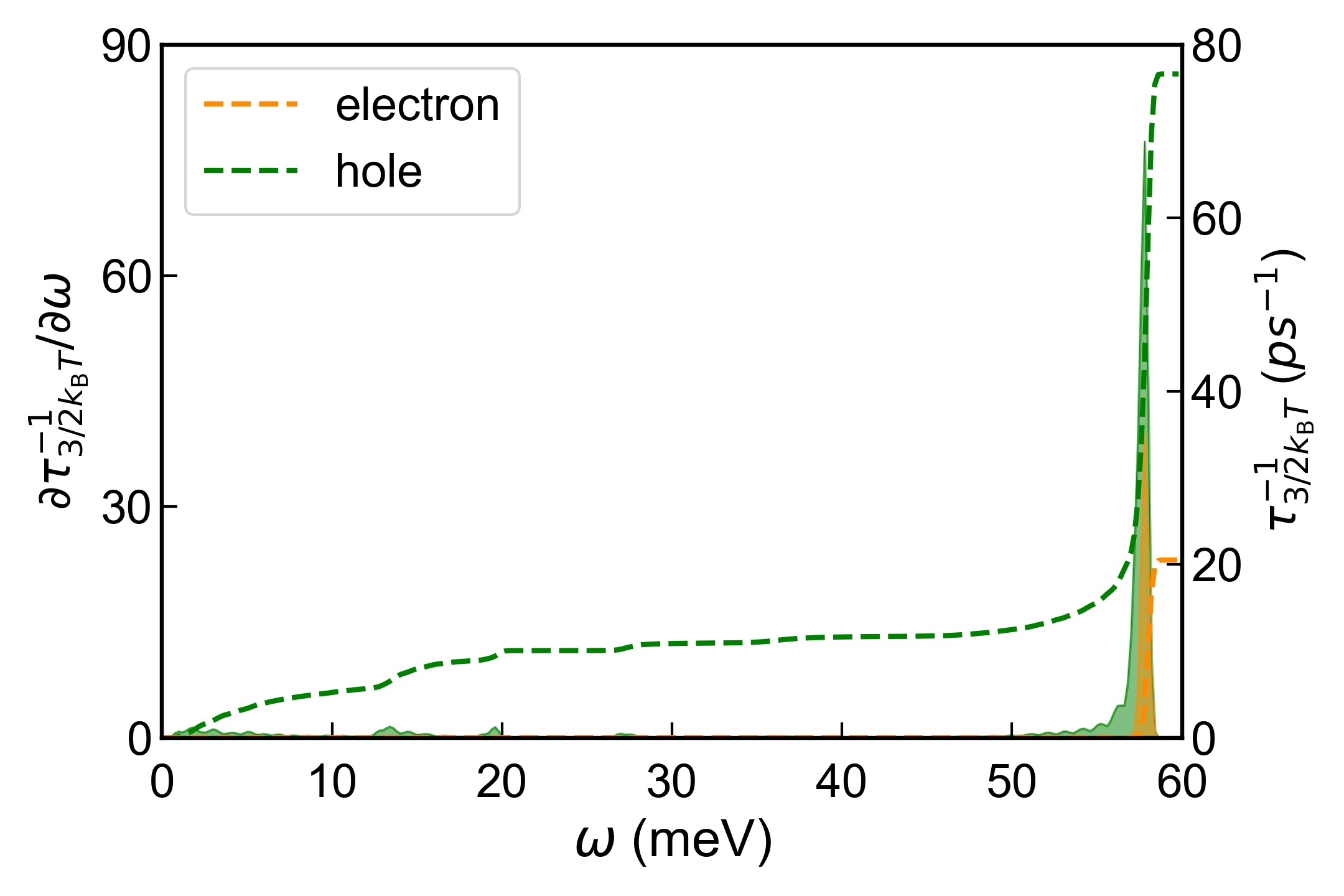}
	\end{center}
        \caption{\label{fig:CdO_spd} At 300 K, the spectral decomposition of electron (orange) and hole (green) scattering rates for CdO in the left ordinate is shown as a function of phonon energy ($\omega$). The rates are computed at an energy of $3k_BT/2$ away from the band edges. The dashed lines indicate the cumulative integrals of the calculated rates, adding up to the carrier scattering rate ($\tau^{-1}_{3/2k_{B}T}$)
        in the right ordinate.}
\end{figure}

\begin{figure}[b]
	\begin{center}
		\includegraphics[width=0.8\columnwidth]{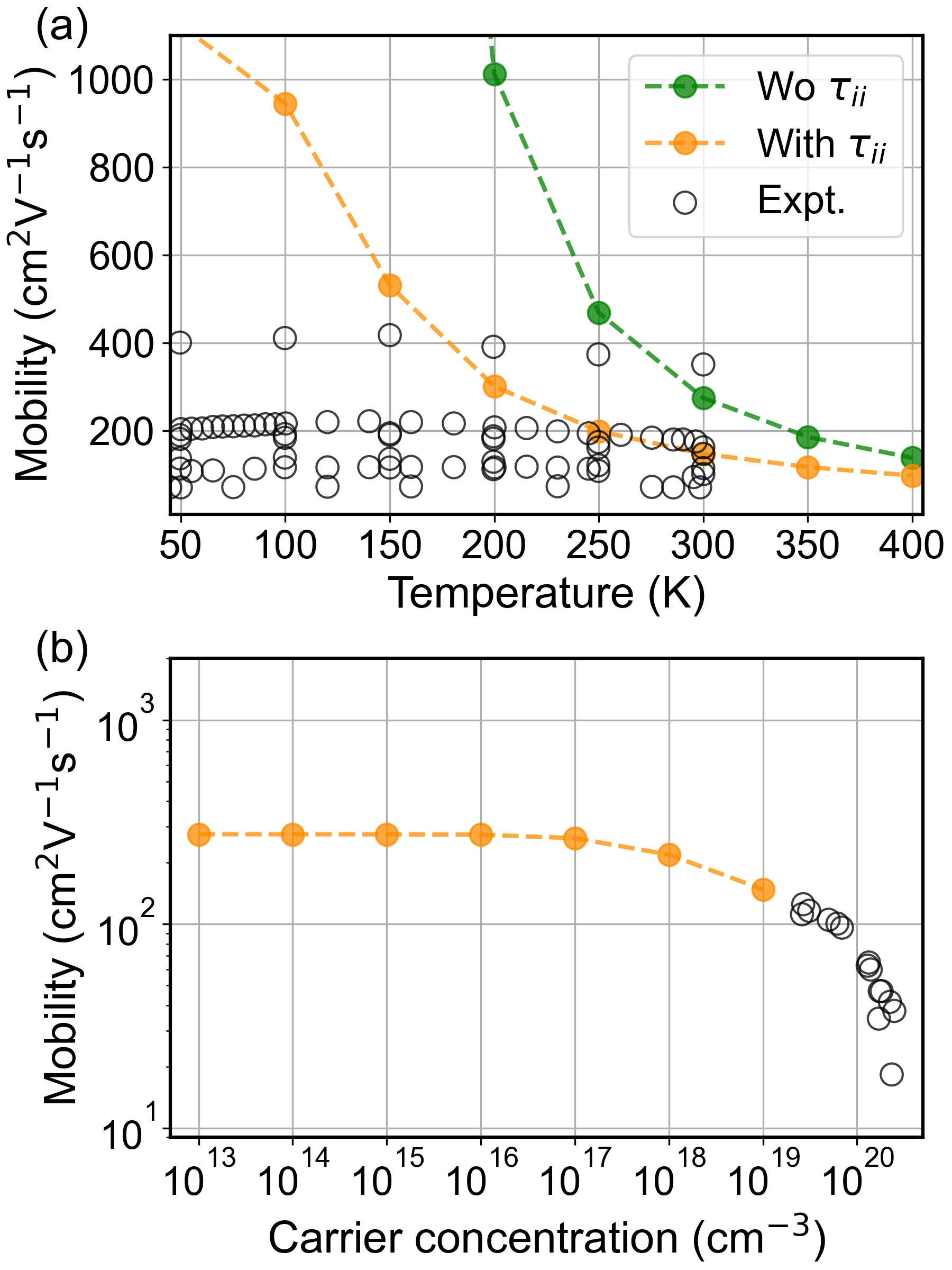}
	\end{center}
        \caption{\label{fig:CdO_mobility} (a) Comparison of the theoretical electron drift mobilities with the experimental values~\cite{Adhikari2023tsf,Vasheghani2013apl} as a function of temperature. The green circles represent the mobility without the ionized-impurity scattering mechanism, while the orange circles include it. Both calculations are basedon a carrier concentration of $10^{18}$ cm$^{-3}$. (b) At 300 K, comparison between the theoretical electron drift mobilities and the experimental values~\cite{Vasheghani2011jap} as a function of carrier concentration.} 
\end{figure}

To identify the phonons that contribute to mobility, we plot the scattering rate as a function of the energy spectrum as shown in Fig.~\ref{fig:CdO_spd}. The calculations correspond to carriers at an energy of $\frac{3}{2}k_BT$ away from the band edge and room temperature, following values commonly used in previous decomposition analyses~\cite{Ponce2019acsel,Ponce2021prr}. We observe that phonons near 60 meV have a large contribution to the scattering rates, indicating that the LO mode plays a significant role for the mobility of CdO. This also highlights the importance of proper treatment of its polar characteristics using $U$.

\begin{figure}[b]
	\begin{center}
		\includegraphics[width=0.8\columnwidth]{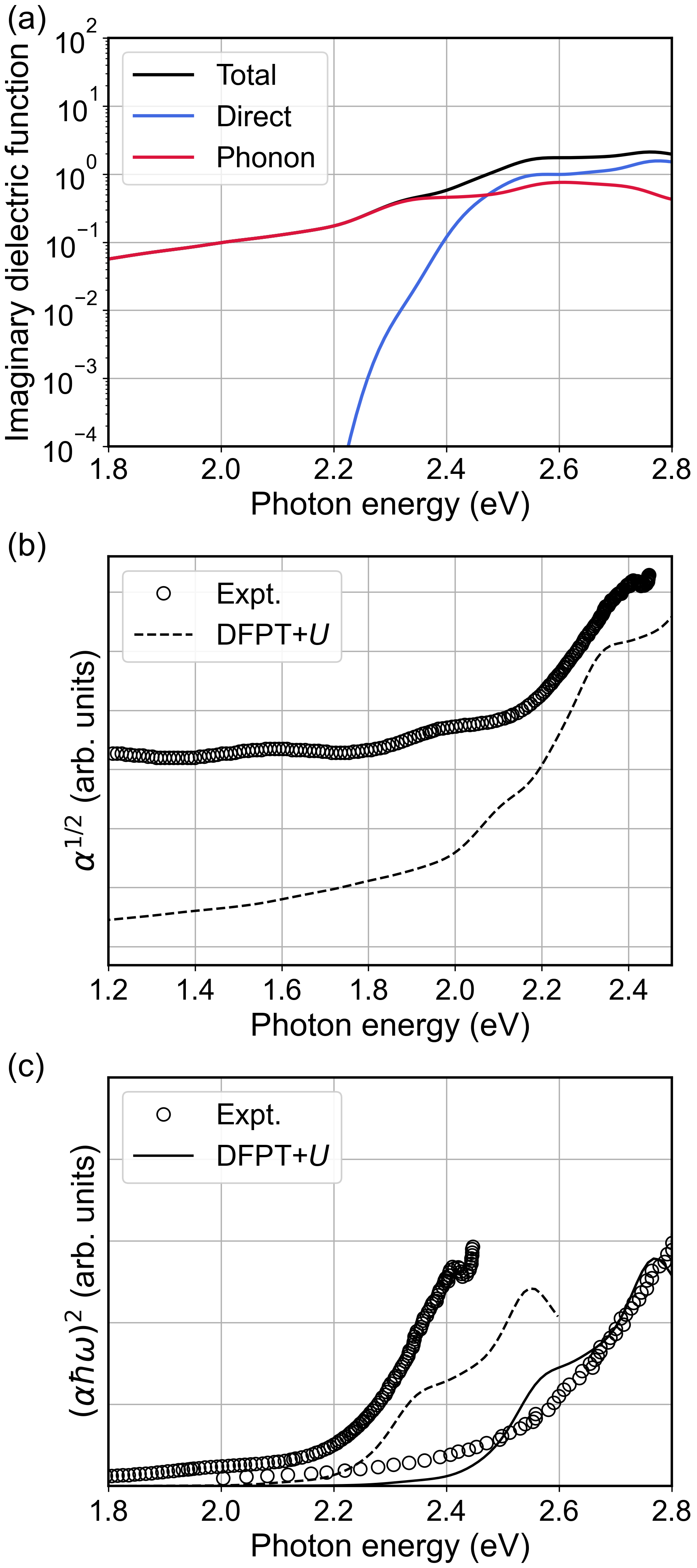}
	\end{center}
        \caption{\label{fig:CdO_optics} (a) The total imaginary dielectric function (black) of CdO is decomposed into contributions from direct transitions (blue) and phonon-assisted transitions (red). (b) Tauc plot for the indirect gap region ($\alpha^{1/2}$) and (c) for the direct gap region ($(\alpha\hbar\omega)^2$) of CdO at 300 K. Open circles denote the experimental values~\cite{Kose2009ijhe,Vasheghani2013apl}. Calculated results are shown both without (solid lines) and with (dotted lines) an applied scissor shift, the latter chosen to align the calculated onset with one of the experimental datasets.}
\end{figure}

Next, we compare the mobility values from our DFPT+$U$ calculations with the experimentally measured ones. As shown in Fig.~\ref{fig:CdO_mobility} (a), for its mobility as a function of temperature, we found that calculations including the impurity scattering mechanism match the experimental trends more closely. Since CdO frequently exhibits strong non-stoichiometry
and features abundant intrinsic defects yielding large carrier concentrations~\cite{Burbano2011jacs,Li2004appliedss}, the experimental measurements are largely limited to samples with high carrier concentration levels. 
So, as can be seen in Fig.~\ref{fig:CdO_mobility}(b), available experimental mobilities are in high carrier concentrations, setting a limitation for a direct one-to-one comparison with our calculations performed across varying carrier concentrations. Unfortunately, our calculation is limited below the carrier concentration of $10^{19}~{\rm cm}^{-3}$, because when the Fermi level crosses the conduction band minimum at higher carrier concentrations, the linear variation of the occupation number induced by the \textbf{E}-field in Eq.~\eqref{Eq:aiBTE} does not converge well~\cite{Leveillee2023prb}.
Notwithstanding this, the extrapolated trend of the computed mobility approaches the measured ones well.
We also note that the mobility of CdO with low carrier concentration is more or less saturated as shown in Fig.~\ref{fig:CdO_mobility}.

To verify transparency, which is another important property of TCOs, we performed optical absorption calculations for CdO. While standard LDA or GGA calculations typically underestimate the band gap, often necessitating a scissor shift for comparison with experiments~\cite{Zacharias2016prb,Tiwari2024prb}, we found that DFT+$U$ significantly improves the electronic structure prediction of  CdO. The direct and indirect band gaps obtained at the DFT+$U$ level are in good agreement with the range of the experimental values (Table~\ref{tabular:CdO_str}). 

Therefore, we first analyze the computed optical absorption based on this improved electronic structure without applying an ad hoc scissor shift. Since the indirect band gap is the minimum gap in CdO, we note that up to photon energies around 2.4 eV, the phonon-assisted contribution is larger than the direct contribution as shown in Fig.~\ref{fig:CdO_optics}(a). Starting from the region where the photon energy matches the direct band gap value, the direct contribution becomes larger. However, the phonon-assisted contribution still provides a contribution of the same order of magnitude, as confirmed in the imaginary part of the dielectric function shown in Fig.~\ref{fig:CdO_optics}(a). 

For a more detailed comparison with specific experimental datasets, which exhibit some variation in the reported band gaps (Table~\ref{tabular:CdO_str}), we generated Tauc plots for direct comparison with experiments by plotting $\alpha^{1/2}$ (indirect) and $(\alpha\hbar\omega)^2$ (direct) against photon energy in Fig.~\ref{fig:CdO_optics}(b) and (c). Our unshifted calculations align well with the onset reported in one data set~\cite{Kose2009ijhe}. For the comparison with the other dataset~\cite{Vasheghani2013apl}, a scissor shift was applied. Despite the variation in experimental onset energies, the overall trends calculated at the DFT+$U$ level are similar to the experimental data~\cite{Kose2009ijhe,Vasheghani2013apl}. On the basis of these findings, we reproduced the result that CdO transmits most of the visible light.

\subsection{ZnO}

\begin{figure}[b]
	\begin{center}
		\includegraphics[width=0.8\columnwidth]{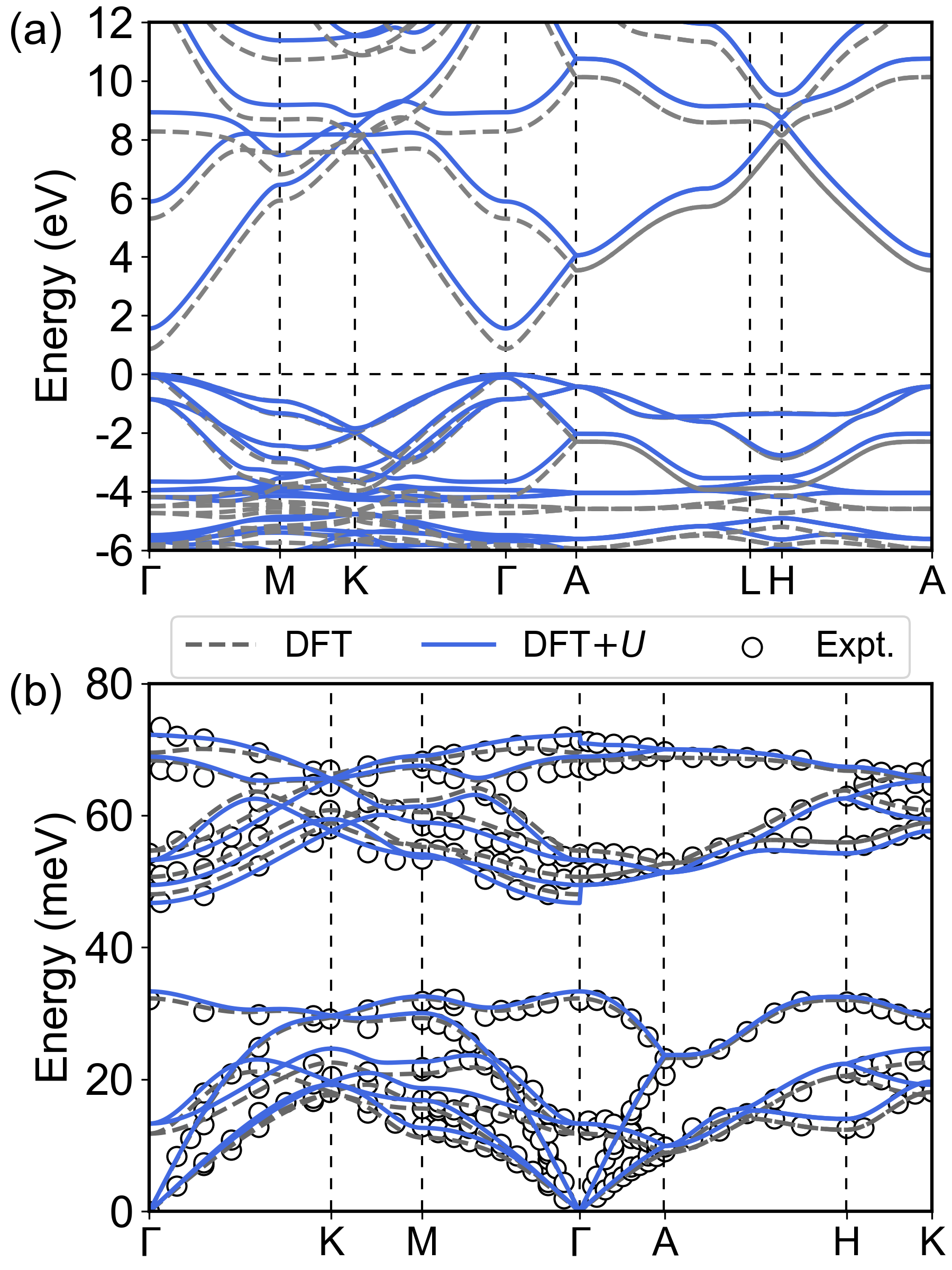}
	\end{center}
	\caption{(a) DFT (dotted grey) and DFT+$U$ (blue) band structures of ZnO plotted along the high-symmetry lines of the Brillouin zone, with the valence band maximum set to zero. (b) Phonon band structures of ZnO obtained with DFPT (dotted grey) and DFPT+$U$ (blue), plotted along the high-symmetry lines of the Brillouin zone. Open circles denote the experimental values~\cite{Hewat1970ssc,Thoma1974ssc,Serrano2010prb}.
	}
	\label{fig:ZnO_el_ph}
\end{figure}

To evaluate whether calculations at the DFT+$U$ level can enhance accuracy  beyond improving the electronic ground state, we applied the Hubbard corrections to compute phonon dispersions as well as electron-phonon interactions, incorporating the self-consistent $U$ parameter on the O-2$p$ state of ZnO. In the case of ZnO, however, the phonon frequencies show a pronounced dependence on the relaxed atomic structures, which vary according to choice of functional (Appendix~\ref{app:ZnO_PBE}). To address this, we performed DFT(+$U$) and DFPT(+$U$) calculations based on the relaxed structure obtained using LDA. Including the Hubbard $U$ correction increased the band gap of ZnO from 0.86 eV to 1.56 eV (Fig.~\ref{fig:ZnO_el_ph}(a)), but this value is still lower than the experimentally measured value of 3.3-3.4 eV~\cite{Mang1995ssc,Reynolds1999prb,Dong2004prb,Chen1998jap}. We note that this situation improves if on-site and inter-site Hubbard interactions are included simultaneously~\cite{Lee2020PRR}.

\begin{figure}[b]
	\begin{center}
		\includegraphics[width=0.95\columnwidth]{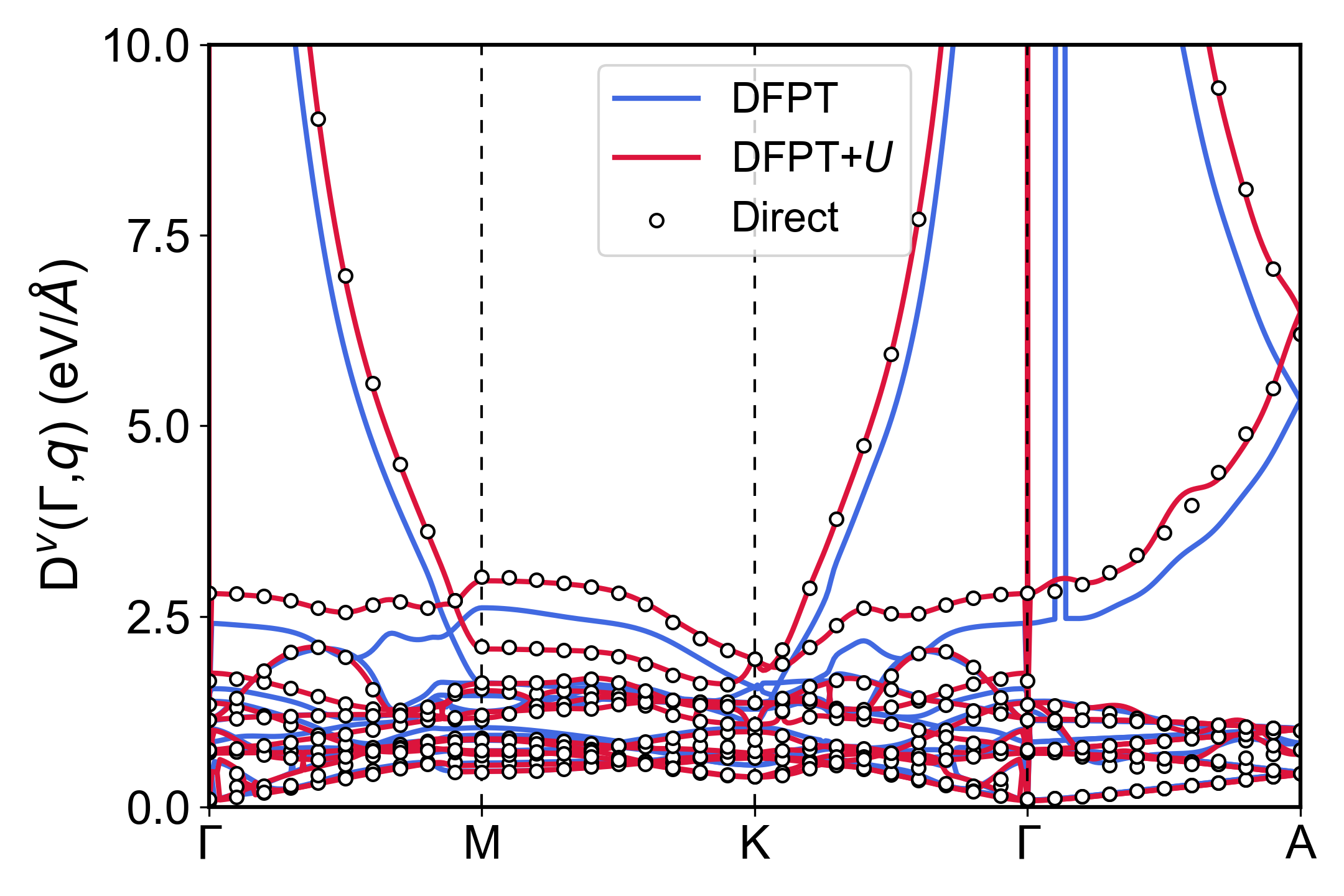}
	\end{center}
        \caption{\label{fig:ZnO_elphmat} Comparison of the deformation potential of the valence band manifold for ZnO, obtained from the direct DFPT+$U$ calculations (dots) and the Wannier interpolations based on DFT (blue) and DFT+$U$ calculations (red). The \textbf{k} point is set to the $\Gamma$ point. }
\end{figure}

Nevertheless, as shown in Fig.~\ref{fig:ZnO_el_ph}(b) for the phonon dispersion of ZnO, the blue solid line obtained from the DFPT+$U$ method 
shows a closer agreement with experimental values (indicated by circle marks) compared to the grey dotted line from DFPT. This improvement is especially evident for the upper optical modes above 40 meV. 
These differences also manifest quantitatively in the deformation potential plots derived from the el-ph matrix, as shown in Fig.~\ref{fig:ZnO_elphmat}. Although the overall trends remain consistent, we observe that the interpolated el-ph matrix calculated at the the DFT+$U$ level consistently exhibits higher values than those obtained from the DFT level, indicating a stronger el-ph interaction.

\begin{figure}[t]
	\begin{center}
		\includegraphics[width=1\columnwidth]{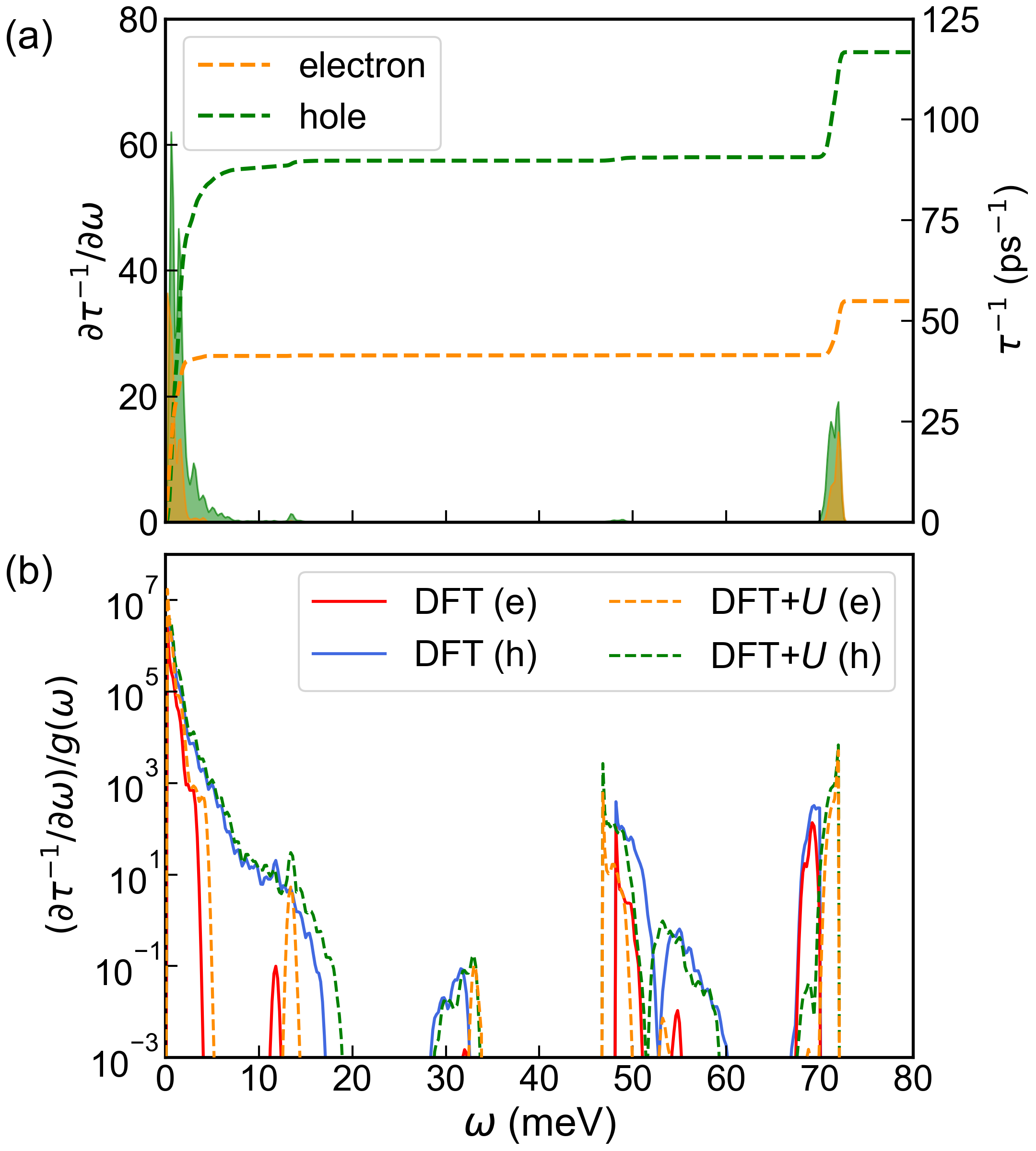}
	\end{center}
        \caption{\label{fig:ZnO_spd} (a) Spectral decomposition of electron (orange area) and hole (green area) scattering rates for ZnO at 300 K, calculated $3k_BT/2$ away from the band edges computed with DFT+$U$ level. The cumulative integral (dashed lines) give the carrier scattering rate $\tau^{-1}_{3/2k_{B}T}$ (right axis). (b) Comparison of scattering rates per state, $[\partial \tau^{-1}/ \partial \omega]/g(\omega)$. Results are shown for electron (e) and hole (h) calculated using both standard DFT (solid lines) and DFT+$U$ (dashed lines). For each method, the spectral scattering rate is normalized by the phonon DOS obtained at the corresponding level.}
\end{figure}

Based on these results, we calculated the electron and hole mobility and conductivity effective mass of ZnO using DFT and DFT+$U$. As shown on Table~\ref{tabular:mobility}, similar to the case of CdO, the mobility of electrons is significantly higher than that of holes, by more than a factor of ten. This difference in mobility between carrier type can be understood by examining the conductivity effective masses calculated using both DFT and DFT+$U$, which consistently show the hole effective mass to be approximately an order of magnitude larger than that of electron. Near the Fermi energy, the valence band has O-2$p$ character, and the conduction band exhibits Zn-4$s$ character. To identify the phonon energy regions contributing to this mobility, we plotted Fig.~\ref{fig:ZnO_spd}(a) for the scattering rate as a function of the energy spectrum. It shows that while the interaction with phonons near 70 meV is significant, the interaction with acoustic modes accounts for more than half of the total scattering. 

\begin{table}[t]
\caption{Theoretical phonon-limited drift ($\mu^d$) and Hall ($\mu^H$) mobility (cm$^2$V$^{-1}$s$^{-1}$) and conductivity effective mass ($m^*$, in units of $m_e$) at 300 K of electrons and holes in ZnO computed with the DFT+$U$ framework. }
\label{tabular:mobility}
\centering
\begin{ruledtabular}
\begin{tabular}{ccccccc} 
\multirow{2}{*}{Carrier type} & \multicolumn{2}{c}{$\mu^d$ } & \multicolumn{2}{c}{$\mu^H$} & \multicolumn{2}{c}{$m^*$}  \\
                              & \textbf{\textit{a}}          & \textbf{\textit{c}}            & \textbf{\textit{a}}     & \textbf{\textit{c}}     & \textbf{\textit{a}} & \textbf{\textit{c}}             \\ 
\hline
& \multicolumn{6}{c}{DFT} \\
Electron                      & 376.3  & 354.1 & 376.7  & 395.0 & 0.174 & 0.177\\
Hole                          & 18.0 & 10.6 & 25.6 & 15.7 & 1.571 & 2.615 \\ 
\hline
& \multicolumn{6}{c}{DFT+$U$} \\
Electron                      & 181.5  & 169.8 & 191.8  & 195.0 & 0.276 & 0.272 \\ 
Hole                          & 15.9 & 9.9 & 20.9 &  14.4 & 1.610 & 2.714\\ 
\end{tabular}
\end{ruledtabular}
\end{table}

\begin{figure}[htb]
	\begin{center}
		\includegraphics[width=0.8\columnwidth]{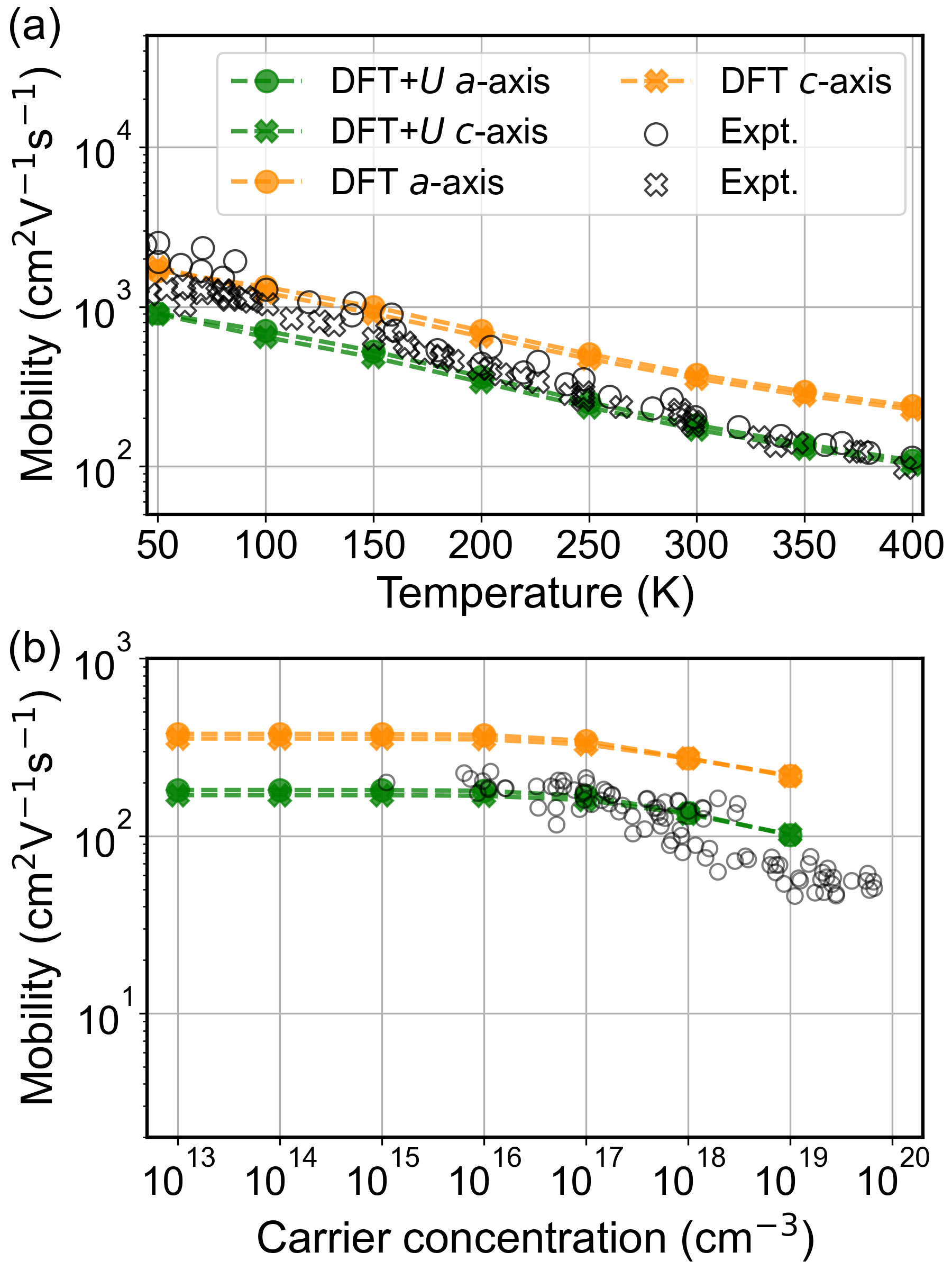}
	\end{center}
        \caption{\label{fig:ZnO_mobility} (a) Comparison of the theoretical electron drift mobilities obtained using DFT (orange) and DFT+$U$ (green) along $a$-axis (circle) and $c$-axis (cross), respectively, with the experimental Hall mobility values~\cite{Look2998ssc,Wagner1974jpcs,Hutson1957PR} as a function of temperature. Both calculations are based on a carrier concentration of $10^{16}$ cm$^{-3}$. (b) At 300 K, comparison of the theoretical electron drift mobilities, including the ionized-impurity scattering mechanism, obtained using DFT (orange) and DFT+$U$ (green) along $a$-axis (circle) and $c$-axis (cross), respectively, with the experimental values~\cite{Ellmer2001jpd,Ellmer2008tsf} as a function of carrier concentration.}
\end{figure}

Next, in Fig.~\ref{fig:ZnO_mobility}, we compared the mobility calculated using DFT and DFT+$U$ as a function of temperature and carrier concentration levels with experimental results~\cite{Look2998ssc,Ellmer2001jpd,Ellmer2008tsf,Wagner1974jpcs,Hutson1957PR}. 
The calculated mobility based on the DFT+$U$ level shows good agreement with experimental values, with no significant discrepancies. In contrast, the DFT results showed differences at temperatures above 250 K.  
The reduced electron mobility in the DFT+$U$ results compared to DFT can be attributed primarily to two factors. First, in calculations of the phonon-limited charge carrier mobility~\cite{ponce2020rpp,Ponce2021prr,Fiorentini2016prb,Ma2018prb,Ponce2018prb,Ponce2019prl,Park2020prb,Lee2020nc,Protik2020prb,Brunin2020prb,Jhalani2020prl}, the mobility is predominantly determined by states near the conduction band edges.  
Applying the Hubbard correction reduces the curvature of the conduction band edge, which corresponds to an increase in the electron effective mass as presented in Table~\ref{tabular:mobility}. 

Second, DFT+$U$ modifies the material's dielectric response and electron-phonon interactions. The static dielectric constant obtained from DFPT+$U$ ($\varepsilon_0$ along $a=4.013$, $\varepsilon_0$ along $c=4.064$) decreases by approximately 25\% compared to the DFPT results ($\varepsilon_0$ along $a=4.984$, $\varepsilon_0$ along $c=5.014$), indicating reduced electronic screening. To further quantify the difference in electron-phonon coupling strength, we estimated the average rate per state. This was done by first calculating the spectral scattering rate, $\partial \tau^{-1}/ \partial \omega$, for both DFT and DFT+$U$ (similar to Figs.~\ref{fig:CdO_spd} and~\ref{fig:ZnO_spd}(a)), and then dividing by the phonon density of states (DOS), $g(\omega)$, at each energy $\omega$ as shown in Fig.~\ref{fig:ZnO_spd}(b). Averaging these rate-per-state values $[\partial \tau^{-1}/ \partial \omega]/g(\omega)$ up to the maximum phonon energy yielded the following results. For DFT, the average for electron carrier was 10214.83 ps$^{-1}$/states and for hole carrier was 26410.07 ps$^{-1}$/states. For DFT+$U$, the average for electron carriers was significantly higher at 64945.05 ps$^{-1}$/states, while for hole carriers it was 37028.49 ps$^{-1}$/states. Notably, in the DFT+$U$ analysis for electrons, a particularly high scattering rate per state is observed near zero phonon energy where the phonon DOS is low (Fig.~\ref{fig:ZnO_spd}(b)), suggesting strong coupling to low-energy modes within this framework and contributing substantially to the high overall average rate per state for electrons. 

\begin{figure}[t]
	\begin{center}
		\includegraphics[width=0.8\columnwidth]{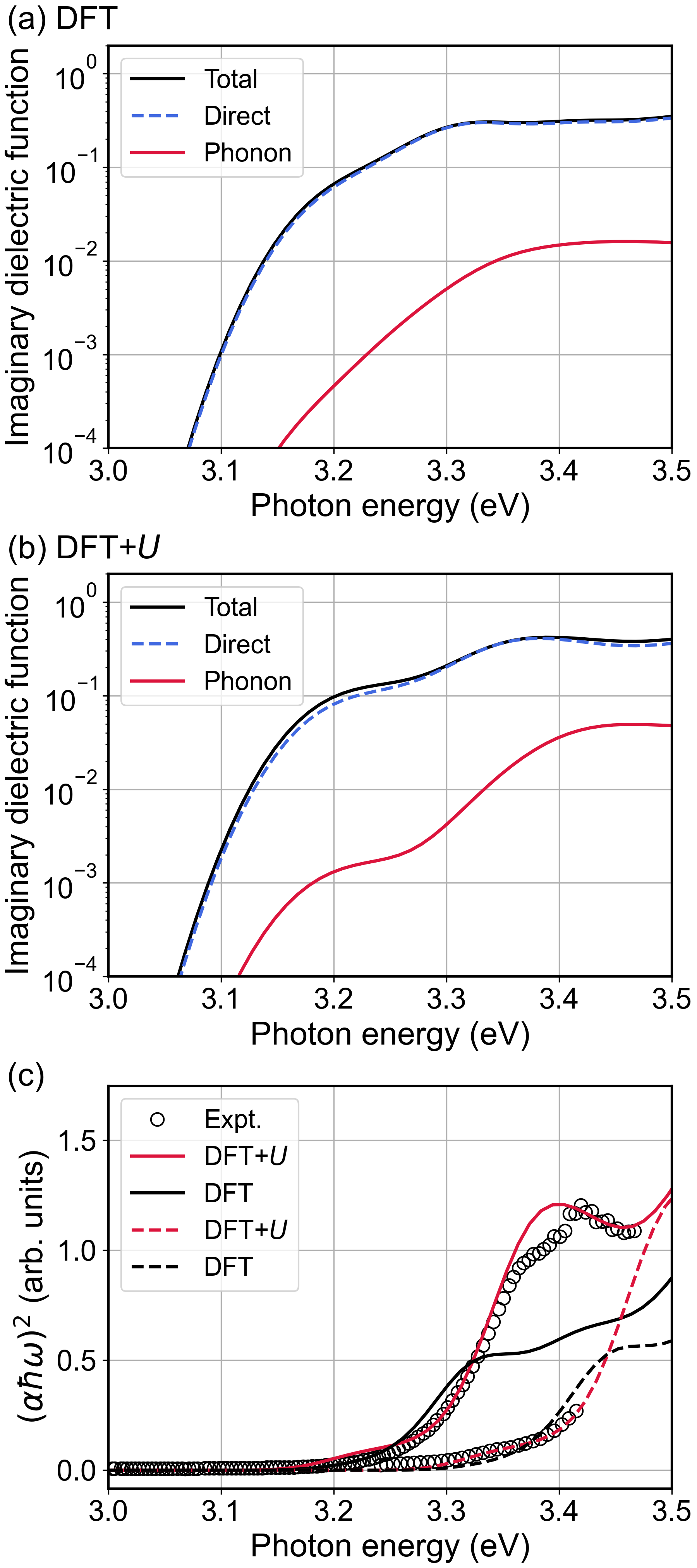}
	\end{center}
        \caption{\label{fig:ZnO_optics} The total imaginary dielectric function (black) of ZnO obtained using DFT (a) and DFT+$U$ (b) is decomposed into contributions from direct transitions (blue) and phonon-assisted transitions (red). (c) The squared product of the optical absorption spectrum and photon energy $(\alpha\hbar\omega)^2$ of ZnO obtained using DFT (black) and DFT+$U$ (red) at 300 K. Open circles denote the experimental value~\cite{Wang2008cpl,Singh2021om}.}
\end{figure}

To verify the transparency, we calculated the optical absorption of ZnO. In the case of ZnO, since the band gap size obtained from DFT+$U$ was still smaller than the experimental values, we applied a scissor shift before plotting~\cite{Zacharias2016prb}. From the imaginary dielectric function as shown in Fig.~\ref{fig:ZnO_optics}(a,b), we observed that the direct contribution is dominant in ZnO due to the minimum direct band gap. The DFT+$U$ result, unlike the DFT result, shows a sharp increase in slope near the photon energy corresponding to the direct band gap. This leads to a more pronounced shoulder feature near 3.2 eV compared to the DFT result. Furthermore, to directly compare with the experimental results, we plotted the optical absorption spectra as shown in Fig.~\ref{fig:ZnO_optics}(c). We confirmed that near 3.4 eV, the DFT+$U$ result exhibits a clearer peak than the DFT result, and this feature matches well with the experimental data~\cite{Wang2008cpl}. The calculated spectral shape remained consistent with additional experimental absorption data~\cite{Singh2021om} when calculated using a different value of the scissor shift.

\section{Conclusion\label{sec:con}}

We have developed a robust method for calculating electron-phonon (el-ph) interactions that integrates the self-consistently obtained Hubbard $U$ correction. 
This approach enhances accuracy for our simulations of both transport properties and optical absorption spectra. 
By incorporating the $U$ correction directly into the el-ph interaction calculation, our method effectively addresses the limitations of standard DFT when dealing with systems where electron correlation plays a significant role.

To validate the effectiveness of our method, we applied it to representative TCO materials, ZnO and CdO. 
These materials were selected due to their unique electronic structures and the critical role of el-ph interactions in their transport and optical properties. 
Our calculations successfully reproduced the Fröhlich interaction, demonstrating that the inclusion of the $U$ correction accurately captures the long-range polar interactions characteristic of these materials.

Moreover, by carefully considering changes in both band curvature and dielectric constant induced by the Hubbard correction, we were able to obtain mobility and optical absorption spectra that are in good agreement with experimental data. The inclusion of the Hubbard $U$ significantly alters the electronic band structure, particularly around the conduction band edges, leading to modifications in the effective mass of charge carriers. 
Additionally, the reduction in the dielectric constant compared to standard DFPT calculations directly influences the strength of the el-ph coupling. These factors collectively result in a more accurate representation of mobility and absorption characteristics, highlighting the importance of incorporating electron correlation effects into el-ph interaction calculations.

Beyond the specific cases of ZnO and CdO, the DFPT+$U$ method demonstrates its versatility and robustness across a broader range of materials. In particular, it shows significant potential for application to strongly correlated systems characterized by localized $d$ or $f$ orbitals~\cite{Floris2011prb,Floris2020prb}. This capability is especially crucial for accurately describing Mott insulators, such as transition metal oxides, where strong el-ph coupling can give rise to polaronic effects. In these materials, the formation of small polarons due to localized charge carriers interacting with lattice vibrations can significantly impact both electrical and thermal transport properties~\cite{Sio2019prb,Sio2019prl,Franchini2021nrm,Dai2024pnas}. 

In the interest of ensuring reproducibility and enabling subsequent studies, our current methodology will be included in the forthcoming public release of EPW~\cite{Lee2023npj}.

Considering a recent study on el-ph interactions of doped BaBiO$_3$~\cite{Jang2023PRL}, the extended Hubbard-corrected DFT approach, known as DFT+$U$+$V$, is capable of accurately capturing the complex interactions involving strong $p$-$d$ coupling. 
This method represents a critical advancement over standard DFT or DFT+$U$, which often fails to account for the nuanced interplay between localized and itinerant states in such materials. 
Additionally, previous GWPT calculations have shown that conventional DFT is inadequate for accurately modeling el-ph coupling in materials like cuprates or infinite-layer nickelates~\cite{Zhenglu2021prl,Zhenglu2024prl}. These materials exhibit strong correlations and unconventional superconductivity, where accurate modeling of el-ph interactions is essential for understanding their physical properties. 
So, we anticipate addressing a variety of complex scenarios that traditional methods have struggled to handle by employing the DFT+$U$+$V$ method to advanced DFPT to describe el-ph interactions in near future.

\begin{acknowledgements}
Y.-W.S. was supported by KIAS individual Grant (No. CG031509). W.Y. was supported by KIAS individual Grant (No. QP090102). Computations were supported by the CAC of KIAS.
This research was partly supported by SUPREME, one of seven centers in JUMP 2.0, a Semiconductor Research Corporation (SRC) program sponsored by DARPA (Mobility calculations in oxides); by the Computational Materials Science program of U.S. Department of Energy, Office of Science, Basic Energy Sciences under Award DE-SC0020129 (DFPT+U development in EPW); and by the U.S. National Science Foundation, DMREF Grant No. 2119555 (Calculations of optical spectra with QDPT). The authors acknowledge the Texas Advanced Computing Center (TACC) at The University of Texas at Austin for providing HPC resources, including the Frontera and Lonestar6 systems, that have contributed to the research results reported within this paper. URL: http://www.tacc.utexas.edu. This research used resources of the National Energy Research Scientific Computing Center, a DOE Office of Science User Facility supported by the Office of Science of the U.S. Department of Energy under Contract No. DE-AC02-05CH11231. This research used resources of the Argonne Leadership Computing Facility, which is a DOE Office of Science User Facility supported under Contract DE-AC02-06CH11357.
\end{acknowledgements}

\appendix

\begin{table}[t]
\caption{Lattice parameters ($a$ and $c$) in angstrom and band gap ($E_g$) in eV.}
\label{tabular:ZnO_str}
\begin{ruledtabular}
\begin{tabular}{lccc} 
      & $a$     & $c$    & $E_g$            \\ \hline
DFT   & 3.197     & 5.134    & 0.86          \\
Zn-3$d$ & 3.301 & 5.296 & 1.53         \\
O-2$p$ & 3.262 & 5.237 & 1.58         \\
Zn-3$d$, O-2$p$ & 3.249 & 5.192 & 4.33         \\
Expt \footnote{\label{exp}Reference~\cite{Chen1998jap}}  & 3.249 & 5.207 & 3.37 
\end{tabular}
\end{ruledtabular}
\end{table}

\begin{figure}[t]
	\begin{center}
		\includegraphics[width=0.95\columnwidth]{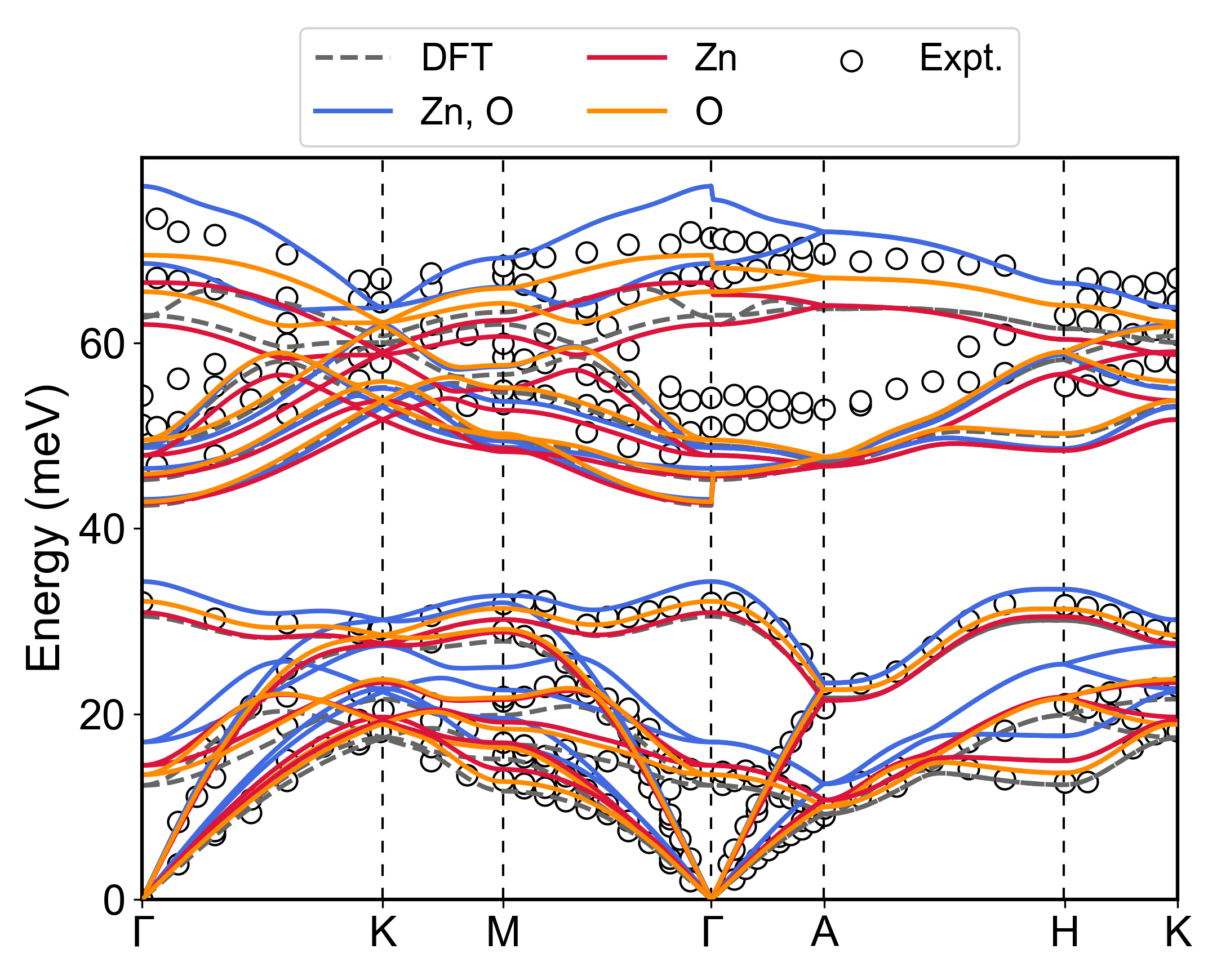}
	\end{center}
	\caption{Phonon band structures of ZnO obtained with DFPT (dotted grey) and DFPT+$U$ with Zn-3$d$ and O-2$p$ (blue), Zn-3$d$ only (red), and O-2$p$ only (Orange), plotted along the high-symmetry lines of the Brillouin zone. Open circles denote the experimental values~\cite{Hewat1970ssc,Thoma1974ssc,Serrano2010prb}
	}
	\label{fig:ZnO_ph_comp}
\end{figure}

\section{\label{app:ZnO_PBE}ZnO calculated with PBE}

We performed DFT calculations for ZnO using the PBE functional, as well as several DFT+$U$ calculations under different conditions. Specifically, we applied the Hubbard correction to three cases: 1) the Zn-3$d$ orbitals, 2) the O-2$p$ orbitals, and 3) both the Zn-3$d$ and O-2$p$ orbitals. Except for the exchange-correlation functional, all other computational parameters were identical to those described in the computational details section. For the DFPT(+$U$) calculations of phonon frequencies, the Brillouin zone is sampled on a $8\times8\times5$ $q$-point grid.

As shown in Table~\ref{tabular:ZnO_str}, applying the Hubbard correction to both atomic species produced converged structural parameters that were closest to the experimental values. However, in this case, the band gap was overestimated by about 1 eV. We also noted that the band gap size varied somewhat depending on the type of the pseudopotential~\cite{Lee2020PRR}. In contrast, when applying the Hubbard correction only to the Zn-3$d$ or O-2$p$ orbitals, the structural parameters were slightly overestimated, and the resulting band gap was similar to that obtained with the LDA approach.

Based on these results, we performed DFPT(+$U$) calculations for each structure to obtain phonon frequencies and compared them with experimentally observed data, as shown in Fig.~\ref{fig:ZnO_ph_comp}. The DFPT results using PBE show that while the phonons in the acoustic mode region generally agree well with experiment, the overall optical phonon modes above 40 meV are underestimated compared to experimental values. When we applied the Hubbard correction to both Zn-3$d$ and O-2$p$ orbitals, the acoustic phonon modes were overestimated relative to experiment, and the optical phonon modes above 40 meV exhibited a broader dispersion and larger deviation from the experimental data.

Applying the Hubbard correction only to Zn-3$d$ resulted in a slight upward shift of the acoustic phonon modes, but a slight downward shift in the optical phonon modes above 40 meV. On the other hand, applying the Hubbard correction only to O-2$p$ caused a small upward shift in the acoustic phonon modes—less than that observed when applying it to Zn-3$d$—and also shifted the $\sim$30 meV optical phonon dispersion upward, bringing it into much closer agreement with the experimental results. In addition, for optical phonon modes above 40 meV, only the highest optical phonon modes were shifted upward, reducing the difference from experiment compared to the PBE results, though still showing the less accuracy than the LDA results. Therefore, for ZnO, we inevitably adopted the LDA functional. Notably, applying the Hubbard correction solely to O-2$p$ orbitals produced the best agreement with the experimental results(Fig. ~\ref{fig:ZnO_el_ph}).

\bibliography{dfptu}

\begin{thebibliography}{127}%
\makeatletter
\providecommand \@ifxundefined [1]{%
 \@ifx{#1\undefined}
}%
\providecommand \@ifnum [1]{%
 \ifnum #1\expandafter \@firstoftwo
 \else \expandafter \@secondoftwo
 \fi
}%
\providecommand \@ifx [1]{%
 \ifx #1\expandafter \@firstoftwo
 \else \expandafter \@secondoftwo
 \fi
}%
\providecommand \natexlab [1]{#1}%
\providecommand \enquote  [1]{``#1''}%
\providecommand \bibnamefont  [1]{#1}%
\providecommand \bibfnamefont [1]{#1}%
\providecommand \citenamefont [1]{#1}%
\providecommand \href@noop [0]{\@secondoftwo}%
\providecommand \href [0]{\begingroup \@sanitize@url \@href}%
\providecommand \@href[1]{\@@startlink{#1}\@@href}%
\providecommand \@@href[1]{\endgroup#1\@@endlink}%
\providecommand \@sanitize@url [0]{\catcode `\\12\catcode `\$12\catcode `\&12\catcode `\#12\catcode `\^12\catcode `\_12\catcode `\%12\relax}%
\providecommand \@@startlink[1]{}%
\providecommand \@@endlink[0]{}%
\providecommand \url  [0]{\begingroup\@sanitize@url \@url }%
\providecommand \@url [1]{\endgroup\@href {#1}{\urlprefix }}%
\providecommand \urlprefix  [0]{URL }%
\providecommand \Eprint [0]{\href }%
\providecommand \doibase [0]{https://doi.org/}%
\providecommand \selectlanguage [0]{\@gobble}%
\providecommand \bibinfo  [0]{\@secondoftwo}%
\providecommand \bibfield  [0]{\@secondoftwo}%
\providecommand \translation [1]{[#1]}%
\providecommand \BibitemOpen [0]{}%
\providecommand \bibitemStop [0]{}%
\providecommand \bibitemNoStop [0]{.\EOS\space}%
\providecommand \EOS [0]{\spacefactor3000\relax}%
\providecommand \BibitemShut  [1]{\csname bibitem#1\endcsname}%
\let\auto@bib@innerbib\@empty
\bibitem [{\citenamefont {Grimvall}(1969)}]{Grimvall1969phys}%
  \BibitemOpen
  \bibfield  {author} {\bibinfo {author} {\bibfnamefont {G.}~\bibnamefont {Grimvall}},\ }\bibfield  {title} {\bibinfo {title} {New aspects on the electron-phonon system at finite temperatures with an application on lead and mercury},\ }\href {https://doi.org/10.1007/BF02422572} {\bibfield  {journal} {\bibinfo  {journal} {Phys. Kondens. Materie}\ }\textbf {\bibinfo {volume} {9}},\ \bibinfo {pages} {283} (\bibinfo {year} {1969})}\BibitemShut {NoStop}%
\bibitem [{\citenamefont {Giustino}(2017)}]{Giustino2017RMP}%
  \BibitemOpen
  \bibfield  {author} {\bibinfo {author} {\bibfnamefont {F.}~\bibnamefont {Giustino}},\ }\bibfield  {title} {\bibinfo {title} {Electron-phonon interactions from first principles},\ }\href {https://doi.org/10.1103/RevModPhys.89.015003} {\bibfield  {journal} {\bibinfo  {journal} {Rev. Mod. Phys.}\ }\textbf {\bibinfo {volume} {89}},\ \bibinfo {pages} {015003} (\bibinfo {year} {2017})}\BibitemShut {NoStop}%
\bibitem [{\citenamefont {Ginley}\ and\ \citenamefont {Perkins}(2010)}]{Ginley2010transparent}%
  \BibitemOpen
  \bibfield  {author} {\bibinfo {author} {\bibfnamefont {D.~S.}\ \bibnamefont {Ginley}}\ and\ \bibinfo {author} {\bibfnamefont {J.~D.}\ \bibnamefont {Perkins}},\ }\bibfield  {title} {\bibinfo {title} {Transparent conductors},\ }in\ \href@noop {} {\emph {\bibinfo {booktitle} {Handbook of transparent conductors}}}\ (\bibinfo  {publisher} {Springer},\ \bibinfo {year} {2010})\ pp.\ \bibinfo {pages} {1--25}\BibitemShut {NoStop}%
\bibitem [{\citenamefont {Ellmer}(2012)}]{Ellmer2012np}%
  \BibitemOpen
  \bibfield  {author} {\bibinfo {author} {\bibfnamefont {K.}~\bibnamefont {Ellmer}},\ }\bibfield  {title} {\bibinfo {title} {Past achievements and future challenges in the development of optically transparent electrodes},\ }\href {https://doi.org/10.1038/nphoton.2012.282} {\bibfield  {journal} {\bibinfo  {journal} {Nat. Photonics}\ }\textbf {\bibinfo {volume} {6}},\ \bibinfo {pages} {809} (\bibinfo {year} {2012})}\BibitemShut {NoStop}%
\bibitem [{\citenamefont {Schmidt}\ \emph {et~al.}(2015)\citenamefont {Schmidt}, \citenamefont {Giustiniano},\ and\ \citenamefont {Eda}}]{Schmidt2015chem}%
  \BibitemOpen
  \bibfield  {author} {\bibinfo {author} {\bibfnamefont {H.}~\bibnamefont {Schmidt}}, \bibinfo {author} {\bibfnamefont {F.}~\bibnamefont {Giustiniano}},\ and\ \bibinfo {author} {\bibfnamefont {G.}~\bibnamefont {Eda}},\ }\bibfield  {title} {\bibinfo {title} {Electronic transport properties of transition metal dichalcogenide field-effect devices: surface and interface effects},\ }\href {https://doi.org/10.1039/C5CS00275C} {\bibfield  {journal} {\bibinfo  {journal} {Chem. Soc. Rev.}\ }\textbf {\bibinfo {volume} {44}},\ \bibinfo {pages} {7715} (\bibinfo {year} {2015})}\BibitemShut {NoStop}%
\bibitem [{\citenamefont {Huo}\ and\ \citenamefont {Konstantatos}(2018)}]{Huo2018Advmat}%
  \BibitemOpen
  \bibfield  {author} {\bibinfo {author} {\bibfnamefont {N.}~\bibnamefont {Huo}}\ and\ \bibinfo {author} {\bibfnamefont {G.}~\bibnamefont {Konstantatos}},\ }\bibfield  {title} {\bibinfo {title} {Recent progress and future prospects of {2D}-based photodetectors},\ }\href {https://doi.org/https://doi.org/10.1002/adma.201801164} {\bibfield  {journal} {\bibinfo  {journal} {Adv. Mater.}\ }\textbf {\bibinfo {volume} {30}},\ \bibinfo {pages} {1801164} (\bibinfo {year} {2018})}\BibitemShut {NoStop}%
\bibitem [{\citenamefont {Hohenberg}\ and\ \citenamefont {Kohn}(1964)}]{HK}%
  \BibitemOpen
  \bibfield  {author} {\bibinfo {author} {\bibfnamefont {P.}~\bibnamefont {Hohenberg}}\ and\ \bibinfo {author} {\bibfnamefont {W.}~\bibnamefont {Kohn}},\ }\bibfield  {title} {\bibinfo {title} {Inhomogeneous electron gas},\ }\href {https://doi.org/10.1103/PhysRev.136.B864} {\bibfield  {journal} {\bibinfo  {journal} {Phys. Rev.}\ }\textbf {\bibinfo {volume} {136}},\ \bibinfo {pages} {B864} (\bibinfo {year} {1964})}\BibitemShut {NoStop}%
\bibitem [{\citenamefont {Kohn}\ and\ \citenamefont {Sham}(1965)}]{KS}%
  \BibitemOpen
  \bibfield  {author} {\bibinfo {author} {\bibfnamefont {W.}~\bibnamefont {Kohn}}\ and\ \bibinfo {author} {\bibfnamefont {L.~J.}\ \bibnamefont {Sham}},\ }\bibfield  {title} {\bibinfo {title} {Self-consistent equations including exchange and correlation effects},\ }\href {https://doi.org/10.1103/PhysRev.140.A1133} {\bibfield  {journal} {\bibinfo  {journal} {Phys. Rev.}\ }\textbf {\bibinfo {volume} {140}},\ \bibinfo {pages} {A1133} (\bibinfo {year} {1965})}\BibitemShut {NoStop}%
\bibitem [{\citenamefont {Ihm}(1988)}]{Ihm1988RPP}%
  \BibitemOpen
  \bibfield  {author} {\bibinfo {author} {\bibfnamefont {J.}~\bibnamefont {Ihm}},\ }\bibfield  {title} {\bibinfo {title} {Total energy calculations in solid state physics},\ }\href {https://doi.org/10.1088/0034-4885/51/1/003} {\bibfield  {journal} {\bibinfo  {journal} {Rep. Prog. Phys.}\ }\textbf {\bibinfo {volume} {51}},\ \bibinfo {pages} {105} (\bibinfo {year} {1988})}\BibitemShut {NoStop}%
\bibitem [{\citenamefont {Baroni}\ \emph {et~al.}(1987)\citenamefont {Baroni}, \citenamefont {Giannozzi},\ and\ \citenamefont {Testa}}]{Baroni1987prl}%
  \BibitemOpen
  \bibfield  {author} {\bibinfo {author} {\bibfnamefont {S.}~\bibnamefont {Baroni}}, \bibinfo {author} {\bibfnamefont {P.}~\bibnamefont {Giannozzi}},\ and\ \bibinfo {author} {\bibfnamefont {A.}~\bibnamefont {Testa}},\ }\bibfield  {title} {\bibinfo {title} {Green's-function approach to linear response in solids},\ }\href {https://doi.org/10.1103/PhysRevLett.58.1861} {\bibfield  {journal} {\bibinfo  {journal} {Phys. Rev. Lett.}\ }\textbf {\bibinfo {volume} {58}},\ \bibinfo {pages} {1861} (\bibinfo {year} {1987})}\BibitemShut {NoStop}%
\bibitem [{\citenamefont {Giannozzi}\ \emph {et~al.}(1991)\citenamefont {Giannozzi}, \citenamefont {de~Gironcoli}, \citenamefont {Pavone},\ and\ \citenamefont {Baroni}}]{Giannozzi1991prb}%
  \BibitemOpen
  \bibfield  {author} {\bibinfo {author} {\bibfnamefont {P.}~\bibnamefont {Giannozzi}}, \bibinfo {author} {\bibfnamefont {S.}~\bibnamefont {de~Gironcoli}}, \bibinfo {author} {\bibfnamefont {P.}~\bibnamefont {Pavone}},\ and\ \bibinfo {author} {\bibfnamefont {S.}~\bibnamefont {Baroni}},\ }\bibfield  {title} {\bibinfo {title} {Ab initio calculation of phonon dispersions in semiconductors},\ }\href {https://doi.org/10.1103/PhysRevB.43.7231} {\bibfield  {journal} {\bibinfo  {journal} {Phys. Rev. B}\ }\textbf {\bibinfo {volume} {43}},\ \bibinfo {pages} {7231} (\bibinfo {year} {1991})}\BibitemShut {NoStop}%
\bibitem [{\citenamefont {Gonze}(1995)}]{Gonze1995pra}%
  \BibitemOpen
  \bibfield  {author} {\bibinfo {author} {\bibfnamefont {X.}~\bibnamefont {Gonze}},\ }\bibfield  {title} {\bibinfo {title} {Adiabatic density-functional perturbation theory},\ }\href {https://doi.org/10.1103/PhysRevA.52.1096} {\bibfield  {journal} {\bibinfo  {journal} {Phys. Rev. A}\ }\textbf {\bibinfo {volume} {52}},\ \bibinfo {pages} {1096} (\bibinfo {year} {1995})}\BibitemShut {NoStop}%
\bibitem [{\citenamefont {Gonze}(1996)}]{Gonze1996pra}%
  \BibitemOpen
  \bibfield  {author} {\bibinfo {author} {\bibfnamefont {X.}~\bibnamefont {Gonze}},\ }\bibfield  {title} {\bibinfo {title} {Erratum: Adiabatic density-functional perturbation theory},\ }\href {https://doi.org/10.1103/PhysRevA.54.4591} {\bibfield  {journal} {\bibinfo  {journal} {Phys. Rev. A}\ }\textbf {\bibinfo {volume} {54}},\ \bibinfo {pages} {4591} (\bibinfo {year} {1996})}\BibitemShut {NoStop}%
\bibitem [{\citenamefont {Baroni}\ \emph {et~al.}(2001)\citenamefont {Baroni}, \citenamefont {de~Gironcoli}, \citenamefont {Dal~Corso},\ and\ \citenamefont {Giannozzi}}]{Baroni2001RMP}%
  \BibitemOpen
  \bibfield  {author} {\bibinfo {author} {\bibfnamefont {S.}~\bibnamefont {Baroni}}, \bibinfo {author} {\bibfnamefont {S.}~\bibnamefont {de~Gironcoli}}, \bibinfo {author} {\bibfnamefont {A.}~\bibnamefont {Dal~Corso}},\ and\ \bibinfo {author} {\bibfnamefont {P.}~\bibnamefont {Giannozzi}},\ }\bibfield  {title} {\bibinfo {title} {Phonons and related crystal properties from density-functional perturbation theory},\ }\href {https://doi.org/10.1103/RevModPhys.73.515} {\bibfield  {journal} {\bibinfo  {journal} {Rev. Mod. Phys.}\ }\textbf {\bibinfo {volume} {73}},\ \bibinfo {pages} {515} (\bibinfo {year} {2001})}\BibitemShut {NoStop}%
\bibitem [{\citenamefont {Jones}(2015)}]{Jones2015RMP}%
  \BibitemOpen
  \bibfield  {author} {\bibinfo {author} {\bibfnamefont {R.~O.}\ \bibnamefont {Jones}},\ }\bibfield  {title} {\bibinfo {title} {Density functional theory: Its origins, rise to prominence, and future},\ }\href {https://doi.org/10.1103/RevModPhys.87.897} {\bibfield  {journal} {\bibinfo  {journal} {Rev. Mod. Phys.}\ }\textbf {\bibinfo {volume} {87}},\ \bibinfo {pages} {897} (\bibinfo {year} {2015})}\BibitemShut {NoStop}%
\bibitem [{\citenamefont {Lejaeghere}\ \emph {et~al.}(2016)\citenamefont {Lejaeghere}, \citenamefont {Bihlmayer}, \citenamefont {Björkman}, \citenamefont {Blaha}, \citenamefont {Blügel}, \citenamefont {Blum}, \citenamefont {Caliste}, \citenamefont {Castelli}, \citenamefont {Clark}, \citenamefont {Corso}, \citenamefont {de~Gironcoli}, \citenamefont {Deutsch}, \citenamefont {Dewhurst}, \citenamefont {Marco}, \citenamefont {Draxl}, \citenamefont {Dułak}, \citenamefont {Eriksson}, \citenamefont {Flores-Livas}, \citenamefont {Garrity}, \citenamefont {Genovese}, \citenamefont {Giannozzi}, \citenamefont {Giantomassi}, \citenamefont {Goedecker}, \citenamefont {Gonze}, \citenamefont {Grånäs}, \citenamefont {Gross}, \citenamefont {Gulans}, \citenamefont {Gygi}, \citenamefont {Hamann}, \citenamefont {Hasnip}, \citenamefont {Holzwarth}, \citenamefont {Iuşan}, \citenamefont {Jochym}, \citenamefont {Jollet}, \citenamefont {Jones}, \citenamefont {Kresse}, \citenamefont {Koepernik}, \citenamefont {Küçükbenli},
  \citenamefont {Kvashnin}, \citenamefont {Locht}, \citenamefont {Lubeck}, \citenamefont {Marsman}, \citenamefont {Marzari}, \citenamefont {Nitzsche}, \citenamefont {Nordström}, \citenamefont {Ozaki}, \citenamefont {Paulatto}, \citenamefont {Pickard}, \citenamefont {Poelmans}, \citenamefont {Probert}, \citenamefont {Refson}, \citenamefont {Richter}, \citenamefont {Rignanese}, \citenamefont {Saha}, \citenamefont {Scheffler}, \citenamefont {Schlipf}, \citenamefont {Schwarz}, \citenamefont {Sharma}, \citenamefont {Tavazza}, \citenamefont {Thunström}, \citenamefont {Tkatchenko}, \citenamefont {Torrent}, \citenamefont {Vanderbilt}, \citenamefont {van Setten}, \citenamefont {Speybroeck}, \citenamefont {Wills}, \citenamefont {Yates}, \citenamefont {Zhang},\ and\ \citenamefont {Cottenier}}]{Kurt2016Science}%
  \BibitemOpen
  \bibfield  {author} {\bibinfo {author} {\bibfnamefont {K.}~\bibnamefont {Lejaeghere}}, \bibinfo {author} {\bibfnamefont {G.}~\bibnamefont {Bihlmayer}}, \bibinfo {author} {\bibfnamefont {T.}~\bibnamefont {Björkman}}, \bibinfo {author} {\bibfnamefont {P.}~\bibnamefont {Blaha}}, \bibinfo {author} {\bibfnamefont {S.}~\bibnamefont {Blügel}}, \bibinfo {author} {\bibfnamefont {V.}~\bibnamefont {Blum}}, \bibinfo {author} {\bibfnamefont {D.}~\bibnamefont {Caliste}}, \bibinfo {author} {\bibfnamefont {I.~E.}\ \bibnamefont {Castelli}}, \bibinfo {author} {\bibfnamefont {S.~J.}\ \bibnamefont {Clark}}, \bibinfo {author} {\bibfnamefont {A.~D.}\ \bibnamefont {Corso}}, \bibinfo {author} {\bibfnamefont {S.}~\bibnamefont {de~Gironcoli}}, \bibinfo {author} {\bibfnamefont {T.}~\bibnamefont {Deutsch}}, \bibinfo {author} {\bibfnamefont {J.~K.}\ \bibnamefont {Dewhurst}}, \bibinfo {author} {\bibfnamefont {I.~D.}\ \bibnamefont {Marco}}, \bibinfo {author} {\bibfnamefont {C.}~\bibnamefont {Draxl}}, \bibinfo {author} {\bibfnamefont
  {M.}~\bibnamefont {Dułak}}, \bibinfo {author} {\bibfnamefont {O.}~\bibnamefont {Eriksson}}, \bibinfo {author} {\bibfnamefont {J.~A.}\ \bibnamefont {Flores-Livas}}, \bibinfo {author} {\bibfnamefont {K.~F.}\ \bibnamefont {Garrity}}, \bibinfo {author} {\bibfnamefont {L.}~\bibnamefont {Genovese}}, \bibinfo {author} {\bibfnamefont {P.}~\bibnamefont {Giannozzi}}, \bibinfo {author} {\bibfnamefont {M.}~\bibnamefont {Giantomassi}}, \bibinfo {author} {\bibfnamefont {S.}~\bibnamefont {Goedecker}}, \bibinfo {author} {\bibfnamefont {X.}~\bibnamefont {Gonze}}, \bibinfo {author} {\bibfnamefont {O.}~\bibnamefont {Grånäs}}, \bibinfo {author} {\bibfnamefont {E.~K.~U.}\ \bibnamefont {Gross}}, \bibinfo {author} {\bibfnamefont {A.}~\bibnamefont {Gulans}}, \bibinfo {author} {\bibfnamefont {F.}~\bibnamefont {Gygi}}, \bibinfo {author} {\bibfnamefont {D.~R.}\ \bibnamefont {Hamann}}, \bibinfo {author} {\bibfnamefont {P.~J.}\ \bibnamefont {Hasnip}}, \bibinfo {author} {\bibfnamefont {N.~A.~W.}\ \bibnamefont {Holzwarth}}, \bibinfo
  {author} {\bibfnamefont {D.}~\bibnamefont {Iuşan}}, \bibinfo {author} {\bibfnamefont {D.~B.}\ \bibnamefont {Jochym}}, \bibinfo {author} {\bibfnamefont {F.}~\bibnamefont {Jollet}}, \bibinfo {author} {\bibfnamefont {D.}~\bibnamefont {Jones}}, \bibinfo {author} {\bibfnamefont {G.}~\bibnamefont {Kresse}}, \bibinfo {author} {\bibfnamefont {K.}~\bibnamefont {Koepernik}}, \bibinfo {author} {\bibfnamefont {E.}~\bibnamefont {Küçükbenli}}, \bibinfo {author} {\bibfnamefont {Y.~O.}\ \bibnamefont {Kvashnin}}, \bibinfo {author} {\bibfnamefont {I.~L.~M.}\ \bibnamefont {Locht}}, \bibinfo {author} {\bibfnamefont {S.}~\bibnamefont {Lubeck}}, \bibinfo {author} {\bibfnamefont {M.}~\bibnamefont {Marsman}}, \bibinfo {author} {\bibfnamefont {N.}~\bibnamefont {Marzari}}, \bibinfo {author} {\bibfnamefont {U.}~\bibnamefont {Nitzsche}}, \bibinfo {author} {\bibfnamefont {L.}~\bibnamefont {Nordström}}, \bibinfo {author} {\bibfnamefont {T.}~\bibnamefont {Ozaki}}, \bibinfo {author} {\bibfnamefont {L.}~\bibnamefont {Paulatto}},
  \bibinfo {author} {\bibfnamefont {C.~J.}\ \bibnamefont {Pickard}}, \bibinfo {author} {\bibfnamefont {W.}~\bibnamefont {Poelmans}}, \bibinfo {author} {\bibfnamefont {M.~I.~J.}\ \bibnamefont {Probert}}, \bibinfo {author} {\bibfnamefont {K.}~\bibnamefont {Refson}}, \bibinfo {author} {\bibfnamefont {M.}~\bibnamefont {Richter}}, \bibinfo {author} {\bibfnamefont {G.-M.}\ \bibnamefont {Rignanese}}, \bibinfo {author} {\bibfnamefont {S.}~\bibnamefont {Saha}}, \bibinfo {author} {\bibfnamefont {M.}~\bibnamefont {Scheffler}}, \bibinfo {author} {\bibfnamefont {M.}~\bibnamefont {Schlipf}}, \bibinfo {author} {\bibfnamefont {K.}~\bibnamefont {Schwarz}}, \bibinfo {author} {\bibfnamefont {S.}~\bibnamefont {Sharma}}, \bibinfo {author} {\bibfnamefont {F.}~\bibnamefont {Tavazza}}, \bibinfo {author} {\bibfnamefont {P.}~\bibnamefont {Thunström}}, \bibinfo {author} {\bibfnamefont {A.}~\bibnamefont {Tkatchenko}}, \bibinfo {author} {\bibfnamefont {M.}~\bibnamefont {Torrent}}, \bibinfo {author} {\bibfnamefont {D.}~\bibnamefont
  {Vanderbilt}}, \bibinfo {author} {\bibfnamefont {M.~J.}\ \bibnamefont {van Setten}}, \bibinfo {author} {\bibfnamefont {V.~V.}\ \bibnamefont {Speybroeck}}, \bibinfo {author} {\bibfnamefont {J.~M.}\ \bibnamefont {Wills}}, \bibinfo {author} {\bibfnamefont {J.~R.}\ \bibnamefont {Yates}}, \bibinfo {author} {\bibfnamefont {G.-X.}\ \bibnamefont {Zhang}},\ and\ \bibinfo {author} {\bibfnamefont {S.}~\bibnamefont {Cottenier}},\ }\bibfield  {title} {\bibinfo {title} {Reproducibility in density functional theory calculations of solids},\ }\href {https://doi.org/10.1126/science.aad3000} {\bibfield  {journal} {\bibinfo  {journal} {Science}\ }\textbf {\bibinfo {volume} {351}},\ \bibinfo {pages} {aad3000} (\bibinfo {year} {2016})}\BibitemShut {NoStop}%
\bibitem [{\citenamefont {Perdew}\ and\ \citenamefont {Zunger}(1981)}]{LDApz}%
  \BibitemOpen
  \bibfield  {author} {\bibinfo {author} {\bibfnamefont {J.~P.}\ \bibnamefont {Perdew}}\ and\ \bibinfo {author} {\bibfnamefont {A.}~\bibnamefont {Zunger}},\ }\bibfield  {title} {\bibinfo {title} {{Self-interaction correction to density-functional approximations for many-electron systems}},\ }\href {https://doi.org/10.1103/PhysRevB.23.5048} {\bibfield  {journal} {\bibinfo  {journal} {Phys. Rev. B}\ }\textbf {\bibinfo {volume} {23}},\ \bibinfo {pages} {5048} (\bibinfo {year} {1981})}\BibitemShut {NoStop}%
\bibitem [{\citenamefont {Perdew}\ \emph {et~al.}(1996)\citenamefont {Perdew}, \citenamefont {Burke},\ and\ \citenamefont {Ernzerhof}}]{PBE}%
  \BibitemOpen
  \bibfield  {author} {\bibinfo {author} {\bibfnamefont {J.~P.}\ \bibnamefont {Perdew}}, \bibinfo {author} {\bibfnamefont {K.}~\bibnamefont {Burke}},\ and\ \bibinfo {author} {\bibfnamefont {M.}~\bibnamefont {Ernzerhof}},\ }\bibfield  {title} {\bibinfo {title} {{Generalized Gradient Approximation Made Simple}},\ }\href {https://doi.org/10.1103/PhysRevLett.77.3865} {\bibfield  {journal} {\bibinfo  {journal} {Phys. Rev. Lett.}\ }\textbf {\bibinfo {volume} {77}},\ \bibinfo {pages} {3865} (\bibinfo {year} {1996})}\BibitemShut {NoStop}%
\bibitem [{\citenamefont {Cohen}\ \emph {et~al.}(2008)\citenamefont {Cohen}, \citenamefont {Mori-Sánchez},\ and\ \citenamefont {Yang}}]{Cohen2008Science}%
  \BibitemOpen
  \bibfield  {author} {\bibinfo {author} {\bibfnamefont {A.~J.}\ \bibnamefont {Cohen}}, \bibinfo {author} {\bibfnamefont {P.}~\bibnamefont {Mori-Sánchez}},\ and\ \bibinfo {author} {\bibfnamefont {W.}~\bibnamefont {Yang}},\ }\bibfield  {title} {\bibinfo {title} {Insights into current limitations of density functional theory},\ }\href {https://doi.org/10.1126/science.1158722} {\bibfield  {journal} {\bibinfo  {journal} {Science}\ }\textbf {\bibinfo {volume} {321}},\ \bibinfo {pages} {792} (\bibinfo {year} {2008})}\BibitemShut {NoStop}%
\bibitem [{\citenamefont {Anisimov}\ \emph {et~al.}(1997)\citenamefont {Anisimov}, \citenamefont {Aryasetiawan},\ and\ \citenamefont {Lichtenstein}}]{Anisimov1997JPC}%
  \BibitemOpen
  \bibfield  {author} {\bibinfo {author} {\bibfnamefont {V.~I.}\ \bibnamefont {Anisimov}}, \bibinfo {author} {\bibfnamefont {F.}~\bibnamefont {Aryasetiawan}},\ and\ \bibinfo {author} {\bibfnamefont {A.~I.}\ \bibnamefont {Lichtenstein}},\ }\bibfield  {title} {\bibinfo {title} {{First-principles calculations of the electronic structure and spectra of strongly correlated systems: the LDA$+ U$ method}},\ }\href {https://doi.org/10.1088/0953-8984/9/4/002} {\bibfield  {journal} {\bibinfo  {journal} {J. Phys.: Condens. Matter}\ }\textbf {\bibinfo {volume} {9}},\ \bibinfo {pages} {767} (\bibinfo {year} {1997})}\BibitemShut {NoStop}%
\bibitem [{\citenamefont {Dudarev}\ \emph {et~al.}(1998)\citenamefont {Dudarev}, \citenamefont {Botton}, \citenamefont {Savrasov}, \citenamefont {Humphreys},\ and\ \citenamefont {Sutton}}]{Dudarev1998PRB}%
  \BibitemOpen
  \bibfield  {author} {\bibinfo {author} {\bibfnamefont {S.~L.}\ \bibnamefont {Dudarev}}, \bibinfo {author} {\bibfnamefont {G.~A.}\ \bibnamefont {Botton}}, \bibinfo {author} {\bibfnamefont {S.~Y.}\ \bibnamefont {Savrasov}}, \bibinfo {author} {\bibfnamefont {C.~J.}\ \bibnamefont {Humphreys}},\ and\ \bibinfo {author} {\bibfnamefont {A.~P.}\ \bibnamefont {Sutton}},\ }\bibfield  {title} {\bibinfo {title} {{Electron-energy-loss spectra and the structural stability of nickel oxide: An LSDA+U study}},\ }\href {https://doi.org/10.1103/PhysRevB.57.1505} {\bibfield  {journal} {\bibinfo  {journal} {Phys. Rev. B}\ }\textbf {\bibinfo {volume} {57}},\ \bibinfo {pages} {1505} (\bibinfo {year} {1998})}\BibitemShut {NoStop}%
\bibitem [{\citenamefont {Himmetoglu}\ \emph {et~al.}(2014)\citenamefont {Himmetoglu}, \citenamefont {Floris}, \citenamefont {de~Gironcoli},\ and\ \citenamefont {Cococcioni}}]{Himmetoglu2014IJQC}%
  \BibitemOpen
  \bibfield  {author} {\bibinfo {author} {\bibfnamefont {B.}~\bibnamefont {Himmetoglu}}, \bibinfo {author} {\bibfnamefont {A.}~\bibnamefont {Floris}}, \bibinfo {author} {\bibfnamefont {S.}~\bibnamefont {de~Gironcoli}},\ and\ \bibinfo {author} {\bibfnamefont {M.}~\bibnamefont {Cococcioni}},\ }\bibfield  {title} {\bibinfo {title} {{Hubbard-corrected \text{DFT} energy functionals: The $\text{LDA}+\text{U}$ description of correlated systems}},\ }\href {https://doi.org/https://doi.org/10.1002/qua.24521} {\bibfield  {journal} {\bibinfo  {journal} {Int. J. Quant. Chem.}\ }\textbf {\bibinfo {volume} {114}},\ \bibinfo {pages} {14} (\bibinfo {year} {2014})}\BibitemShut {NoStop}%
\bibitem [{\citenamefont {Anisimov}\ \emph {et~al.}(1991)\citenamefont {Anisimov}, \citenamefont {Zaanen},\ and\ \citenamefont {Andersen}}]{Anisimov1991PRB}%
  \BibitemOpen
  \bibfield  {author} {\bibinfo {author} {\bibfnamefont {V.~I.}\ \bibnamefont {Anisimov}}, \bibinfo {author} {\bibfnamefont {J.}~\bibnamefont {Zaanen}},\ and\ \bibinfo {author} {\bibfnamefont {O.~K.}\ \bibnamefont {Andersen}},\ }\bibfield  {title} {\bibinfo {title} {{Band theory and Mott insulators: Hubbard U instead of Stoner I}},\ }\href {https://doi.org/10.1103/PhysRevB.44.943} {\bibfield  {journal} {\bibinfo  {journal} {Phys. Rev. B}\ }\textbf {\bibinfo {volume} {44}},\ \bibinfo {pages} {943} (\bibinfo {year} {1991})}\BibitemShut {NoStop}%
\bibitem [{\citenamefont {Anisimov}\ \emph {et~al.}(1993)\citenamefont {Anisimov}, \citenamefont {Solovyev}, \citenamefont {Korotin}, \citenamefont {Czy\ifmmode~\dot{z}\else \.{z}\fi{}yk},\ and\ \citenamefont {Sawatzky}}]{Anisimov1993PRB}%
  \BibitemOpen
  \bibfield  {author} {\bibinfo {author} {\bibfnamefont {V.~I.}\ \bibnamefont {Anisimov}}, \bibinfo {author} {\bibfnamefont {I.~V.}\ \bibnamefont {Solovyev}}, \bibinfo {author} {\bibfnamefont {M.~A.}\ \bibnamefont {Korotin}}, \bibinfo {author} {\bibfnamefont {M.~T.}\ \bibnamefont {Czy\ifmmode~\dot{z}\else \.{z}\fi{}yk}},\ and\ \bibinfo {author} {\bibfnamefont {G.~A.}\ \bibnamefont {Sawatzky}},\ }\bibfield  {title} {\bibinfo {title} {{Density-functional theory and NiO photoemission spectra}},\ }\href {https://doi.org/10.1103/PhysRevB.48.16929} {\bibfield  {journal} {\bibinfo  {journal} {Phys. Rev. B}\ }\textbf {\bibinfo {volume} {48}},\ \bibinfo {pages} {16929} (\bibinfo {year} {1993})}\BibitemShut {NoStop}%
\bibitem [{\citenamefont {Liechtenstein}\ \emph {et~al.}(1995)\citenamefont {Liechtenstein}, \citenamefont {Anisimov},\ and\ \citenamefont {Zaanen}}]{Liechtenstein1995PRB}%
  \BibitemOpen
  \bibfield  {author} {\bibinfo {author} {\bibfnamefont {A.~I.}\ \bibnamefont {Liechtenstein}}, \bibinfo {author} {\bibfnamefont {V.~I.}\ \bibnamefont {Anisimov}},\ and\ \bibinfo {author} {\bibfnamefont {J.}~\bibnamefont {Zaanen}},\ }\bibfield  {title} {\bibinfo {title} {{Density-functional theory and strong interactions: Orbital ordering in Mott-Hubbard insulators}},\ }\href {https://doi.org/10.1103/PhysRevB.52.R5467} {\bibfield  {journal} {\bibinfo  {journal} {Phys. Rev. B}\ }\textbf {\bibinfo {volume} {52}},\ \bibinfo {pages} {R5467} (\bibinfo {year} {1995})}\BibitemShut {NoStop}%
\bibitem [{\citenamefont {Korotin}\ \emph {et~al.}(2012)\citenamefont {Korotin}, \citenamefont {Kukolev}, \citenamefont {Kozhevnikov}, \citenamefont {Novoselov},\ and\ \citenamefont {Anisimov}}]{Korotin2012JPCM}%
  \BibitemOpen
  \bibfield  {author} {\bibinfo {author} {\bibfnamefont {D.}~\bibnamefont {Korotin}}, \bibinfo {author} {\bibfnamefont {V.}~\bibnamefont {Kukolev}}, \bibinfo {author} {\bibfnamefont {A.~V.}\ \bibnamefont {Kozhevnikov}}, \bibinfo {author} {\bibfnamefont {D.}~\bibnamefont {Novoselov}},\ and\ \bibinfo {author} {\bibfnamefont {V.~I.}\ \bibnamefont {Anisimov}},\ }\bibfield  {title} {\bibinfo {title} {{Electronic correlations and crystal structure distortions in BaBiO$_3$}},\ }\href {https://doi.org/10.1088/0953-8984/24/41/415603} {\bibfield  {journal} {\bibinfo  {journal} {J. Phys. Condens. Matter}\ }\textbf {\bibinfo {volume} {24}},\ \bibinfo {pages} {415603} (\bibinfo {year} {2012})}\BibitemShut {NoStop}%
\bibitem [{\citenamefont {Agapito}\ \emph {et~al.}(2015)\citenamefont {Agapito}, \citenamefont {Curtarolo},\ and\ \citenamefont {{Buongiorno Nardelli}}}]{Agapito2015PRX}%
  \BibitemOpen
  \bibfield  {author} {\bibinfo {author} {\bibfnamefont {L.~A.}\ \bibnamefont {Agapito}}, \bibinfo {author} {\bibfnamefont {S.}~\bibnamefont {Curtarolo}},\ and\ \bibinfo {author} {\bibfnamefont {M.}~\bibnamefont {{Buongiorno Nardelli}}},\ }\bibfield  {title} {\bibinfo {title} {Reformulation of {$\mathrm{DFT}+U$} as a pseudohybrid {Hubbard} density functional for accelerated materials discovery},\ }\href {https://doi.org/10.1103/PhysRevX.5.011006} {\bibfield  {journal} {\bibinfo  {journal} {Phys. Rev. X}\ }\textbf {\bibinfo {volume} {5}},\ \bibinfo {pages} {011006} (\bibinfo {year} {2015})}\BibitemShut {NoStop}%
\bibitem [{\citenamefont {Cococcioni}\ and\ \citenamefont {de~Gironcoli}(2005)}]{Cococcioni2005PRB}%
  \BibitemOpen
  \bibfield  {author} {\bibinfo {author} {\bibfnamefont {M.}~\bibnamefont {Cococcioni}}\ and\ \bibinfo {author} {\bibfnamefont {S.}~\bibnamefont {de~Gironcoli}},\ }\bibfield  {title} {\bibinfo {title} {{Linear response approach to the calculation of the effective interaction parameters in the $\mathrm{LDA}+\mathrm{U}$ method}},\ }\href {https://doi.org/10.1103/PhysRevB.71.035105} {\bibfield  {journal} {\bibinfo  {journal} {Phys. Rev. B}\ }\textbf {\bibinfo {volume} {71}},\ \bibinfo {pages} {035105} (\bibinfo {year} {2005})}\BibitemShut {NoStop}%
\bibitem [{\citenamefont {Kulik}\ \emph {et~al.}(2006)\citenamefont {Kulik}, \citenamefont {Cococcioni}, \citenamefont {Scherlis},\ and\ \citenamefont {Marzari}}]{Kulik2006prl}%
  \BibitemOpen
  \bibfield  {author} {\bibinfo {author} {\bibfnamefont {H.~J.}\ \bibnamefont {Kulik}}, \bibinfo {author} {\bibfnamefont {M.}~\bibnamefont {Cococcioni}}, \bibinfo {author} {\bibfnamefont {D.~A.}\ \bibnamefont {Scherlis}},\ and\ \bibinfo {author} {\bibfnamefont {N.}~\bibnamefont {Marzari}},\ }\bibfield  {title} {\bibinfo {title} {Density functional theory in transition-metal chemistry: A self-consistent \text{Hubbard} $\text{U}$ approach},\ }\href {https://doi.org/10.1103/PhysRevLett.97.103001} {\bibfield  {journal} {\bibinfo  {journal} {Phys. Rev. Lett.}\ }\textbf {\bibinfo {volume} {97}},\ \bibinfo {pages} {103001} (\bibinfo {year} {2006})}\BibitemShut {NoStop}%
\bibitem [{\citenamefont {Timrov}\ \emph {et~al.}(2018)\citenamefont {Timrov}, \citenamefont {Marzari},\ and\ \citenamefont {Cococcioni}}]{Timrov2018prb}%
  \BibitemOpen
  \bibfield  {author} {\bibinfo {author} {\bibfnamefont {I.}~\bibnamefont {Timrov}}, \bibinfo {author} {\bibfnamefont {N.}~\bibnamefont {Marzari}},\ and\ \bibinfo {author} {\bibfnamefont {M.}~\bibnamefont {Cococcioni}},\ }\bibfield  {title} {\bibinfo {title} {Hubbard parameters from density-functional perturbation theory},\ }\href {https://doi.org/10.1103/PhysRevB.98.085127} {\bibfield  {journal} {\bibinfo  {journal} {Phys. Rev. B}\ }\textbf {\bibinfo {volume} {98}},\ \bibinfo {pages} {085127} (\bibinfo {year} {2018})}\BibitemShut {NoStop}%
\bibitem [{\citenamefont {Miyake}\ and\ \citenamefont {Aryasetiawan}(2008)}]{Miyake2008prb}%
  \BibitemOpen
  \bibfield  {author} {\bibinfo {author} {\bibfnamefont {T.}~\bibnamefont {Miyake}}\ and\ \bibinfo {author} {\bibfnamefont {F.}~\bibnamefont {Aryasetiawan}},\ }\bibfield  {title} {\bibinfo {title} {Screened coulomb interaction in the maximally localized \text{Wannier} basis},\ }\href {https://doi.org/10.1103/PhysRevB.77.085122} {\bibfield  {journal} {\bibinfo  {journal} {Phys. Rev. B}\ }\textbf {\bibinfo {volume} {77}},\ \bibinfo {pages} {085122} (\bibinfo {year} {2008})}\BibitemShut {NoStop}%
\bibitem [{\citenamefont {Miyake}\ \emph {et~al.}(2009)\citenamefont {Miyake}, \citenamefont {Aryasetiawan},\ and\ \citenamefont {Imada}}]{Miyake2009prb}%
  \BibitemOpen
  \bibfield  {author} {\bibinfo {author} {\bibfnamefont {T.}~\bibnamefont {Miyake}}, \bibinfo {author} {\bibfnamefont {F.}~\bibnamefont {Aryasetiawan}},\ and\ \bibinfo {author} {\bibfnamefont {M.}~\bibnamefont {Imada}},\ }\bibfield  {title} {\bibinfo {title} {Ab initio procedure for constructing effective models of correlated materials with entangled band structure},\ }\href {https://doi.org/10.1103/PhysRevB.80.155134} {\bibfield  {journal} {\bibinfo  {journal} {Phys. Rev. B}\ }\textbf {\bibinfo {volume} {80}},\ \bibinfo {pages} {155134} (\bibinfo {year} {2009})}\BibitemShut {NoStop}%
\bibitem [{\citenamefont {Aichhorn}\ \emph {et~al.}(2009)\citenamefont {Aichhorn}, \citenamefont {Pourovskii}, \citenamefont {Vildosola}, \citenamefont {Ferrero}, \citenamefont {Parcollet}, \citenamefont {Miyake}, \citenamefont {Georges},\ and\ \citenamefont {Biermann}}]{Aichhorn2009prb}%
  \BibitemOpen
  \bibfield  {author} {\bibinfo {author} {\bibfnamefont {M.}~\bibnamefont {Aichhorn}}, \bibinfo {author} {\bibfnamefont {L.}~\bibnamefont {Pourovskii}}, \bibinfo {author} {\bibfnamefont {V.}~\bibnamefont {Vildosola}}, \bibinfo {author} {\bibfnamefont {M.}~\bibnamefont {Ferrero}}, \bibinfo {author} {\bibfnamefont {O.}~\bibnamefont {Parcollet}}, \bibinfo {author} {\bibfnamefont {T.}~\bibnamefont {Miyake}}, \bibinfo {author} {\bibfnamefont {A.}~\bibnamefont {Georges}},\ and\ \bibinfo {author} {\bibfnamefont {S.}~\bibnamefont {Biermann}},\ }\bibfield  {title} {\bibinfo {title} {Dynamical mean-field theory within an augmented plane-wave framework: Assessing electronic correlations in the iron pnictide {LaFeAsO}},\ }\href {https://doi.org/10.1103/PhysRevB.80.085101} {\bibfield  {journal} {\bibinfo  {journal} {Phys. Rev. B}\ }\textbf {\bibinfo {volume} {80}},\ \bibinfo {pages} {085101} (\bibinfo {year} {2009})}\BibitemShut {NoStop}%
\bibitem [{\citenamefont {Mosey}\ and\ \citenamefont {Carter}(2007)}]{Mosey2007prb}%
  \BibitemOpen
  \bibfield  {author} {\bibinfo {author} {\bibfnamefont {N.~J.}\ \bibnamefont {Mosey}}\ and\ \bibinfo {author} {\bibfnamefont {E.~A.}\ \bibnamefont {Carter}},\ }\bibfield  {title} {\bibinfo {title} {Ab initio evaluation of coulomb and exchange parameters for $\text{DFT}+\text{U}$ calculations},\ }\href {https://doi.org/10.1103/PhysRevB.76.155123} {\bibfield  {journal} {\bibinfo  {journal} {Phys. Rev. B}\ }\textbf {\bibinfo {volume} {76}},\ \bibinfo {pages} {155123} (\bibinfo {year} {2007})}\BibitemShut {NoStop}%
\bibitem [{\citenamefont {Mosey}\ \emph {et~al.}(2008)\citenamefont {Mosey}, \citenamefont {Liao},\ and\ \citenamefont {Carter}}]{Mosey2008jcp}%
  \BibitemOpen
  \bibfield  {author} {\bibinfo {author} {\bibfnamefont {N.~J.}\ \bibnamefont {Mosey}}, \bibinfo {author} {\bibfnamefont {P.}~\bibnamefont {Liao}},\ and\ \bibinfo {author} {\bibfnamefont {E.~A.}\ \bibnamefont {Carter}},\ }\bibfield  {title} {\bibinfo {title} {Rotationally invariant ab initio evaluation of coulomb and exchange parameters for $\text{DFT}+\text{U}$ calculations},\ }\href {https://doi.org/10.1063/1.2943142} {\bibfield  {journal} {\bibinfo  {journal} {J. Chem. Phys.}\ }\textbf {\bibinfo {volume} {129}},\ \bibinfo {pages} {014103} (\bibinfo {year} {2008})}\BibitemShut {NoStop}%
\bibitem [{\citenamefont {Campo~Jr}\ and\ \citenamefont {Cococcioni}(2010)}]{Campo2010JPCM}%
  \BibitemOpen
  \bibfield  {author} {\bibinfo {author} {\bibfnamefont {V.~L.}\ \bibnamefont {Campo~Jr}}\ and\ \bibinfo {author} {\bibfnamefont {M.}~\bibnamefont {Cococcioni}},\ }\bibfield  {title} {\bibinfo {title} {{Extended DFT$+U+V$ method with on-site and inter-site electronic interactions}},\ }\href {https://doi.org/10.1088/0953-8984/22/5/055602} {\bibfield  {journal} {\bibinfo  {journal} {J. Phys. Condens. Matter}\ }\textbf {\bibinfo {volume} {22}},\ \bibinfo {pages} {055602} (\bibinfo {year} {2010})}\BibitemShut {NoStop}%
\bibitem [{\citenamefont {Lee}\ and\ \citenamefont {Son}(2020)}]{Lee2020PRR}%
  \BibitemOpen
  \bibfield  {author} {\bibinfo {author} {\bibfnamefont {S.-H.}\ \bibnamefont {Lee}}\ and\ \bibinfo {author} {\bibfnamefont {Y.-W.}\ \bibnamefont {Son}},\ }\bibfield  {title} {\bibinfo {title} {{First-principles approach with a pseudohybrid density functional for extended Hubbard interactions}},\ }\href {https://doi.org/10.1103/PhysRevResearch.2.043410} {\bibfield  {journal} {\bibinfo  {journal} {Phys. Rev. Research}\ }\textbf {\bibinfo {volume} {2}},\ \bibinfo {pages} {043410} (\bibinfo {year} {2020})}\BibitemShut {NoStop}%
\bibitem [{\citenamefont {Tancogne-Dejean}\ and\ \citenamefont {Rubio}(2020)}]{Rubio2020PRB}%
  \BibitemOpen
  \bibfield  {author} {\bibinfo {author} {\bibfnamefont {N.}~\bibnamefont {Tancogne-Dejean}}\ and\ \bibinfo {author} {\bibfnamefont {A.}~\bibnamefont {Rubio}},\ }\bibfield  {title} {\bibinfo {title} {Parameter-free hybridlike functional based on an extended \text{Hubbard} model: $\mathrm{DFT}+\text{U}+\text{V}$},\ }\href {https://doi.org/10.1103/PhysRevB.102.155117} {\bibfield  {journal} {\bibinfo  {journal} {Phys. Rev. B}\ }\textbf {\bibinfo {volume} {102}},\ \bibinfo {pages} {155117} (\bibinfo {year} {2020})}\BibitemShut {NoStop}%
\bibitem [{\citenamefont {Yang}\ \emph {et~al.}(2021)\citenamefont {Yang}, \citenamefont {Jhi}, \citenamefont {Lee},\ and\ \citenamefont {Son}}]{Yang2021PRB}%
  \BibitemOpen
  \bibfield  {author} {\bibinfo {author} {\bibfnamefont {W.}~\bibnamefont {Yang}}, \bibinfo {author} {\bibfnamefont {S.-H.}\ \bibnamefont {Jhi}}, \bibinfo {author} {\bibfnamefont {S.-H.}\ \bibnamefont {Lee}},\ and\ \bibinfo {author} {\bibfnamefont {Y.-W.}\ \bibnamefont {Son}},\ }\bibfield  {title} {\bibinfo {title} {{Ab initio study of lattice dynamics of group {IV} semiconductors using pseudohybrid functionals for extended Hubbard interactions}},\ }\href {https://doi.org/10.1103/PhysRevB.104.104313} {\bibfield  {journal} {\bibinfo  {journal} {Phys. Rev. B}\ }\textbf {\bibinfo {volume} {104}},\ \bibinfo {pages} {104313} (\bibinfo {year} {2021})}\BibitemShut {NoStop}%
\bibitem [{\citenamefont {Yang}\ \emph {et~al.}(2022)\citenamefont {Yang}, \citenamefont {Jang}, \citenamefont {Son},\ and\ \citenamefont {Jhi}}]{Yang2022JPC}%
  \BibitemOpen
  \bibfield  {author} {\bibinfo {author} {\bibfnamefont {W.}~\bibnamefont {Yang}}, \bibinfo {author} {\bibfnamefont {B.~G.}\ \bibnamefont {Jang}}, \bibinfo {author} {\bibfnamefont {Y.-W.}\ \bibnamefont {Son}},\ and\ \bibinfo {author} {\bibfnamefont {S.-H.}\ \bibnamefont {Jhi}},\ }\bibfield  {title} {\bibinfo {title} {{Lattice dynamical properties of antiferromagnetic oxides calculated using self-consistent extended Hubbard functional method}},\ }\href {https://doi.org/10.1088/1361-648X/ac6c69} {\bibfield  {journal} {\bibinfo  {journal} {J. Phys.: Condens. Matter}\ }\textbf {\bibinfo {volume} {34}},\ \bibinfo {pages} {295601} (\bibinfo {year} {2022})}\BibitemShut {NoStop}%
\bibitem [{\citenamefont {Yang}\ and\ \citenamefont {Son}(2024)}]{Yang2024prb}%
  \BibitemOpen
  \bibfield  {author} {\bibinfo {author} {\bibfnamefont {W.}~\bibnamefont {Yang}}\ and\ \bibinfo {author} {\bibfnamefont {Y.-W.}\ \bibnamefont {Son}},\ }\bibfield  {title} {\bibinfo {title} {Effects of self-consistent extended hubbard interactions and spin-orbit couplings on energy bands of semiconductors and topological insulators},\ }\href {https://doi.org/10.1103/PhysRevB.110.155133} {\bibfield  {journal} {\bibinfo  {journal} {Phys. Rev. B}\ }\textbf {\bibinfo {volume} {110}},\ \bibinfo {pages} {155133} (\bibinfo {year} {2024})}\BibitemShut {NoStop}%
\bibitem [{\citenamefont {Jang}\ \emph {et~al.}(2023)\citenamefont {Jang}, \citenamefont {Kim}, \citenamefont {Lee}, \citenamefont {Yang}, \citenamefont {Jhi},\ and\ \citenamefont {Son}}]{Jang2023PRL}%
  \BibitemOpen
  \bibfield  {author} {\bibinfo {author} {\bibfnamefont {B.~G.}\ \bibnamefont {Jang}}, \bibinfo {author} {\bibfnamefont {M.}~\bibnamefont {Kim}}, \bibinfo {author} {\bibfnamefont {S.-H.}\ \bibnamefont {Lee}}, \bibinfo {author} {\bibfnamefont {W.}~\bibnamefont {Yang}}, \bibinfo {author} {\bibfnamefont {S.-H.}\ \bibnamefont {Jhi}},\ and\ \bibinfo {author} {\bibfnamefont {Y.-W.}\ \bibnamefont {Son}},\ }\bibfield  {title} {\bibinfo {title} {Intersite coulomb interactions in charge-ordered systems},\ }\href {https://doi.org/10.1103/PhysRevLett.130.136401} {\bibfield  {journal} {\bibinfo  {journal} {Phys. Rev. Lett.}\ }\textbf {\bibinfo {volume} {130}},\ \bibinfo {pages} {136401} (\bibinfo {year} {2023})}\BibitemShut {NoStop}%
\bibitem [{\citenamefont {Floris}\ \emph {et~al.}(2011)\citenamefont {Floris}, \citenamefont {de~Gironcoli}, \citenamefont {Gross},\ and\ \citenamefont {Cococcioni}}]{Floris2011prb}%
  \BibitemOpen
  \bibfield  {author} {\bibinfo {author} {\bibfnamefont {A.}~\bibnamefont {Floris}}, \bibinfo {author} {\bibfnamefont {S.}~\bibnamefont {de~Gironcoli}}, \bibinfo {author} {\bibfnamefont {E.~K.~U.}\ \bibnamefont {Gross}},\ and\ \bibinfo {author} {\bibfnamefont {M.}~\bibnamefont {Cococcioni}},\ }\bibfield  {title} {\bibinfo {title} {{Vibrational properties of MnO and NiO from DFT $+U$-based density functional perturbation theory}},\ }\href {https://doi.org/10.1103/PhysRevB.84.161102} {\bibfield  {journal} {\bibinfo  {journal} {Phys. Rev. B}\ }\textbf {\bibinfo {volume} {84}},\ \bibinfo {pages} {161102} (\bibinfo {year} {2011})}\BibitemShut {NoStop}%
\bibitem [{\citenamefont {Floris}\ \emph {et~al.}(2020)\citenamefont {Floris}, \citenamefont {Timrov}, \citenamefont {Himmetoglu}, \citenamefont {Marzari}, \citenamefont {de~Gironcoli},\ and\ \citenamefont {Cococcioni}}]{Floris2020prb}%
  \BibitemOpen
  \bibfield  {author} {\bibinfo {author} {\bibfnamefont {A.}~\bibnamefont {Floris}}, \bibinfo {author} {\bibfnamefont {I.}~\bibnamefont {Timrov}}, \bibinfo {author} {\bibfnamefont {B.}~\bibnamefont {Himmetoglu}}, \bibinfo {author} {\bibfnamefont {N.}~\bibnamefont {Marzari}}, \bibinfo {author} {\bibfnamefont {S.}~\bibnamefont {de~Gironcoli}},\ and\ \bibinfo {author} {\bibfnamefont {M.}~\bibnamefont {Cococcioni}},\ }\bibfield  {title} {\bibinfo {title} {Hubbard-corrected density functional perturbation theory with ultrasoft pseudopotentials},\ }\href {https://doi.org/10.1103/PhysRevB.101.064305} {\bibfield  {journal} {\bibinfo  {journal} {Phys. Rev. B}\ }\textbf {\bibinfo {volume} {101}},\ \bibinfo {pages} {064305} (\bibinfo {year} {2020})}\BibitemShut {NoStop}%
\bibitem [{\citenamefont {Zhou}\ \emph {et~al.}(2021)\citenamefont {Zhou}, \citenamefont {Park}, \citenamefont {Timrov}, \citenamefont {Floris}, \citenamefont {Cococcioni}, \citenamefont {Marzari},\ and\ \citenamefont {Bernardi}}]{Zhou2021prl}%
  \BibitemOpen
  \bibfield  {author} {\bibinfo {author} {\bibfnamefont {J.-J.}\ \bibnamefont {Zhou}}, \bibinfo {author} {\bibfnamefont {J.}~\bibnamefont {Park}}, \bibinfo {author} {\bibfnamefont {I.}~\bibnamefont {Timrov}}, \bibinfo {author} {\bibfnamefont {A.}~\bibnamefont {Floris}}, \bibinfo {author} {\bibfnamefont {M.}~\bibnamefont {Cococcioni}}, \bibinfo {author} {\bibfnamefont {N.}~\bibnamefont {Marzari}},\ and\ \bibinfo {author} {\bibfnamefont {M.}~\bibnamefont {Bernardi}},\ }\bibfield  {title} {\bibinfo {title} {Ab initio electron-phonon interactions in correlated electron systems},\ }\href {https://doi.org/10.1103/PhysRevLett.127.126404} {\bibfield  {journal} {\bibinfo  {journal} {Phys. Rev. Lett.}\ }\textbf {\bibinfo {volume} {127}},\ \bibinfo {pages} {126404} (\bibinfo {year} {2021})}\BibitemShut {NoStop}%
\bibitem [{\citenamefont {Hedin}(1965)}]{Hedin1965pr}%
  \BibitemOpen
  \bibfield  {author} {\bibinfo {author} {\bibfnamefont {L.}~\bibnamefont {Hedin}},\ }\bibfield  {title} {\bibinfo {title} {New method for calculating the one-particle green's function with application to the electron-gas problem},\ }\href {https://doi.org/10.1103/PhysRev.139.A796} {\bibfield  {journal} {\bibinfo  {journal} {Phys. Rev.}\ }\textbf {\bibinfo {volume} {139}},\ \bibinfo {pages} {A796} (\bibinfo {year} {1965})}\BibitemShut {NoStop}%
\bibitem [{\citenamefont {Hybertsen}\ and\ \citenamefont {Louie}(1986)}]{Hybertsen1986prb}%
  \BibitemOpen
  \bibfield  {author} {\bibinfo {author} {\bibfnamefont {M.~S.}\ \bibnamefont {Hybertsen}}\ and\ \bibinfo {author} {\bibfnamefont {S.~G.}\ \bibnamefont {Louie}},\ }\bibfield  {title} {\bibinfo {title} {Electron correlation in semiconductors and insulators: Band gaps and quasiparticle energies},\ }\href {https://doi.org/10.1103/PhysRevB.34.5390} {\bibfield  {journal} {\bibinfo  {journal} {Phys. Rev. B}\ }\textbf {\bibinfo {volume} {34}},\ \bibinfo {pages} {5390} (\bibinfo {year} {1986})}\BibitemShut {NoStop}%
\bibitem [{\citenamefont {Shishkin}\ \emph {et~al.}(2007)\citenamefont {Shishkin}, \citenamefont {Marsman},\ and\ \citenamefont {Kresse}}]{Shishkin2007prl}%
  \BibitemOpen
  \bibfield  {author} {\bibinfo {author} {\bibfnamefont {M.}~\bibnamefont {Shishkin}}, \bibinfo {author} {\bibfnamefont {M.}~\bibnamefont {Marsman}},\ and\ \bibinfo {author} {\bibfnamefont {G.}~\bibnamefont {Kresse}},\ }\bibfield  {title} {\bibinfo {title} {Accurate quasiparticle spectra from self-consistent \text{GW} calculations with vertex corrections},\ }\href {https://doi.org/10.1103/PhysRevLett.99.246403} {\bibfield  {journal} {\bibinfo  {journal} {Phys. Rev. Lett.}\ }\textbf {\bibinfo {volume} {99}},\ \bibinfo {pages} {246403} (\bibinfo {year} {2007})}\BibitemShut {NoStop}%
\bibitem [{\citenamefont {Li}\ \emph {et~al.}(2019)\citenamefont {Li}, \citenamefont {Antonius}, \citenamefont {Wu}, \citenamefont {da~Jornada},\ and\ \citenamefont {Louie}}]{Zhenglu2019prl}%
  \BibitemOpen
  \bibfield  {author} {\bibinfo {author} {\bibfnamefont {Z.}~\bibnamefont {Li}}, \bibinfo {author} {\bibfnamefont {G.}~\bibnamefont {Antonius}}, \bibinfo {author} {\bibfnamefont {M.}~\bibnamefont {Wu}}, \bibinfo {author} {\bibfnamefont {F.~H.}\ \bibnamefont {da~Jornada}},\ and\ \bibinfo {author} {\bibfnamefont {S.~G.}\ \bibnamefont {Louie}},\ }\bibfield  {title} {\bibinfo {title} {Electron-phonon coupling from ab initio linear-response theory within the $gw$ method: Correlation-enhanced interactions and superconductivity in ${\mathrm{ba}}_{1\ensuremath{-}x}{\mathrm{k}}_{x}{\mathrm{bio}}_{3}$},\ }\href {https://doi.org/10.1103/PhysRevLett.122.186402} {\bibfield  {journal} {\bibinfo  {journal} {Phys. Rev. Lett.}\ }\textbf {\bibinfo {volume} {122}},\ \bibinfo {pages} {186402} (\bibinfo {year} {2019})}\BibitemShut {NoStop}%
\bibitem [{\citenamefont {Li}\ \emph {et~al.}(2021)\citenamefont {Li}, \citenamefont {Wu}, \citenamefont {Chan},\ and\ \citenamefont {Louie}}]{Zhenglu2021prl}%
  \BibitemOpen
  \bibfield  {author} {\bibinfo {author} {\bibfnamefont {Z.}~\bibnamefont {Li}}, \bibinfo {author} {\bibfnamefont {M.}~\bibnamefont {Wu}}, \bibinfo {author} {\bibfnamefont {Y.-H.}\ \bibnamefont {Chan}},\ and\ \bibinfo {author} {\bibfnamefont {S.~G.}\ \bibnamefont {Louie}},\ }\bibfield  {title} {\bibinfo {title} {Unmasking the origin of kinks in the photoemission spectra of cuprate superconductors},\ }\href {https://doi.org/10.1103/PhysRevLett.126.146401} {\bibfield  {journal} {\bibinfo  {journal} {Phys. Rev. Lett.}\ }\textbf {\bibinfo {volume} {126}},\ \bibinfo {pages} {146401} (\bibinfo {year} {2021})}\BibitemShut {NoStop}%
\bibitem [{\citenamefont {Li}\ and\ \citenamefont {Louie}(2024)}]{Zhenglu2024prl}%
  \BibitemOpen
  \bibfield  {author} {\bibinfo {author} {\bibfnamefont {Z.}~\bibnamefont {Li}}\ and\ \bibinfo {author} {\bibfnamefont {S.~G.}\ \bibnamefont {Louie}},\ }\bibfield  {title} {\bibinfo {title} {Two-gap superconductivity and the decisive role of rare-earth $d$ electrons in infinite-layer nickelates},\ }\href {https://doi.org/10.1103/PhysRevLett.133.126401} {\bibfield  {journal} {\bibinfo  {journal} {Phys. Rev. Lett.}\ }\textbf {\bibinfo {volume} {133}},\ \bibinfo {pages} {126401} (\bibinfo {year} {2024})}\BibitemShut {NoStop}%
\bibitem [{\citenamefont {Li}\ \emph {et~al.}(2024)\citenamefont {Li}, \citenamefont {Antonius}, \citenamefont {Chan},\ and\ \citenamefont {Louie}}]{Zhanglu2024CPC}%
  \BibitemOpen
  \bibfield  {author} {\bibinfo {author} {\bibfnamefont {Z.}~\bibnamefont {Li}}, \bibinfo {author} {\bibfnamefont {G.}~\bibnamefont {Antonius}}, \bibinfo {author} {\bibfnamefont {Y.-H.}\ \bibnamefont {Chan}},\ and\ \bibinfo {author} {\bibfnamefont {S.~G.}\ \bibnamefont {Louie}},\ }\bibfield  {title} {\bibinfo {title} {{Electron-phonon coupling from GW perturbation theory: Practical workflow combining BerkeleyGW, ABINIT, and EPW}},\ }\href {https://doi.org/https://doi.org/10.1016/j.cpc.2023.109003} {\bibfield  {journal} {\bibinfo  {journal} {Comp. Phys. Commun.}\ }\textbf {\bibinfo {volume} {295}},\ \bibinfo {pages} {109003} (\bibinfo {year} {2024})}\BibitemShut {NoStop}%
\bibitem [{\citenamefont {Restrepo}\ \emph {et~al.}(2009)\citenamefont {Restrepo}, \citenamefont {Varga},\ and\ \citenamefont {Pantelides}}]{Restrepo2009apl}%
  \BibitemOpen
  \bibfield  {author} {\bibinfo {author} {\bibfnamefont {O.~D.}\ \bibnamefont {Restrepo}}, \bibinfo {author} {\bibfnamefont {K.}~\bibnamefont {Varga}},\ and\ \bibinfo {author} {\bibfnamefont {S.~T.}\ \bibnamefont {Pantelides}},\ }\bibfield  {title} {\bibinfo {title} {First-principles calculations of electron mobilities in silicon: Phonon and coulomb scattering},\ }\href {https://doi.org/10.1063/1.3147189} {\bibfield  {journal} {\bibinfo  {journal} {Appl. Phys. Lett.}\ }\textbf {\bibinfo {volume} {94}},\ \bibinfo {pages} {212103} (\bibinfo {year} {2009})}\BibitemShut {NoStop}%
\bibitem [{\citenamefont {Fiorentini}\ and\ \citenamefont {Bonini}(2016)}]{Fiorentini2016prb}%
  \BibitemOpen
  \bibfield  {author} {\bibinfo {author} {\bibfnamefont {M.}~\bibnamefont {Fiorentini}}\ and\ \bibinfo {author} {\bibfnamefont {N.}~\bibnamefont {Bonini}},\ }\bibfield  {title} {\bibinfo {title} {Thermoelectric coefficients of $n$-doped silicon from first principles via the solution of the boltzmann transport equation},\ }\href {https://doi.org/10.1103/PhysRevB.94.085204} {\bibfield  {journal} {\bibinfo  {journal} {Phys. Rev. B}\ }\textbf {\bibinfo {volume} {94}},\ \bibinfo {pages} {085204} (\bibinfo {year} {2016})}\BibitemShut {NoStop}%
\bibitem [{\citenamefont {Poncé}\ \emph {et~al.}(2016)\citenamefont {Poncé}, \citenamefont {Margine}, \citenamefont {Verdi},\ and\ \citenamefont {Giustino}}]{Ponce2016EPW}%
  \BibitemOpen
  \bibfield  {author} {\bibinfo {author} {\bibfnamefont {S.}~\bibnamefont {Poncé}}, \bibinfo {author} {\bibfnamefont {E.}~\bibnamefont {Margine}}, \bibinfo {author} {\bibfnamefont {C.}~\bibnamefont {Verdi}},\ and\ \bibinfo {author} {\bibfnamefont {F.}~\bibnamefont {Giustino}},\ }\bibfield  {title} {\bibinfo {title} {{EPW: Electron–phonon coupling, transport and superconducting properties using maximally localized Wannier functions}},\ }\href {https://doi.org/https://doi.org/10.1016/j.cpc.2016.07.028} {\bibfield  {journal} {\bibinfo  {journal} {Comp. Phys. Commun.}\ }\textbf {\bibinfo {volume} {209}},\ \bibinfo {pages} {116} (\bibinfo {year} {2016})}\BibitemShut {NoStop}%
\bibitem [{\citenamefont {Ponc{\'e}}\ \emph {et~al.}(2020)\citenamefont {Ponc{\'e}}, \citenamefont {Li}, \citenamefont {Reichardt},\ and\ \citenamefont {Giustino}}]{ponce2020rpp}%
  \BibitemOpen
  \bibfield  {author} {\bibinfo {author} {\bibfnamefont {S.}~\bibnamefont {Ponc{\'e}}}, \bibinfo {author} {\bibfnamefont {W.}~\bibnamefont {Li}}, \bibinfo {author} {\bibfnamefont {S.}~\bibnamefont {Reichardt}},\ and\ \bibinfo {author} {\bibfnamefont {F.}~\bibnamefont {Giustino}},\ }\bibfield  {title} {\bibinfo {title} {First-principles calculations of charge carrier mobility and conductivity in bulk semiconductors and two-dimensional materials},\ }\href {https://doi.org/10.1088/1361-6633/ab6a43} {\bibfield  {journal} {\bibinfo  {journal} {Rep. Prog. Phys.}\ }\textbf {\bibinfo {volume} {83}},\ \bibinfo {pages} {036501} (\bibinfo {year} {2020})}\BibitemShut {NoStop}%
\bibitem [{\citenamefont {Lee}\ \emph {et~al.}(2023)\citenamefont {Lee}, \citenamefont {Ponc{\'e}}, \citenamefont {Bushick}, \citenamefont {Hajinazar}, \citenamefont {Lafuente-Bartolome}, \citenamefont {Leveillee}, \citenamefont {Lian}, \citenamefont {Lihm}, \citenamefont {Macheda}, \citenamefont {Mori}, \citenamefont {Paudyal}, \citenamefont {Sio}, \citenamefont {Tiwari}, \citenamefont {Zacharias}, \citenamefont {Zhang}, \citenamefont {Bonini}, \citenamefont {Kioupakis}, \citenamefont {Margine},\ and\ \citenamefont {Giustino}}]{Lee2023npj}%
  \BibitemOpen
  \bibfield  {author} {\bibinfo {author} {\bibfnamefont {H.}~\bibnamefont {Lee}}, \bibinfo {author} {\bibfnamefont {S.}~\bibnamefont {Ponc{\'e}}}, \bibinfo {author} {\bibfnamefont {K.}~\bibnamefont {Bushick}}, \bibinfo {author} {\bibfnamefont {S.}~\bibnamefont {Hajinazar}}, \bibinfo {author} {\bibfnamefont {J.}~\bibnamefont {Lafuente-Bartolome}}, \bibinfo {author} {\bibfnamefont {J.}~\bibnamefont {Leveillee}}, \bibinfo {author} {\bibfnamefont {C.}~\bibnamefont {Lian}}, \bibinfo {author} {\bibfnamefont {J.-M.}\ \bibnamefont {Lihm}}, \bibinfo {author} {\bibfnamefont {F.}~\bibnamefont {Macheda}}, \bibinfo {author} {\bibfnamefont {H.}~\bibnamefont {Mori}}, \bibinfo {author} {\bibfnamefont {H.}~\bibnamefont {Paudyal}}, \bibinfo {author} {\bibfnamefont {W.~H.}\ \bibnamefont {Sio}}, \bibinfo {author} {\bibfnamefont {S.}~\bibnamefont {Tiwari}}, \bibinfo {author} {\bibfnamefont {M.}~\bibnamefont {Zacharias}}, \bibinfo {author} {\bibfnamefont {X.}~\bibnamefont {Zhang}}, \bibinfo {author} {\bibfnamefont
  {N.}~\bibnamefont {Bonini}}, \bibinfo {author} {\bibfnamefont {E.}~\bibnamefont {Kioupakis}}, \bibinfo {author} {\bibfnamefont {E.~R.}\ \bibnamefont {Margine}},\ and\ \bibinfo {author} {\bibfnamefont {F.}~\bibnamefont {Giustino}},\ }\bibfield  {title} {\bibinfo {title} {Electron--phonon physics from first principles using the {EPW} code},\ }\href {https://doi.org/10.1038/s41524-023-01107-3} {\bibfield  {journal} {\bibinfo  {journal} {Npj Comput. Mater.}\ }\textbf {\bibinfo {volume} {9}},\ \bibinfo {pages} {156} (\bibinfo {year} {2023})}\BibitemShut {NoStop}%
\bibitem [{\citenamefont {L\"owdin}(1950)}]{Lowdin1950jcp}%
  \BibitemOpen
  \bibfield  {author} {\bibinfo {author} {\bibfnamefont {P.}~\bibnamefont {L\"owdin}},\ }\bibfield  {title} {\bibinfo {title} {On the non‐orthogonality problem connected with the use of atomic wave functions in the theory of molecules and crystals},\ }\href {https://doi.org/10.1063/1.1747632} {\bibfield  {journal} {\bibinfo  {journal} {J. Chem. Phys.}\ }\textbf {\bibinfo {volume} {18}},\ \bibinfo {pages} {365} (\bibinfo {year} {1950})}\BibitemShut {NoStop}%
\bibitem [{\citenamefont {Ponc\'e}\ \emph {et~al.}(2021)\citenamefont {Ponc\'e}, \citenamefont {Macheda}, \citenamefont {Margine}, \citenamefont {Marzari}, \citenamefont {Bonini},\ and\ \citenamefont {Giustino}}]{Ponce2021prr}%
  \BibitemOpen
  \bibfield  {author} {\bibinfo {author} {\bibfnamefont {S.}~\bibnamefont {Ponc\'e}}, \bibinfo {author} {\bibfnamefont {F.}~\bibnamefont {Macheda}}, \bibinfo {author} {\bibfnamefont {E.~R.}\ \bibnamefont {Margine}}, \bibinfo {author} {\bibfnamefont {N.}~\bibnamefont {Marzari}}, \bibinfo {author} {\bibfnamefont {N.}~\bibnamefont {Bonini}},\ and\ \bibinfo {author} {\bibfnamefont {F.}~\bibnamefont {Giustino}},\ }\bibfield  {title} {\bibinfo {title} {First-principles predictions of hall and drift mobilities in semiconductors},\ }\href {https://doi.org/10.1103/PhysRevResearch.3.043022} {\bibfield  {journal} {\bibinfo  {journal} {Phys. Rev. Res.}\ }\textbf {\bibinfo {volume} {3}},\ \bibinfo {pages} {043022} (\bibinfo {year} {2021})}\BibitemShut {NoStop}%
\bibitem [{\citenamefont {Tiwari}\ \emph {et~al.}(2024)\citenamefont {Tiwari}, \citenamefont {Kioupakis}, \citenamefont {Menendez},\ and\ \citenamefont {Giustino}}]{Tiwari2024prb}%
  \BibitemOpen
  \bibfield  {author} {\bibinfo {author} {\bibfnamefont {S.}~\bibnamefont {Tiwari}}, \bibinfo {author} {\bibfnamefont {E.}~\bibnamefont {Kioupakis}}, \bibinfo {author} {\bibfnamefont {J.}~\bibnamefont {Menendez}},\ and\ \bibinfo {author} {\bibfnamefont {F.}~\bibnamefont {Giustino}},\ }\bibfield  {title} {\bibinfo {title} {Unified theory of optical absorption and luminescence including both direct and phonon-assisted processes},\ }\href {https://doi.org/10.1103/PhysRevB.109.195127} {\bibfield  {journal} {\bibinfo  {journal} {Phys. Rev. B}\ }\textbf {\bibinfo {volume} {109}},\ \bibinfo {pages} {195127} (\bibinfo {year} {2024})}\BibitemShut {NoStop}%
\bibitem [{\citenamefont {Giustino}\ \emph {et~al.}(2007)\citenamefont {Giustino}, \citenamefont {Cohen},\ and\ \citenamefont {Louie}}]{Giustino2007prb}%
  \BibitemOpen
  \bibfield  {author} {\bibinfo {author} {\bibfnamefont {F.}~\bibnamefont {Giustino}}, \bibinfo {author} {\bibfnamefont {M.~L.}\ \bibnamefont {Cohen}},\ and\ \bibinfo {author} {\bibfnamefont {S.~G.}\ \bibnamefont {Louie}},\ }\bibfield  {title} {\bibinfo {title} {Electron-phonon interaction using {Wannier} functions},\ }\href {https://doi.org/10.1103/PhysRevB.76.165108} {\bibfield  {journal} {\bibinfo  {journal} {Phys. Rev. B}\ }\textbf {\bibinfo {volume} {76}},\ \bibinfo {pages} {165108} (\bibinfo {year} {2007})}\BibitemShut {NoStop}%
\bibitem [{\citenamefont {Timrov}\ \emph {et~al.}(2020)\citenamefont {Timrov}, \citenamefont {Aquilante}, \citenamefont {Binci}, \citenamefont {Cococcioni},\ and\ \citenamefont {Marzari}}]{Timrov2020prb}%
  \BibitemOpen
  \bibfield  {author} {\bibinfo {author} {\bibfnamefont {I.}~\bibnamefont {Timrov}}, \bibinfo {author} {\bibfnamefont {F.}~\bibnamefont {Aquilante}}, \bibinfo {author} {\bibfnamefont {L.}~\bibnamefont {Binci}}, \bibinfo {author} {\bibfnamefont {M.}~\bibnamefont {Cococcioni}},\ and\ \bibinfo {author} {\bibfnamefont {N.}~\bibnamefont {Marzari}},\ }\bibfield  {title} {\bibinfo {title} {Pulay forces in density-functional theory with extended hubbard functionals: From nonorthogonalized to orthogonalized manifolds},\ }\href {https://doi.org/10.1103/PhysRevB.102.235159} {\bibfield  {journal} {\bibinfo  {journal} {Phys. Rev. B}\ }\textbf {\bibinfo {volume} {102}},\ \bibinfo {pages} {235159} (\bibinfo {year} {2020})}\BibitemShut {NoStop}%
\bibitem [{\citenamefont {Wu}\ and\ \citenamefont {Van~Voorhis}(2006)}]{Wu2006jpca}%
  \BibitemOpen
  \bibfield  {author} {\bibinfo {author} {\bibfnamefont {Q.}~\bibnamefont {Wu}}\ and\ \bibinfo {author} {\bibfnamefont {T.}~\bibnamefont {Van~Voorhis}},\ }\bibfield  {title} {\bibinfo {title} {Direct calculation of electron transfer parameters through constrained density functional theory},\ }\href {https://doi.org/10.1021/jp061848y} {\bibfield  {journal} {\bibinfo  {journal} {J. Phys. Chem. A}\ }\textbf {\bibinfo {volume} {110}},\ \bibinfo {pages} {9212} (\bibinfo {year} {2006})}\BibitemShut {NoStop}%
\bibitem [{\citenamefont {Maradudin}\ and\ \citenamefont {Vosko}(1968)}]{Maradudin1968rmp}%
  \BibitemOpen
  \bibfield  {author} {\bibinfo {author} {\bibfnamefont {A.~A.}\ \bibnamefont {Maradudin}}\ and\ \bibinfo {author} {\bibfnamefont {S.~H.}\ \bibnamefont {Vosko}},\ }\bibfield  {title} {\bibinfo {title} {Symmetry properties of the normal vibrations of a crystal},\ }\href {https://doi.org/10.1103/RevModPhys.40.1} {\bibfield  {journal} {\bibinfo  {journal} {Rev. Mod. Phys.}\ }\textbf {\bibinfo {volume} {40}},\ \bibinfo {pages} {1} (\bibinfo {year} {1968})}\BibitemShut {NoStop}%
\bibitem [{\citenamefont {Leveillee}\ \emph {et~al.}(2023)\citenamefont {Leveillee}, \citenamefont {Zhang}, \citenamefont {Kioupakis},\ and\ \citenamefont {Giustino}}]{Leveillee2023prb}%
  \BibitemOpen
  \bibfield  {author} {\bibinfo {author} {\bibfnamefont {J.}~\bibnamefont {Leveillee}}, \bibinfo {author} {\bibfnamefont {X.}~\bibnamefont {Zhang}}, \bibinfo {author} {\bibfnamefont {E.}~\bibnamefont {Kioupakis}},\ and\ \bibinfo {author} {\bibfnamefont {F.}~\bibnamefont {Giustino}},\ }\bibfield  {title} {\bibinfo {title} {Ab initio calculation of carrier mobility in semiconductors including ionized-impurity scattering},\ }\href {https://doi.org/10.1103/PhysRevB.107.125207} {\bibfield  {journal} {\bibinfo  {journal} {Phys. Rev. B}\ }\textbf {\bibinfo {volume} {107}},\ \bibinfo {pages} {125207} (\bibinfo {year} {2023})}\BibitemShut {NoStop}%
\bibitem [{\citenamefont {Giustino}(2014)}]{GiustinoBook}%
  \BibitemOpen
  \bibfield  {author} {\bibinfo {author} {\bibfnamefont {F.}~\bibnamefont {Giustino}},\ }\href@noop {} {\emph {\bibinfo {title} {Materials Modelling Using Density Functional Theory: Properties and Predictions}}}\ (\bibinfo  {publisher} {Oxford University Press},\ \bibinfo {year} {2014})\BibitemShut {NoStop}%
\bibitem [{\citenamefont {Giannozzi}\ \emph {et~al.}(2009)\citenamefont {Giannozzi}, \citenamefont {Baroni}, \citenamefont {Bonini}, \citenamefont {Calandra}, \citenamefont {Car}, \citenamefont {Cavazzoni}, \citenamefont {Ceresoli}, \citenamefont {Chiarotti}, \citenamefont {Cococcioni}, \citenamefont {Dabo}, \citenamefont {Corso}, \citenamefont {de~Gironcoli}, \citenamefont {Fabris}, \citenamefont {Fratesi}, \citenamefont {Gebauer}, \citenamefont {Gerstmann}, \citenamefont {Gougoussis}, \citenamefont {Kokalj}, \citenamefont {Lazzeri}, \citenamefont {Martin-Samos}, \citenamefont {Marzari}, \citenamefont {Mauri}, \citenamefont {Mazzarello}, \citenamefont {Paolini}, \citenamefont {Pasquarello}, \citenamefont {Paulatto}, \citenamefont {Sbraccia}, \citenamefont {Scandolo}, \citenamefont {Sclauzero}, \citenamefont {Seitsonen}, \citenamefont {Smogunov}, \citenamefont {Umari},\ and\ \citenamefont {Wentzcovitch}}]{Giannozzi2009JPC}%
  \BibitemOpen
  \bibfield  {author} {\bibinfo {author} {\bibfnamefont {P.}~\bibnamefont {Giannozzi}}, \bibinfo {author} {\bibfnamefont {S.}~\bibnamefont {Baroni}}, \bibinfo {author} {\bibfnamefont {N.}~\bibnamefont {Bonini}}, \bibinfo {author} {\bibfnamefont {M.}~\bibnamefont {Calandra}}, \bibinfo {author} {\bibfnamefont {R.}~\bibnamefont {Car}}, \bibinfo {author} {\bibfnamefont {C.}~\bibnamefont {Cavazzoni}}, \bibinfo {author} {\bibfnamefont {D.}~\bibnamefont {Ceresoli}}, \bibinfo {author} {\bibfnamefont {G.~L.}\ \bibnamefont {Chiarotti}}, \bibinfo {author} {\bibfnamefont {M.}~\bibnamefont {Cococcioni}}, \bibinfo {author} {\bibfnamefont {I.}~\bibnamefont {Dabo}}, \bibinfo {author} {\bibfnamefont {A.~D.}\ \bibnamefont {Corso}}, \bibinfo {author} {\bibfnamefont {S.}~\bibnamefont {de~Gironcoli}}, \bibinfo {author} {\bibfnamefont {S.}~\bibnamefont {Fabris}}, \bibinfo {author} {\bibfnamefont {G.}~\bibnamefont {Fratesi}}, \bibinfo {author} {\bibfnamefont {R.}~\bibnamefont {Gebauer}}, \bibinfo {author} {\bibfnamefont
  {U.}~\bibnamefont {Gerstmann}}, \bibinfo {author} {\bibfnamefont {C.}~\bibnamefont {Gougoussis}}, \bibinfo {author} {\bibfnamefont {A.}~\bibnamefont {Kokalj}}, \bibinfo {author} {\bibfnamefont {M.}~\bibnamefont {Lazzeri}}, \bibinfo {author} {\bibfnamefont {L.}~\bibnamefont {Martin-Samos}}, \bibinfo {author} {\bibfnamefont {N.}~\bibnamefont {Marzari}}, \bibinfo {author} {\bibfnamefont {F.}~\bibnamefont {Mauri}}, \bibinfo {author} {\bibfnamefont {R.}~\bibnamefont {Mazzarello}}, \bibinfo {author} {\bibfnamefont {S.}~\bibnamefont {Paolini}}, \bibinfo {author} {\bibfnamefont {A.}~\bibnamefont {Pasquarello}}, \bibinfo {author} {\bibfnamefont {L.}~\bibnamefont {Paulatto}}, \bibinfo {author} {\bibfnamefont {C.}~\bibnamefont {Sbraccia}}, \bibinfo {author} {\bibfnamefont {S.}~\bibnamefont {Scandolo}}, \bibinfo {author} {\bibfnamefont {G.}~\bibnamefont {Sclauzero}}, \bibinfo {author} {\bibfnamefont {A.~P.}\ \bibnamefont {Seitsonen}}, \bibinfo {author} {\bibfnamefont {A.}~\bibnamefont {Smogunov}}, \bibinfo {author}
  {\bibfnamefont {P.}~\bibnamefont {Umari}},\ and\ \bibinfo {author} {\bibfnamefont {R.~M.}\ \bibnamefont {Wentzcovitch}},\ }\bibfield  {title} {\bibinfo {title} {{QUANTUM ESPRESSO: a modular and open-source software project for quantum simulations of materials}},\ }\href {https://doi.org/10.1088/0953-8984/21/39/395502} {\bibfield  {journal} {\bibinfo  {journal} {J. Phys.: Condens. Matter}\ }\textbf {\bibinfo {volume} {21}},\ \bibinfo {pages} {395502} (\bibinfo {year} {2009})}\BibitemShut {NoStop}%
\bibitem [{\citenamefont {Giannozzi}\ \emph {et~al.}(2017)\citenamefont {Giannozzi}, \citenamefont {Andreussi}, \citenamefont {Brumme}, \citenamefont {Bunau}, \citenamefont {Nardelli}, \citenamefont {Calandra}, \citenamefont {Car}, \citenamefont {Cavazzoni}, \citenamefont {Ceresoli}, \citenamefont {Cococcioni}, \citenamefont {Colonna}, \citenamefont {Carnimeo}, \citenamefont {Corso}, \citenamefont {de~Gironcoli}, \citenamefont {Delugas}, \citenamefont {DiStasio}, \citenamefont {Ferretti}, \citenamefont {Floris}, \citenamefont {Fratesi}, \citenamefont {Fugallo}, \citenamefont {Gebauer}, \citenamefont {Gerstmann}, \citenamefont {Giustino}, \citenamefont {Gorni}, \citenamefont {Jia}, \citenamefont {Kawamura}, \citenamefont {Ko}, \citenamefont {Kokalj}, \citenamefont {Küçükbenli}, \citenamefont {Lazzeri}, \citenamefont {Marsili}, \citenamefont {Marzari}, \citenamefont {Mauri}, \citenamefont {Nguyen}, \citenamefont {Nguyen}, \citenamefont {de-la Roza}, \citenamefont {Paulatto}, \citenamefont {Poncé},
  \citenamefont {Rocca}, \citenamefont {Sabatini}, \citenamefont {Santra}, \citenamefont {Schlipf}, \citenamefont {Seitsonen}, \citenamefont {Smogunov}, \citenamefont {Timrov}, \citenamefont {Thonhauser}, \citenamefont {Umari}, \citenamefont {Vast}, \citenamefont {Wu},\ and\ \citenamefont {Baroni}}]{Giannozzi2017JPC}%
  \BibitemOpen
  \bibfield  {author} {\bibinfo {author} {\bibfnamefont {P.}~\bibnamefont {Giannozzi}}, \bibinfo {author} {\bibfnamefont {O.}~\bibnamefont {Andreussi}}, \bibinfo {author} {\bibfnamefont {T.}~\bibnamefont {Brumme}}, \bibinfo {author} {\bibfnamefont {O.}~\bibnamefont {Bunau}}, \bibinfo {author} {\bibfnamefont {M.~B.}\ \bibnamefont {Nardelli}}, \bibinfo {author} {\bibfnamefont {M.}~\bibnamefont {Calandra}}, \bibinfo {author} {\bibfnamefont {R.}~\bibnamefont {Car}}, \bibinfo {author} {\bibfnamefont {C.}~\bibnamefont {Cavazzoni}}, \bibinfo {author} {\bibfnamefont {D.}~\bibnamefont {Ceresoli}}, \bibinfo {author} {\bibfnamefont {M.}~\bibnamefont {Cococcioni}}, \bibinfo {author} {\bibfnamefont {N.}~\bibnamefont {Colonna}}, \bibinfo {author} {\bibfnamefont {I.}~\bibnamefont {Carnimeo}}, \bibinfo {author} {\bibfnamefont {A.~D.}\ \bibnamefont {Corso}}, \bibinfo {author} {\bibfnamefont {S.}~\bibnamefont {de~Gironcoli}}, \bibinfo {author} {\bibfnamefont {P.}~\bibnamefont {Delugas}}, \bibinfo {author} {\bibfnamefont
  {R.~A.}\ \bibnamefont {DiStasio}}, \bibinfo {author} {\bibfnamefont {A.}~\bibnamefont {Ferretti}}, \bibinfo {author} {\bibfnamefont {A.}~\bibnamefont {Floris}}, \bibinfo {author} {\bibfnamefont {G.}~\bibnamefont {Fratesi}}, \bibinfo {author} {\bibfnamefont {G.}~\bibnamefont {Fugallo}}, \bibinfo {author} {\bibfnamefont {R.}~\bibnamefont {Gebauer}}, \bibinfo {author} {\bibfnamefont {U.}~\bibnamefont {Gerstmann}}, \bibinfo {author} {\bibfnamefont {F.}~\bibnamefont {Giustino}}, \bibinfo {author} {\bibfnamefont {T.}~\bibnamefont {Gorni}}, \bibinfo {author} {\bibfnamefont {J.}~\bibnamefont {Jia}}, \bibinfo {author} {\bibfnamefont {M.}~\bibnamefont {Kawamura}}, \bibinfo {author} {\bibfnamefont {H.-Y.}\ \bibnamefont {Ko}}, \bibinfo {author} {\bibfnamefont {A.}~\bibnamefont {Kokalj}}, \bibinfo {author} {\bibfnamefont {E.}~\bibnamefont {Küçükbenli}}, \bibinfo {author} {\bibfnamefont {M.}~\bibnamefont {Lazzeri}}, \bibinfo {author} {\bibfnamefont {M.}~\bibnamefont {Marsili}}, \bibinfo {author} {\bibfnamefont
  {N.}~\bibnamefont {Marzari}}, \bibinfo {author} {\bibfnamefont {F.}~\bibnamefont {Mauri}}, \bibinfo {author} {\bibfnamefont {N.~L.}\ \bibnamefont {Nguyen}}, \bibinfo {author} {\bibfnamefont {H.-V.}\ \bibnamefont {Nguyen}}, \bibinfo {author} {\bibfnamefont {A.~O.}\ \bibnamefont {de-la Roza}}, \bibinfo {author} {\bibfnamefont {L.}~\bibnamefont {Paulatto}}, \bibinfo {author} {\bibfnamefont {S.}~\bibnamefont {Poncé}}, \bibinfo {author} {\bibfnamefont {D.}~\bibnamefont {Rocca}}, \bibinfo {author} {\bibfnamefont {R.}~\bibnamefont {Sabatini}}, \bibinfo {author} {\bibfnamefont {B.}~\bibnamefont {Santra}}, \bibinfo {author} {\bibfnamefont {M.}~\bibnamefont {Schlipf}}, \bibinfo {author} {\bibfnamefont {A.~P.}\ \bibnamefont {Seitsonen}}, \bibinfo {author} {\bibfnamefont {A.}~\bibnamefont {Smogunov}}, \bibinfo {author} {\bibfnamefont {I.}~\bibnamefont {Timrov}}, \bibinfo {author} {\bibfnamefont {T.}~\bibnamefont {Thonhauser}}, \bibinfo {author} {\bibfnamefont {P.}~\bibnamefont {Umari}}, \bibinfo {author}
  {\bibfnamefont {N.}~\bibnamefont {Vast}}, \bibinfo {author} {\bibfnamefont {X.}~\bibnamefont {Wu}},\ and\ \bibinfo {author} {\bibfnamefont {S.}~\bibnamefont {Baroni}},\ }\bibfield  {title} {\bibinfo {title} {{Advanced capabilities for materials modelling with Quantum ESPRESSO}},\ }\href {https://doi.org/10.1088/1361-648X/aa8f79} {\bibfield  {journal} {\bibinfo  {journal} {J. Phys.: Condens. Matter}\ }\textbf {\bibinfo {volume} {29}},\ \bibinfo {pages} {465901} (\bibinfo {year} {2017})}\BibitemShut {NoStop}%
\bibitem [{\citenamefont {{van Setten}}\ \emph {et~al.}(2018)\citenamefont {{van Setten}}, \citenamefont {Giantomassi}, \citenamefont {Bousquet}, \citenamefont {Verstraete}, \citenamefont {Hamann}, \citenamefont {Gonze},\ and\ \citenamefont {Rignanese}}]{Setten2018CPC}%
  \BibitemOpen
  \bibfield  {author} {\bibinfo {author} {\bibfnamefont {M.}~\bibnamefont {{van Setten}}}, \bibinfo {author} {\bibfnamefont {M.}~\bibnamefont {Giantomassi}}, \bibinfo {author} {\bibfnamefont {E.}~\bibnamefont {Bousquet}}, \bibinfo {author} {\bibfnamefont {M.}~\bibnamefont {Verstraete}}, \bibinfo {author} {\bibfnamefont {D.}~\bibnamefont {Hamann}}, \bibinfo {author} {\bibfnamefont {X.}~\bibnamefont {Gonze}},\ and\ \bibinfo {author} {\bibfnamefont {G.-M.}\ \bibnamefont {Rignanese}},\ }\bibfield  {title} {\bibinfo {title} {{The PseudoDojo: Training and grading a 85 element optimized norm-conserving pseudopotential table}},\ }\href {https://doi.org/https://doi.org/10.1016/j.cpc.2018.01.012} {\bibfield  {journal} {\bibinfo  {journal} {Comput. Phys. Commun.}\ }\textbf {\bibinfo {volume} {226}},\ \bibinfo {pages} {39} (\bibinfo {year} {2018})}\BibitemShut {NoStop}%
\bibitem [{\citenamefont {Hamann}(2013)}]{hamann2013prb}%
  \BibitemOpen
  \bibfield  {author} {\bibinfo {author} {\bibfnamefont {D.~R.}\ \bibnamefont {Hamann}},\ }\bibfield  {title} {\bibinfo {title} {{Optimized norm-conserving Vanderbilt pseudopotentials}},\ }\href {https://doi.org/10.1103/PhysRevB.88.085117} {\bibfield  {journal} {\bibinfo  {journal} {Phys. Rev. B}\ }\textbf {\bibinfo {volume} {88}},\ \bibinfo {pages} {085117} (\bibinfo {year} {2013})}\BibitemShut {NoStop}%
\bibitem [{\citenamefont {Gonze}\ and\ \citenamefont {Lee}(1997)}]{Gonze1997PRB}%
  \BibitemOpen
  \bibfield  {author} {\bibinfo {author} {\bibfnamefont {X.}~\bibnamefont {Gonze}}\ and\ \bibinfo {author} {\bibfnamefont {C.}~\bibnamefont {Lee}},\ }\bibfield  {title} {\bibinfo {title} {Dynamical matrices, born effective charges, dielectric permittivity tensors, and interatomic force constants from density-functional perturbation theory},\ }\href {https://doi.org/10.1103/PhysRevB.55.10355} {\bibfield  {journal} {\bibinfo  {journal} {Phys. Rev. B}\ }\textbf {\bibinfo {volume} {55}},\ \bibinfo {pages} {10355} (\bibinfo {year} {1997})}\BibitemShut {NoStop}%
\bibitem [{\citenamefont {Pizzi}\ \emph {et~al.}(2020)\citenamefont {Pizzi}, \citenamefont {Vitale}, \citenamefont {Arita}, \citenamefont {Blügel}, \citenamefont {Freimuth}, \citenamefont {Géranton}, \citenamefont {Gibertini}, \citenamefont {Gresch}, \citenamefont {Johnson}, \citenamefont {Koretsune}, \citenamefont {Ibañez-Azpiroz}, \citenamefont {Lee}, \citenamefont {Lihm}, \citenamefont {Marchand}, \citenamefont {Marrazzo}, \citenamefont {Mokrousov}, \citenamefont {Mustafa}, \citenamefont {Nohara}, \citenamefont {Nomura}, \citenamefont {Paulatto}, \citenamefont {Poncé}, \citenamefont {Ponweiser}, \citenamefont {Qiao}, \citenamefont {Thöle}, \citenamefont {Tsirkin}, \citenamefont {Wierzbowska}, \citenamefont {Marzari}, \citenamefont {Vanderbilt}, \citenamefont {Souza}, \citenamefont {Mostofi},\ and\ \citenamefont {Yates}}]{Pizzi2020jpcm}%
  \BibitemOpen
  \bibfield  {author} {\bibinfo {author} {\bibfnamefont {G.}~\bibnamefont {Pizzi}}, \bibinfo {author} {\bibfnamefont {V.}~\bibnamefont {Vitale}}, \bibinfo {author} {\bibfnamefont {R.}~\bibnamefont {Arita}}, \bibinfo {author} {\bibfnamefont {S.}~\bibnamefont {Blügel}}, \bibinfo {author} {\bibfnamefont {F.}~\bibnamefont {Freimuth}}, \bibinfo {author} {\bibfnamefont {G.}~\bibnamefont {Géranton}}, \bibinfo {author} {\bibfnamefont {M.}~\bibnamefont {Gibertini}}, \bibinfo {author} {\bibfnamefont {D.}~\bibnamefont {Gresch}}, \bibinfo {author} {\bibfnamefont {C.}~\bibnamefont {Johnson}}, \bibinfo {author} {\bibfnamefont {T.}~\bibnamefont {Koretsune}}, \bibinfo {author} {\bibfnamefont {J.}~\bibnamefont {Ibañez-Azpiroz}}, \bibinfo {author} {\bibfnamefont {H.}~\bibnamefont {Lee}}, \bibinfo {author} {\bibfnamefont {J.-M.}\ \bibnamefont {Lihm}}, \bibinfo {author} {\bibfnamefont {D.}~\bibnamefont {Marchand}}, \bibinfo {author} {\bibfnamefont {A.}~\bibnamefont {Marrazzo}}, \bibinfo {author} {\bibfnamefont
  {Y.}~\bibnamefont {Mokrousov}}, \bibinfo {author} {\bibfnamefont {J.~I.}\ \bibnamefont {Mustafa}}, \bibinfo {author} {\bibfnamefont {Y.}~\bibnamefont {Nohara}}, \bibinfo {author} {\bibfnamefont {Y.}~\bibnamefont {Nomura}}, \bibinfo {author} {\bibfnamefont {L.}~\bibnamefont {Paulatto}}, \bibinfo {author} {\bibfnamefont {S.}~\bibnamefont {Poncé}}, \bibinfo {author} {\bibfnamefont {T.}~\bibnamefont {Ponweiser}}, \bibinfo {author} {\bibfnamefont {J.}~\bibnamefont {Qiao}}, \bibinfo {author} {\bibfnamefont {F.}~\bibnamefont {Thöle}}, \bibinfo {author} {\bibfnamefont {S.~S.}\ \bibnamefont {Tsirkin}}, \bibinfo {author} {\bibfnamefont {M.}~\bibnamefont {Wierzbowska}}, \bibinfo {author} {\bibfnamefont {N.}~\bibnamefont {Marzari}}, \bibinfo {author} {\bibfnamefont {D.}~\bibnamefont {Vanderbilt}}, \bibinfo {author} {\bibfnamefont {I.}~\bibnamefont {Souza}}, \bibinfo {author} {\bibfnamefont {A.~A.}\ \bibnamefont {Mostofi}},\ and\ \bibinfo {author} {\bibfnamefont {J.~R.}\ \bibnamefont {Yates}},\ }\bibfield  {title}
  {\bibinfo {title} {{Wannier90 as a community code: new features and applications}},\ }\href {https://doi.org/10.1088/1361-648X/ab51ff} {\bibfield  {journal} {\bibinfo  {journal} {J. Phys. Condens. Matter}\ }\textbf {\bibinfo {volume} {32}},\ \bibinfo {pages} {165902} (\bibinfo {year} {2020})}\BibitemShut {NoStop}%
\bibitem [{\citenamefont {Mostofi}\ \emph {et~al.}(2014)\citenamefont {Mostofi}, \citenamefont {Yates}, \citenamefont {Pizzi}, \citenamefont {Lee}, \citenamefont {Souza}, \citenamefont {Vanderbilt},\ and\ \citenamefont {Marzari}}]{Mostofi2014cpc}%
  \BibitemOpen
  \bibfield  {author} {\bibinfo {author} {\bibfnamefont {A.~A.}\ \bibnamefont {Mostofi}}, \bibinfo {author} {\bibfnamefont {J.~R.}\ \bibnamefont {Yates}}, \bibinfo {author} {\bibfnamefont {G.}~\bibnamefont {Pizzi}}, \bibinfo {author} {\bibfnamefont {Y.-S.}\ \bibnamefont {Lee}}, \bibinfo {author} {\bibfnamefont {I.}~\bibnamefont {Souza}}, \bibinfo {author} {\bibfnamefont {D.}~\bibnamefont {Vanderbilt}},\ and\ \bibinfo {author} {\bibfnamefont {N.}~\bibnamefont {Marzari}},\ }\bibfield  {title} {\bibinfo {title} {{An updated version of Wannier90: A tool for obtaining maximally-localised Wannier functions}},\ }\href {https://doi.org/https://doi.org/10.1016/j.cpc.2014.05.003} {\bibfield  {journal} {\bibinfo  {journal} {Comp. Phys. Commun.}\ }\textbf {\bibinfo {volume} {185}},\ \bibinfo {pages} {2309} (\bibinfo {year} {2014})}\BibitemShut {NoStop}%
\bibitem [{\citenamefont {Lihm}\ \emph {et~al.}(2024)\citenamefont {Lihm}, \citenamefont {Ponc\'e},\ and\ \citenamefont {Park}}]{Lihm2024prb}%
  \BibitemOpen
  \bibfield  {author} {\bibinfo {author} {\bibfnamefont {J.-M.}\ \bibnamefont {Lihm}}, \bibinfo {author} {\bibfnamefont {S.}~\bibnamefont {Ponc\'e}},\ and\ \bibinfo {author} {\bibfnamefont {C.-H.}\ \bibnamefont {Park}},\ }\bibfield  {title} {\bibinfo {title} {Self-consistent electron lifetimes for electron-phonon scattering},\ }\href {https://doi.org/10.1103/PhysRevB.110.L121106} {\bibfield  {journal} {\bibinfo  {journal} {Phys. Rev. B}\ }\textbf {\bibinfo {volume} {110}},\ \bibinfo {pages} {L121106} (\bibinfo {year} {2024})}\BibitemShut {NoStop}%
\bibitem [{\citenamefont {Bädeker}(1907)}]{Badeker1907adp}%
  \BibitemOpen
  \bibfield  {author} {\bibinfo {author} {\bibfnamefont {K.}~\bibnamefont {Bädeker}},\ }\bibfield  {title} {\bibinfo {title} {Über die elektrische leitfähigkeit und die thermoelektrische kraft einiger schwermetallverbindungen},\ }\href {https://doi.org/https://doi.org/10.1002/andp.19073270409} {\bibfield  {journal} {\bibinfo  {journal} {Ann. Phys.}\ }\textbf {\bibinfo {volume} {327}},\ \bibinfo {pages} {749} (\bibinfo {year} {1907})}\BibitemShut {NoStop}%
\bibitem [{\citenamefont {Minami}(2005)}]{Minami2005sst}%
  \BibitemOpen
  \bibfield  {author} {\bibinfo {author} {\bibfnamefont {T.}~\bibnamefont {Minami}},\ }\bibfield  {title} {\bibinfo {title} {Transparent conducting oxide semiconductors for transparent electrodes},\ }\href {https://doi.org/10.1088/0268-1242/20/4/004} {\bibfield  {journal} {\bibinfo  {journal} {Semicond. Sci. Technol.}\ }\textbf {\bibinfo {volume} {20}},\ \bibinfo {pages} {S35} (\bibinfo {year} {2005})}\BibitemShut {NoStop}%
\bibitem [{\citenamefont {Mang}\ \emph {et~al.}(1995)\citenamefont {Mang}, \citenamefont {Reimann},\ and\ \citenamefont {Rübenacke}}]{Mang1995ssc}%
  \BibitemOpen
  \bibfield  {author} {\bibinfo {author} {\bibfnamefont {A.}~\bibnamefont {Mang}}, \bibinfo {author} {\bibfnamefont {K.}~\bibnamefont {Reimann}},\ and\ \bibinfo {author} {\bibfnamefont {S.}~\bibnamefont {Rübenacke}},\ }\bibfield  {title} {\bibinfo {title} {Band gaps, crystal-field splitting, spin-orbit coupling, and exciton binding energies in {ZnO} under hydrostatic pressure},\ }\href {https://doi.org/https://doi.org/10.1016/0038-1098(95)00054-2} {\bibfield  {journal} {\bibinfo  {journal} {Solid State Commun.}\ }\textbf {\bibinfo {volume} {94}},\ \bibinfo {pages} {251} (\bibinfo {year} {1995})}\BibitemShut {NoStop}%
\bibitem [{\citenamefont {Reynolds}\ \emph {et~al.}(1999)\citenamefont {Reynolds}, \citenamefont {Look}, \citenamefont {Jogai}, \citenamefont {Litton}, \citenamefont {Cantwell},\ and\ \citenamefont {Harsch}}]{Reynolds1999prb}%
  \BibitemOpen
  \bibfield  {author} {\bibinfo {author} {\bibfnamefont {D.~C.}\ \bibnamefont {Reynolds}}, \bibinfo {author} {\bibfnamefont {D.~C.}\ \bibnamefont {Look}}, \bibinfo {author} {\bibfnamefont {B.}~\bibnamefont {Jogai}}, \bibinfo {author} {\bibfnamefont {C.~W.}\ \bibnamefont {Litton}}, \bibinfo {author} {\bibfnamefont {G.}~\bibnamefont {Cantwell}},\ and\ \bibinfo {author} {\bibfnamefont {W.~C.}\ \bibnamefont {Harsch}},\ }\bibfield  {title} {\bibinfo {title} {Valence-band ordering in {ZnO}},\ }\href {https://doi.org/10.1103/PhysRevB.60.2340} {\bibfield  {journal} {\bibinfo  {journal} {Phys. Rev. B}\ }\textbf {\bibinfo {volume} {60}},\ \bibinfo {pages} {2340} (\bibinfo {year} {1999})}\BibitemShut {NoStop}%
\bibitem [{\citenamefont {Dong}\ \emph {et~al.}(2004)\citenamefont {Dong}, \citenamefont {Persson}, \citenamefont {Vayssieres}, \citenamefont {Augustsson}, \citenamefont {Schmitt}, \citenamefont {Mattesini}, \citenamefont {Ahuja}, \citenamefont {Chang},\ and\ \citenamefont {Guo}}]{Dong2004prb}%
  \BibitemOpen
  \bibfield  {author} {\bibinfo {author} {\bibfnamefont {C.~L.}\ \bibnamefont {Dong}}, \bibinfo {author} {\bibfnamefont {C.}~\bibnamefont {Persson}}, \bibinfo {author} {\bibfnamefont {L.}~\bibnamefont {Vayssieres}}, \bibinfo {author} {\bibfnamefont {A.}~\bibnamefont {Augustsson}}, \bibinfo {author} {\bibfnamefont {T.}~\bibnamefont {Schmitt}}, \bibinfo {author} {\bibfnamefont {M.}~\bibnamefont {Mattesini}}, \bibinfo {author} {\bibfnamefont {R.}~\bibnamefont {Ahuja}}, \bibinfo {author} {\bibfnamefont {C.~L.}\ \bibnamefont {Chang}},\ and\ \bibinfo {author} {\bibfnamefont {J.-H.}\ \bibnamefont {Guo}},\ }\bibfield  {title} {\bibinfo {title} {Electronic structure of nanostructured {ZnO} from x-ray absorption and emission spectroscopy and the local density approximation},\ }\href {https://doi.org/10.1103/PhysRevB.70.195325} {\bibfield  {journal} {\bibinfo  {journal} {Phys. Rev. B}\ }\textbf {\bibinfo {volume} {70}},\ \bibinfo {pages} {195325} (\bibinfo {year} {2004})}\BibitemShut {NoStop}%
\bibitem [{\citenamefont {Minami}\ \emph {et~al.}(2002)\citenamefont {Minami}, \citenamefont {Ida},\ and\ \citenamefont {Miyata}}]{Minami2002thin}%
  \BibitemOpen
  \bibfield  {author} {\bibinfo {author} {\bibfnamefont {T.}~\bibnamefont {Minami}}, \bibinfo {author} {\bibfnamefont {S.}~\bibnamefont {Ida}},\ and\ \bibinfo {author} {\bibfnamefont {T.}~\bibnamefont {Miyata}},\ }\bibfield  {title} {\bibinfo {title} {High rate deposition of transparent conducting oxide thin films by vacuum arc plasma evaporation},\ }\href {https://doi.org/https://doi.org/10.1016/S0040-6090(02)00706-X} {\bibfield  {journal} {\bibinfo  {journal} {Thin Solid Films}\ }\textbf {\bibinfo {volume} {416}},\ \bibinfo {pages} {92} (\bibinfo {year} {2002})}\BibitemShut {NoStop}%
\bibitem [{\citenamefont {Sato}\ \emph {et~al.}(1993)\citenamefont {Sato}, \citenamefont {Minami}, \citenamefont {Takata}, \citenamefont {Miyata},\ and\ \citenamefont {Ishii}}]{Sato1993thin}%
  \BibitemOpen
  \bibfield  {author} {\bibinfo {author} {\bibfnamefont {H.}~\bibnamefont {Sato}}, \bibinfo {author} {\bibfnamefont {T.}~\bibnamefont {Minami}}, \bibinfo {author} {\bibfnamefont {S.}~\bibnamefont {Takata}}, \bibinfo {author} {\bibfnamefont {T.}~\bibnamefont {Miyata}},\ and\ \bibinfo {author} {\bibfnamefont {M.}~\bibnamefont {Ishii}},\ }\bibfield  {title} {\bibinfo {title} {{Low temperature preparation of transparent conducting ZnO:Al thin films by chemical beam deposition}},\ }\href {https://doi.org/https://doi.org/10.1016/0040-6090(93)90634-2} {\bibfield  {journal} {\bibinfo  {journal} {Thin Solid Films}\ }\textbf {\bibinfo {volume} {236}},\ \bibinfo {pages} {14} (\bibinfo {year} {1993})}\BibitemShut {NoStop}%
\bibitem [{\citenamefont {Minami}\ \emph {et~al.}(1984)\citenamefont {Minami}, \citenamefont {Nanto},\ and\ \citenamefont {Takata}}]{Minami1984jjap}%
  \BibitemOpen
  \bibfield  {author} {\bibinfo {author} {\bibfnamefont {T.}~\bibnamefont {Minami}}, \bibinfo {author} {\bibfnamefont {H.}~\bibnamefont {Nanto}},\ and\ \bibinfo {author} {\bibfnamefont {S.}~\bibnamefont {Takata}},\ }\bibfield  {title} {\bibinfo {title} {Highly conductive and transparent aluminum doped zinc oxide thin films prepared by {RF} magnetron sputtering},\ }\href {https://doi.org/10.1143/JJAP.23.L280} {\bibfield  {journal} {\bibinfo  {journal} {Jpn. J. Appl. Phys.}\ }\textbf {\bibinfo {volume} {23}},\ \bibinfo {pages} {L280} (\bibinfo {year} {1984})}\BibitemShut {NoStop}%
\bibitem [{\citenamefont {Kim}\ \emph {et~al.}(2013)\citenamefont {Kim}, \citenamefont {Youn}, \citenamefont {Seo},\ and\ \citenamefont {Jang}}]{Kim2013jmcc}%
  \BibitemOpen
  \bibfield  {author} {\bibinfo {author} {\bibfnamefont {H.-M.}\ \bibnamefont {Kim}}, \bibinfo {author} {\bibfnamefont {J.-H.}\ \bibnamefont {Youn}}, \bibinfo {author} {\bibfnamefont {G.-J.}\ \bibnamefont {Seo}},\ and\ \bibinfo {author} {\bibfnamefont {J.}~\bibnamefont {Jang}},\ }\bibfield  {title} {\bibinfo {title} {Inverted quantum-dot light-emitting diodes with solution-processed aluminium–zinc oxide as a cathode buffer},\ }\href {https://doi.org/10.1039/C2TC00339B} {\bibfield  {journal} {\bibinfo  {journal} {J. Mater. Chem. C}\ }\textbf {\bibinfo {volume} {1}},\ \bibinfo {pages} {1567} (\bibinfo {year} {2013})}\BibitemShut {NoStop}%
\bibitem [{\citenamefont {Sun}\ \emph {et~al.}(2018)\citenamefont {Sun}, \citenamefont {Wang}, \citenamefont {Zhang}, \citenamefont {Su}, \citenamefont {Wei}, \citenamefont {Liu}, \citenamefont {Chen},\ and\ \citenamefont {Zhang}}]{Sun2018acsam}%
  \BibitemOpen
  \bibfield  {author} {\bibinfo {author} {\bibfnamefont {Y.}~\bibnamefont {Sun}}, \bibinfo {author} {\bibfnamefont {W.}~\bibnamefont {Wang}}, \bibinfo {author} {\bibfnamefont {H.}~\bibnamefont {Zhang}}, \bibinfo {author} {\bibfnamefont {Q.}~\bibnamefont {Su}}, \bibinfo {author} {\bibfnamefont {J.}~\bibnamefont {Wei}}, \bibinfo {author} {\bibfnamefont {P.}~\bibnamefont {Liu}}, \bibinfo {author} {\bibfnamefont {S.}~\bibnamefont {Chen}},\ and\ \bibinfo {author} {\bibfnamefont {S.}~\bibnamefont {Zhang}},\ }\bibfield  {title} {\bibinfo {title} {High-performance quantum dot light-emitting diodes based on {Al}-doped {ZnO} nanoparticles electron transport layer},\ }\href {https://doi.org/10.1021/acsami.8b04754} {\bibfield  {journal} {\bibinfo  {journal} {ACS Appl. Mater. Interfaces}\ }\textbf {\bibinfo {volume} {10}},\ \bibinfo {pages} {18902} (\bibinfo {year} {2018})}\BibitemShut {NoStop}%
\bibitem [{\citenamefont {Song}\ \emph {et~al.}(2015)\citenamefont {Song}, \citenamefont {Kulinich}, \citenamefont {Li}, \citenamefont {Liu},\ and\ \citenamefont {Zeng}}]{Song2015angchem}%
  \BibitemOpen
  \bibfield  {author} {\bibinfo {author} {\bibfnamefont {J.}~\bibnamefont {Song}}, \bibinfo {author} {\bibfnamefont {S.~A.}\ \bibnamefont {Kulinich}}, \bibinfo {author} {\bibfnamefont {J.}~\bibnamefont {Li}}, \bibinfo {author} {\bibfnamefont {Y.}~\bibnamefont {Liu}},\ and\ \bibinfo {author} {\bibfnamefont {H.}~\bibnamefont {Zeng}},\ }\bibfield  {title} {\bibinfo {title} {A general one-pot strategy for the synthesis of high-performance transparent-conducting-oxide nanocrystal inks for all-solution-processed devices},\ }\href {https://doi.org/https://doi.org/10.1002/anie.201408621} {\bibfield  {journal} {\bibinfo  {journal} {Angew. Chem., Int. Ed.}\ }\textbf {\bibinfo {volume} {54}},\ \bibinfo {pages} {462} (\bibinfo {year} {2015})}\BibitemShut {NoStop}%
\bibitem [{\citenamefont {Cao}\ \emph {et~al.}(2017)\citenamefont {Cao}, \citenamefont {Zheng}, \citenamefont {Zhao}, \citenamefont {Yang}, \citenamefont {Li}, \citenamefont {Guan}, \citenamefont {Yang}, \citenamefont {Shang},\ and\ \citenamefont {Wu}}]{Cao2017acsam}%
  \BibitemOpen
  \bibfield  {author} {\bibinfo {author} {\bibfnamefont {S.}~\bibnamefont {Cao}}, \bibinfo {author} {\bibfnamefont {J.}~\bibnamefont {Zheng}}, \bibinfo {author} {\bibfnamefont {J.}~\bibnamefont {Zhao}}, \bibinfo {author} {\bibfnamefont {Z.}~\bibnamefont {Yang}}, \bibinfo {author} {\bibfnamefont {C.}~\bibnamefont {Li}}, \bibinfo {author} {\bibfnamefont {X.}~\bibnamefont {Guan}}, \bibinfo {author} {\bibfnamefont {W.}~\bibnamefont {Yang}}, \bibinfo {author} {\bibfnamefont {M.}~\bibnamefont {Shang}},\ and\ \bibinfo {author} {\bibfnamefont {T.}~\bibnamefont {Wu}},\ }\bibfield  {title} {\bibinfo {title} {Enhancing the performance of quantum dot light-emitting diodes using room-temperature-processed ga-doped {ZnO} nanoparticles as the electron transport layer},\ }\href {https://doi.org/10.1021/acsami.7b03262} {\bibfield  {journal} {\bibinfo  {journal} {ACS Appl. Mater. Interfaces}\ }\textbf {\bibinfo {volume} {9}},\ \bibinfo {pages} {15605} (\bibinfo {year} {2017})}\BibitemShut {NoStop}%
\bibitem [{\citenamefont {Heiba}\ and\ \citenamefont {Mohamed}(2019)}]{Heiba2019jms}%
  \BibitemOpen
  \bibfield  {author} {\bibinfo {author} {\bibfnamefont {Z.~K.}\ \bibnamefont {Heiba}}\ and\ \bibinfo {author} {\bibfnamefont {M.~B.}\ \bibnamefont {Mohamed}},\ }\bibfield  {title} {\bibinfo {title} {{Effect of annealed and Mg-doped nano ZnO on physical properties of PVA}},\ }\href {https://doi.org/https://doi.org/10.1016/j.molstruc.2019.01.008} {\bibfield  {journal} {\bibinfo  {journal} {J. Mol. Struct.}\ }\textbf {\bibinfo {volume} {1181}},\ \bibinfo {pages} {507} (\bibinfo {year} {2019})}\BibitemShut {NoStop}%
\bibitem [{\citenamefont {Şenay~Şen Türkyılmaz}\ \emph {et~al.}(2017)\citenamefont {Şenay~Şen Türkyılmaz}, \citenamefont {Güy},\ and\ \citenamefont {Özacar}}]{Turkyilmaz2017jppa}%
  \BibitemOpen
  \bibfield  {author} {\bibinfo {author} {\bibnamefont {Şenay~Şen Türkyılmaz}}, \bibinfo {author} {\bibfnamefont {N.}~\bibnamefont {Güy}},\ and\ \bibinfo {author} {\bibfnamefont {M.}~\bibnamefont {Özacar}},\ }\bibfield  {title} {\bibinfo {title} {{Photocatalytic efficiencies of Ni, Mn, Fe and Ag doped ZnO nanostructures synthesized by hydrothermal method: The synergistic/antagonistic effect between ZnO and metals}},\ }\href {https://doi.org/https://doi.org/10.1016/j.jphotochem.2017.03.027} {\bibfield  {journal} {\bibinfo  {journal} {J. Photochem. Photobiol. A: Chem.}\ }\textbf {\bibinfo {volume} {341}},\ \bibinfo {pages} {39} (\bibinfo {year} {2017})}\BibitemShut {NoStop}%
\bibitem [{\citenamefont {Joshi}\ \emph {et~al.}(2016)\citenamefont {Joshi}, \citenamefont {Rawat}, \citenamefont {Gautam}, \citenamefont {Singh}, \citenamefont {Ramola},\ and\ \citenamefont {Singh}}]{Joshi2016jac}%
  \BibitemOpen
  \bibfield  {author} {\bibinfo {author} {\bibfnamefont {K.}~\bibnamefont {Joshi}}, \bibinfo {author} {\bibfnamefont {M.}~\bibnamefont {Rawat}}, \bibinfo {author} {\bibfnamefont {S.~K.}\ \bibnamefont {Gautam}}, \bibinfo {author} {\bibfnamefont {R.}~\bibnamefont {Singh}}, \bibinfo {author} {\bibfnamefont {R.}~\bibnamefont {Ramola}},\ and\ \bibinfo {author} {\bibfnamefont {F.}~\bibnamefont {Singh}},\ }\bibfield  {title} {\bibinfo {title} {{Band gap widening and narrowing in Cu-doped ZnO thin films}},\ }\href {https://doi.org/https://doi.org/10.1016/j.jallcom.2016.04.093} {\bibfield  {journal} {\bibinfo  {journal} {J. Alloys Compd.}\ }\textbf {\bibinfo {volume} {680}},\ \bibinfo {pages} {252} (\bibinfo {year} {2016})}\BibitemShut {NoStop}%
\bibitem [{\citenamefont {Lung}\ \emph {et~al.}(2017)\citenamefont {Lung}, \citenamefont {Toma}, \citenamefont {Pop}, \citenamefont {Marconi},\ and\ \citenamefont {Pop}}]{Lung2017jac}%
  \BibitemOpen
  \bibfield  {author} {\bibinfo {author} {\bibfnamefont {C.}~\bibnamefont {Lung}}, \bibinfo {author} {\bibfnamefont {M.}~\bibnamefont {Toma}}, \bibinfo {author} {\bibfnamefont {M.}~\bibnamefont {Pop}}, \bibinfo {author} {\bibfnamefont {D.}~\bibnamefont {Marconi}},\ and\ \bibinfo {author} {\bibfnamefont {A.}~\bibnamefont {Pop}},\ }\bibfield  {title} {\bibinfo {title} {{Characterization of the structural and optical properties of ZnO thin films doped with Ga, Al and (Al+Ga)}},\ }\href {https://doi.org/https://doi.org/10.1016/j.jallcom.2017.07.265} {\bibfield  {journal} {\bibinfo  {journal} {J. Alloys Compd.}\ }\textbf {\bibinfo {volume} {725}},\ \bibinfo {pages} {1238} (\bibinfo {year} {2017})}\BibitemShut {NoStop}%
\bibitem [{\citenamefont {Alexandrov}\ \emph {et~al.}(2020)\citenamefont {Alexandrov}, \citenamefont {Zvaigzne}, \citenamefont {Lypenko}, \citenamefont {Nabiev},\ and\ \citenamefont {Samokhvalov}}]{Alexandrov2020scirep}%
  \BibitemOpen
  \bibfield  {author} {\bibinfo {author} {\bibfnamefont {A.}~\bibnamefont {Alexandrov}}, \bibinfo {author} {\bibfnamefont {M.}~\bibnamefont {Zvaigzne}}, \bibinfo {author} {\bibfnamefont {D.}~\bibnamefont {Lypenko}}, \bibinfo {author} {\bibfnamefont {I.}~\bibnamefont {Nabiev}},\ and\ \bibinfo {author} {\bibfnamefont {P.}~\bibnamefont {Samokhvalov}},\ }\bibfield  {title} {\bibinfo {title} {{Al-, Ga-, Mg-, or Li-doped zinc oxide nanoparticles as electron transport layers for quantum dot light-emitting diodes}},\ }\href {https://doi.org/10.1038/s41598-020-64263-2} {\bibfield  {journal} {\bibinfo  {journal} {Sci. Rep.}\ }\textbf {\bibinfo {volume} {10}},\ \bibinfo {pages} {7496} (\bibinfo {year} {2020})}\BibitemShut {NoStop}%
\bibitem [{\citenamefont {Lu}\ \emph {et~al.}(2022)\citenamefont {Lu}, \citenamefont {Zhou}, \citenamefont {Park},\ and\ \citenamefont {Bernardi}}]{Lu2022prm}%
  \BibitemOpen
  \bibfield  {author} {\bibinfo {author} {\bibfnamefont {I.-T.}\ \bibnamefont {Lu}}, \bibinfo {author} {\bibfnamefont {J.-J.}\ \bibnamefont {Zhou}}, \bibinfo {author} {\bibfnamefont {J.}~\bibnamefont {Park}},\ and\ \bibinfo {author} {\bibfnamefont {M.}~\bibnamefont {Bernardi}},\ }\bibfield  {title} {\bibinfo {title} {First-principles ionized-impurity scattering and charge transport in doped materials},\ }\href {https://doi.org/10.1103/PhysRevMaterials.6.L010801} {\bibfield  {journal} {\bibinfo  {journal} {Phys. Rev. Mater.}\ }\textbf {\bibinfo {volume} {6}},\ \bibinfo {pages} {L010801} (\bibinfo {year} {2022})}\BibitemShut {NoStop}%
\bibitem [{\citenamefont {Burbano}\ \emph {et~al.}(2011)\citenamefont {Burbano}, \citenamefont {Scanlon},\ and\ \citenamefont {Watson}}]{Burbano2011jacs}%
  \BibitemOpen
  \bibfield  {author} {\bibinfo {author} {\bibfnamefont {M.}~\bibnamefont {Burbano}}, \bibinfo {author} {\bibfnamefont {D.~O.}\ \bibnamefont {Scanlon}},\ and\ \bibinfo {author} {\bibfnamefont {G.~W.}\ \bibnamefont {Watson}},\ }\bibfield  {title} {\bibinfo {title} {Sources of conductivity and doping limits in {CdO} from hybrid density functional theory},\ }\href {https://doi.org/10.1021/ja204639y} {\bibfield  {journal} {\bibinfo  {journal} {J. Am. Chem. Soc.}\ }\textbf {\bibinfo {volume} {133}},\ \bibinfo {pages} {15065} (\bibinfo {year} {2011})}\BibitemShut {NoStop}%
\bibitem [{\citenamefont {Dixit}\ \emph {et~al.}(2012)\citenamefont {Dixit}, \citenamefont {Lamoen},\ and\ \citenamefont {Partoens}}]{Dixit2012jpcm}%
  \BibitemOpen
  \bibfield  {author} {\bibinfo {author} {\bibfnamefont {H.}~\bibnamefont {Dixit}}, \bibinfo {author} {\bibfnamefont {D.}~\bibnamefont {Lamoen}},\ and\ \bibinfo {author} {\bibfnamefont {B.}~\bibnamefont {Partoens}},\ }\bibfield  {title} {\bibinfo {title} {{Quasiparticle band structure of rocksalt-CdO determined using maximally localized Wannier functions}},\ }\href {https://doi.org/10.1088/0953-8984/25/3/035501} {\bibfield  {journal} {\bibinfo  {journal} {J. Phys.: Condens. Matter}\ }\textbf {\bibinfo {volume} {25}},\ \bibinfo {pages} {035501} (\bibinfo {year} {2012})}\BibitemShut {NoStop}%
\bibitem [{\citenamefont {Zuñiga-Pérez}\ \emph {et~al.}(2004)\citenamefont {Zuñiga-Pérez}, \citenamefont {Munuera}, \citenamefont {Ocal},\ and\ \citenamefont {Muñoz-Sanjosé}}]{Zuniga2004jcg}%
  \BibitemOpen
  \bibfield  {author} {\bibinfo {author} {\bibfnamefont {J.}~\bibnamefont {Zuñiga-Pérez}}, \bibinfo {author} {\bibfnamefont {C.}~\bibnamefont {Munuera}}, \bibinfo {author} {\bibfnamefont {C.}~\bibnamefont {Ocal}},\ and\ \bibinfo {author} {\bibfnamefont {V.}~\bibnamefont {Muñoz-Sanjosé}},\ }\bibfield  {title} {\bibinfo {title} {{Structural analysis of CdO layers grown on $r$-plane sapphire (01$\bar{1}$2) by metalorganic vapor-phase epitaxy}},\ }\href {https://doi.org/https://doi.org/10.1016/j.jcrysgro.2004.07.069} {\bibfield  {journal} {\bibinfo  {journal} {J. Crystal Growth}\ }\textbf {\bibinfo {volume} {271}},\ \bibinfo {pages} {223} (\bibinfo {year} {2004})}\BibitemShut {NoStop}%
\bibitem [{\citenamefont {Demchenko}\ \emph {et~al.}(2010)\citenamefont {Demchenko}, \citenamefont {Denlinger}, \citenamefont {Chernyshova}, \citenamefont {Yu}, \citenamefont {Speaks}, \citenamefont {Olalde-Velasco}, \citenamefont {Hemmers}, \citenamefont {Walukiewicz}, \citenamefont {Derkachova},\ and\ \citenamefont {Lawniczak-Jablonska}}]{Demchenko2010prb}%
  \BibitemOpen
  \bibfield  {author} {\bibinfo {author} {\bibfnamefont {I.~N.}\ \bibnamefont {Demchenko}}, \bibinfo {author} {\bibfnamefont {J.~D.}\ \bibnamefont {Denlinger}}, \bibinfo {author} {\bibfnamefont {M.}~\bibnamefont {Chernyshova}}, \bibinfo {author} {\bibfnamefont {K.~M.}\ \bibnamefont {Yu}}, \bibinfo {author} {\bibfnamefont {D.~T.}\ \bibnamefont {Speaks}}, \bibinfo {author} {\bibfnamefont {P.}~\bibnamefont {Olalde-Velasco}}, \bibinfo {author} {\bibfnamefont {O.}~\bibnamefont {Hemmers}}, \bibinfo {author} {\bibfnamefont {W.}~\bibnamefont {Walukiewicz}}, \bibinfo {author} {\bibfnamefont {A.}~\bibnamefont {Derkachova}},\ and\ \bibinfo {author} {\bibfnamefont {K.}~\bibnamefont {Lawniczak-Jablonska}},\ }\bibfield  {title} {\bibinfo {title} {{Full multiple scattering analysis of XANES at the $\text{Cd}\text{ }{L}_{3}$ and $\text{O}\text{ }K$ edges in CdO films combined with a soft-x-ray emission investigation}},\ }\href {https://doi.org/10.1103/PhysRevB.82.075107} {\bibfield  {journal} {\bibinfo  {journal} {Phys. Rev.
  B}\ }\textbf {\bibinfo {volume} {82}},\ \bibinfo {pages} {075107} (\bibinfo {year} {2010})}\BibitemShut {NoStop}%
\bibitem [{\citenamefont {Vasheghani~Farahani}\ \emph {et~al.}(2013)\citenamefont {Vasheghani~Farahani}, \citenamefont {Muñoz-Sanjosé}, \citenamefont {Zúñiga-Pérez}, \citenamefont {McConville},\ and\ \citenamefont {Veal}}]{Vasheghani2013apl}%
  \BibitemOpen
  \bibfield  {author} {\bibinfo {author} {\bibfnamefont {S.~K.}\ \bibnamefont {Vasheghani~Farahani}}, \bibinfo {author} {\bibfnamefont {V.}~\bibnamefont {Muñoz-Sanjosé}}, \bibinfo {author} {\bibfnamefont {J.}~\bibnamefont {Zúñiga-Pérez}}, \bibinfo {author} {\bibfnamefont {C.~F.}\ \bibnamefont {McConville}},\ and\ \bibinfo {author} {\bibfnamefont {T.~D.}\ \bibnamefont {Veal}},\ }\bibfield  {title} {\bibinfo {title} {Temperature dependence of the direct bandgap and transport properties of cdo},\ }\href {https://doi.org/10.1063/1.4775691} {\bibfield  {journal} {\bibinfo  {journal} {Appl. Phys. Lett.}\ }\textbf {\bibinfo {volume} {102}},\ \bibinfo {pages} {022102} (\bibinfo {year} {2013})}\BibitemShut {NoStop}%
\bibitem [{\citenamefont {King}\ \emph {et~al.}(2009)\citenamefont {King}, \citenamefont {Veal}, \citenamefont {Schleife}, \citenamefont {Z\'u\~niga P\'erez}, \citenamefont {Martel}, \citenamefont {Jefferson}, \citenamefont {Fuchs}, \citenamefont {Mu\~noz Sanjos\'e}, \citenamefont {Bechstedt},\ and\ \citenamefont {McConville}}]{King2009prb}%
  \BibitemOpen
  \bibfield  {author} {\bibinfo {author} {\bibfnamefont {P.~D.~C.}\ \bibnamefont {King}}, \bibinfo {author} {\bibfnamefont {T.~D.}\ \bibnamefont {Veal}}, \bibinfo {author} {\bibfnamefont {A.}~\bibnamefont {Schleife}}, \bibinfo {author} {\bibfnamefont {J.}~\bibnamefont {Z\'u\~niga P\'erez}}, \bibinfo {author} {\bibfnamefont {B.}~\bibnamefont {Martel}}, \bibinfo {author} {\bibfnamefont {P.~H.}\ \bibnamefont {Jefferson}}, \bibinfo {author} {\bibfnamefont {F.}~\bibnamefont {Fuchs}}, \bibinfo {author} {\bibfnamefont {V.}~\bibnamefont {Mu\~noz Sanjos\'e}}, \bibinfo {author} {\bibfnamefont {F.}~\bibnamefont {Bechstedt}},\ and\ \bibinfo {author} {\bibfnamefont {C.~F.}\ \bibnamefont {McConville}},\ }\bibfield  {title} {\bibinfo {title} {{Valence-band electronic structure of CdO, ZnO, and MgO from x-ray photoemission spectroscopy and quasi-particle-corrected density-functional theory calculations}},\ }\href {https://doi.org/10.1103/PhysRevB.79.205205} {\bibfield  {journal} {\bibinfo  {journal} {Phys. Rev. B}\ }\textbf
  {\bibinfo {volume} {79}},\ \bibinfo {pages} {205205} (\bibinfo {year} {2009})}\BibitemShut {NoStop}%
\bibitem [{\citenamefont {Hewat}(1970)}]{Hewat1970ssc}%
  \BibitemOpen
  \bibfield  {author} {\bibinfo {author} {\bibfnamefont {A.}~\bibnamefont {Hewat}},\ }\bibfield  {title} {\bibinfo {title} {{Lattice dynamics of ZnO and BeO}},\ }\href {https://doi.org/https://doi.org/10.1016/0038-1098(70)90077-3} {\bibfield  {journal} {\bibinfo  {journal} {Solid State Commun.}\ }\textbf {\bibinfo {volume} {8}},\ \bibinfo {pages} {187} (\bibinfo {year} {1970})}\BibitemShut {NoStop}%
\bibitem [{\citenamefont {Thoma}\ \emph {et~al.}(1974)\citenamefont {Thoma}, \citenamefont {Dorner}, \citenamefont {Duesing},\ and\ \citenamefont {Wegener}}]{Thoma1974ssc}%
  \BibitemOpen
  \bibfield  {author} {\bibinfo {author} {\bibfnamefont {K.}~\bibnamefont {Thoma}}, \bibinfo {author} {\bibfnamefont {B.}~\bibnamefont {Dorner}}, \bibinfo {author} {\bibfnamefont {G.}~\bibnamefont {Duesing}},\ and\ \bibinfo {author} {\bibfnamefont {W.}~\bibnamefont {Wegener}},\ }\bibfield  {title} {\bibinfo {title} {{Lattice dynamics of ZnO}},\ }\href {https://doi.org/https://doi.org/10.1016/0038-1098(74)90543-2} {\bibfield  {journal} {\bibinfo  {journal} {Solid State Commun.}\ }\textbf {\bibinfo {volume} {15}},\ \bibinfo {pages} {1111} (\bibinfo {year} {1974})}\BibitemShut {NoStop}%
\bibitem [{\citenamefont {Serrano}\ \emph {et~al.}(2010)\citenamefont {Serrano}, \citenamefont {Manj\'on}, \citenamefont {Romero}, \citenamefont {Ivanov}, \citenamefont {Cardona}, \citenamefont {Lauck}, \citenamefont {Bosak},\ and\ \citenamefont {Krisch}}]{Serrano2010prb}%
  \BibitemOpen
  \bibfield  {author} {\bibinfo {author} {\bibfnamefont {J.}~\bibnamefont {Serrano}}, \bibinfo {author} {\bibfnamefont {F.~J.}\ \bibnamefont {Manj\'on}}, \bibinfo {author} {\bibfnamefont {A.~H.}\ \bibnamefont {Romero}}, \bibinfo {author} {\bibfnamefont {A.}~\bibnamefont {Ivanov}}, \bibinfo {author} {\bibfnamefont {M.}~\bibnamefont {Cardona}}, \bibinfo {author} {\bibfnamefont {R.}~\bibnamefont {Lauck}}, \bibinfo {author} {\bibfnamefont {A.}~\bibnamefont {Bosak}},\ and\ \bibinfo {author} {\bibfnamefont {M.}~\bibnamefont {Krisch}},\ }\bibfield  {title} {\bibinfo {title} {Phonon dispersion relations of zinc oxide: Inelastic neutron scattering and ab initio calculations},\ }\href {https://doi.org/10.1103/PhysRevB.81.174304} {\bibfield  {journal} {\bibinfo  {journal} {Phys. Rev. B}\ }\textbf {\bibinfo {volume} {81}},\ \bibinfo {pages} {174304} (\bibinfo {year} {2010})}\BibitemShut {NoStop}%
\bibitem [{\citenamefont {Adhikari}\ \emph {et~al.}(2023)\citenamefont {Adhikari}, \citenamefont {Wierzbicka}, \citenamefont {Adamus}, \citenamefont {Lysak}, \citenamefont {Sybilski}, \citenamefont {Jarosz},\ and\ \citenamefont {Przezdziecka}}]{Adhikari2023tsf}%
  \BibitemOpen
  \bibfield  {author} {\bibinfo {author} {\bibfnamefont {A.}~\bibnamefont {Adhikari}}, \bibinfo {author} {\bibfnamefont {A.}~\bibnamefont {Wierzbicka}}, \bibinfo {author} {\bibfnamefont {Z.}~\bibnamefont {Adamus}}, \bibinfo {author} {\bibfnamefont {A.}~\bibnamefont {Lysak}}, \bibinfo {author} {\bibfnamefont {P.}~\bibnamefont {Sybilski}}, \bibinfo {author} {\bibfnamefont {D.}~\bibnamefont {Jarosz}},\ and\ \bibinfo {author} {\bibfnamefont {E.}~\bibnamefont {Przezdziecka}},\ }\bibfield  {title} {\bibinfo {title} {Correlated carrier transport and optical phenomena in {CdO} layers grown by plasma-assisted molecular beam epitaxy technique},\ }\href {https://doi.org/https://doi.org/10.1016/j.tsf.2023.139963} {\bibfield  {journal} {\bibinfo  {journal} {Thin Solid Films}\ }\textbf {\bibinfo {volume} {780}},\ \bibinfo {pages} {139963} (\bibinfo {year} {2023})}\BibitemShut {NoStop}%
\bibitem [{\citenamefont {Vasheghani~Farahani}\ \emph {et~al.}(2011)\citenamefont {Vasheghani~Farahani}, \citenamefont {Veal}, \citenamefont {King}, \citenamefont {Zúñiga-Pérez}, \citenamefont {Muñoz-Sanjosé},\ and\ \citenamefont {McConville}}]{Vasheghani2011jap}%
  \BibitemOpen
  \bibfield  {author} {\bibinfo {author} {\bibfnamefont {S.~K.}\ \bibnamefont {Vasheghani~Farahani}}, \bibinfo {author} {\bibfnamefont {T.~D.}\ \bibnamefont {Veal}}, \bibinfo {author} {\bibfnamefont {P.~D.~C.}\ \bibnamefont {King}}, \bibinfo {author} {\bibfnamefont {J.}~\bibnamefont {Zúñiga-Pérez}}, \bibinfo {author} {\bibfnamefont {V.}~\bibnamefont {Muñoz-Sanjosé}},\ and\ \bibinfo {author} {\bibfnamefont {C.~F.}\ \bibnamefont {McConville}},\ }\bibfield  {title} {\bibinfo {title} {Electron mobility in {CdO} films},\ }\href {https://doi.org/10.1063/1.3562141} {\bibfield  {journal} {\bibinfo  {journal} {J. Appl. Phys.}\ }\textbf {\bibinfo {volume} {109}},\ \bibinfo {pages} {073712} (\bibinfo {year} {2011})}\BibitemShut {NoStop}%
\bibitem [{\citenamefont {Ponc{\'e}}\ \emph {et~al.}(2019)\citenamefont {Ponc{\'e}}, \citenamefont {Schlipf},\ and\ \citenamefont {Giustino}}]{Ponce2019acsel}%
  \BibitemOpen
  \bibfield  {author} {\bibinfo {author} {\bibfnamefont {S.}~\bibnamefont {Ponc{\'e}}}, \bibinfo {author} {\bibfnamefont {M.}~\bibnamefont {Schlipf}},\ and\ \bibinfo {author} {\bibfnamefont {F.}~\bibnamefont {Giustino}},\ }\bibfield  {title} {\bibinfo {title} {Origin of low carrier mobilities in halide perovskites},\ }\href {https://doi.org/10.1021/acsenergylett.8b02346} {\bibfield  {journal} {\bibinfo  {journal} {ACS Energy Lett.}\ }\textbf {\bibinfo {volume} {4}},\ \bibinfo {pages} {456} (\bibinfo {year} {2019})}\BibitemShut {NoStop}%
\bibitem [{\citenamefont {Kose}\ \emph {et~al.}(2009)\citenamefont {Kose}, \citenamefont {Atay}, \citenamefont {Bilgin},\ and\ \citenamefont {Akyuz}}]{Kose2009ijhe}%
  \BibitemOpen
  \bibfield  {author} {\bibinfo {author} {\bibfnamefont {S.}~\bibnamefont {Kose}}, \bibinfo {author} {\bibfnamefont {F.}~\bibnamefont {Atay}}, \bibinfo {author} {\bibfnamefont {V.}~\bibnamefont {Bilgin}},\ and\ \bibinfo {author} {\bibfnamefont {I.}~\bibnamefont {Akyuz}},\ }\bibfield  {title} {\bibinfo {title} {In doped {CdO} films: Electrical, optical, structural and surface properties},\ }\href {https://doi.org/https://doi.org/10.1016/j.ijhydene.2008.11.110} {\bibfield  {journal} {\bibinfo  {journal} {Int. J. Hydrogen Energy}\ }\textbf {\bibinfo {volume} {34}},\ \bibinfo {pages} {5260} (\bibinfo {year} {2009})}\BibitemShut {NoStop}%
\bibitem [{\citenamefont {Li}\ \emph {et~al.}(2004)\citenamefont {Li}, \citenamefont {Gessert},\ and\ \citenamefont {Coutts}}]{Li2004appliedss}%
  \BibitemOpen
  \bibfield  {author} {\bibinfo {author} {\bibfnamefont {X.}~\bibnamefont {Li}}, \bibinfo {author} {\bibfnamefont {T.~A.}\ \bibnamefont {Gessert}},\ and\ \bibinfo {author} {\bibfnamefont {T.}~\bibnamefont {Coutts}},\ }\bibfield  {title} {\bibinfo {title} {The properties of cadmium tin oxide thin-film compounds prepared by linear combinatorial synthesis},\ }\href {https://doi.org/https://doi.org/10.1016/S0169-4332(03)00909-7} {\bibfield  {journal} {\bibinfo  {journal} {Appl. Surf. Sci.}\ }\textbf {\bibinfo {volume} {223}},\ \bibinfo {pages} {138} (\bibinfo {year} {2004})}\BibitemShut {NoStop}%
\bibitem [{\citenamefont {Zacharias}\ and\ \citenamefont {Giustino}(2016)}]{Zacharias2016prb}%
  \BibitemOpen
  \bibfield  {author} {\bibinfo {author} {\bibfnamefont {M.}~\bibnamefont {Zacharias}}\ and\ \bibinfo {author} {\bibfnamefont {F.}~\bibnamefont {Giustino}},\ }\bibfield  {title} {\bibinfo {title} {One-shot calculation of temperature-dependent optical spectra and phonon-induced band-gap renormalization},\ }\href {https://doi.org/10.1103/PhysRevB.94.075125} {\bibfield  {journal} {\bibinfo  {journal} {Phys. Rev. B}\ }\textbf {\bibinfo {volume} {94}},\ \bibinfo {pages} {075125} (\bibinfo {year} {2016})}\BibitemShut {NoStop}%
\bibitem [{\citenamefont {Chen}\ \emph {et~al.}(1998)\citenamefont {Chen}, \citenamefont {Bagnall}, \citenamefont {Koh}, \citenamefont {Park}, \citenamefont {Hiraga}, \citenamefont {Zhu},\ and\ \citenamefont {Yao}}]{Chen1998jap}%
  \BibitemOpen
  \bibfield  {author} {\bibinfo {author} {\bibfnamefont {Y.}~\bibnamefont {Chen}}, \bibinfo {author} {\bibfnamefont {D.~M.}\ \bibnamefont {Bagnall}}, \bibinfo {author} {\bibfnamefont {H.-j.}\ \bibnamefont {Koh}}, \bibinfo {author} {\bibfnamefont {K.-t.}\ \bibnamefont {Park}}, \bibinfo {author} {\bibfnamefont {K.}~\bibnamefont {Hiraga}}, \bibinfo {author} {\bibfnamefont {Z.}~\bibnamefont {Zhu}},\ and\ \bibinfo {author} {\bibfnamefont {T.}~\bibnamefont {Yao}},\ }\bibfield  {title} {\bibinfo {title} {Plasma assisted molecular beam epitaxy of zno on $c$-plane sapphire: Growth and characterization},\ }\href {https://doi.org/10.1063/1.368595} {\bibfield  {journal} {\bibinfo  {journal} {J. Appl. Phys.}\ }\textbf {\bibinfo {volume} {84}},\ \bibinfo {pages} {3912} (\bibinfo {year} {1998})}\BibitemShut {NoStop}%
\bibitem [{\citenamefont {Look}\ \emph {et~al.}(1998)\citenamefont {Look}, \citenamefont {Reynolds}, \citenamefont {Sizelove}, \citenamefont {Jones}, \citenamefont {Litton}, \citenamefont {Cantwell},\ and\ \citenamefont {Harsch}}]{Look2998ssc}%
  \BibitemOpen
  \bibfield  {author} {\bibinfo {author} {\bibfnamefont {D.}~\bibnamefont {Look}}, \bibinfo {author} {\bibfnamefont {D.}~\bibnamefont {Reynolds}}, \bibinfo {author} {\bibfnamefont {J.}~\bibnamefont {Sizelove}}, \bibinfo {author} {\bibfnamefont {R.}~\bibnamefont {Jones}}, \bibinfo {author} {\bibfnamefont {C.}~\bibnamefont {Litton}}, \bibinfo {author} {\bibfnamefont {G.}~\bibnamefont {Cantwell}},\ and\ \bibinfo {author} {\bibfnamefont {W.}~\bibnamefont {Harsch}},\ }\bibfield  {title} {\bibinfo {title} {Electrical properties of bulk {ZnO}},\ }\href {https://doi.org/https://doi.org/10.1016/S0038-1098(97)10145-4} {\bibfield  {journal} {\bibinfo  {journal} {Solid State Commun.}\ }\textbf {\bibinfo {volume} {105}},\ \bibinfo {pages} {399} (\bibinfo {year} {1998})}\BibitemShut {NoStop}%
\bibitem [{\citenamefont {Wagner}\ and\ \citenamefont {Helbig}(1974)}]{Wagner1974jpcs}%
  \BibitemOpen
  \bibfield  {author} {\bibinfo {author} {\bibfnamefont {P.}~\bibnamefont {Wagner}}\ and\ \bibinfo {author} {\bibfnamefont {R.}~\bibnamefont {Helbig}},\ }\bibfield  {title} {\bibinfo {title} {Halleffekt und anisotropie der beweglichkeit der elektronen in zno},\ }\href {https://doi.org/https://doi.org/10.1016/S0022-3697(74)80026-0} {\bibfield  {journal} {\bibinfo  {journal} {J. Phys. Chem. Solids}\ }\textbf {\bibinfo {volume} {35}},\ \bibinfo {pages} {327} (\bibinfo {year} {1974})}\BibitemShut {NoStop}%
\bibitem [{\citenamefont {Hutson}(1957)}]{Hutson1957PR}%
  \BibitemOpen
  \bibfield  {author} {\bibinfo {author} {\bibfnamefont {A.~R.}\ \bibnamefont {Hutson}},\ }\bibfield  {title} {\bibinfo {title} {Hall effect studies of doped zinc oxide single crystals},\ }\href {https://doi.org/10.1103/PhysRev.108.222} {\bibfield  {journal} {\bibinfo  {journal} {Phys. Rev.}\ }\textbf {\bibinfo {volume} {108}},\ \bibinfo {pages} {222} (\bibinfo {year} {1957})}\BibitemShut {NoStop}%
\bibitem [{\citenamefont {Ellmer}(2001)}]{Ellmer2001jpd}%
  \BibitemOpen
  \bibfield  {author} {\bibinfo {author} {\bibfnamefont {K.}~\bibnamefont {Ellmer}},\ }\bibfield  {title} {\bibinfo {title} {Resistivity of polycrystalline zinc oxide films: current status and physical limit},\ }\href {https://doi.org/10.1088/0022-3727/34/21/301} {\bibfield  {journal} {\bibinfo  {journal} {J. Phys. D: Appl. Phys.}\ }\textbf {\bibinfo {volume} {34}},\ \bibinfo {pages} {3097} (\bibinfo {year} {2001})}\BibitemShut {NoStop}%
\bibitem [{\citenamefont {Ellmer}\ and\ \citenamefont {Mientus}(2008)}]{Ellmer2008tsf}%
  \BibitemOpen
  \bibfield  {author} {\bibinfo {author} {\bibfnamefont {K.}~\bibnamefont {Ellmer}}\ and\ \bibinfo {author} {\bibfnamefont {R.}~\bibnamefont {Mientus}},\ }\bibfield  {title} {\bibinfo {title} {Carrier transport in polycrystalline transparent conductive oxides: A comparative study of zinc oxide and indium oxide},\ }\href {https://doi.org/https://doi.org/10.1016/j.tsf.2007.05.084} {\bibfield  {journal} {\bibinfo  {journal} {Thin Solid Films}\ }\textbf {\bibinfo {volume} {516}},\ \bibinfo {pages} {4620} (\bibinfo {year} {2008})}\BibitemShut {NoStop}%
\bibitem [{\citenamefont {Ma}\ \emph {et~al.}(2018)\citenamefont {Ma}, \citenamefont {Nissimagoudar},\ and\ \citenamefont {Li}}]{Ma2018prb}%
  \BibitemOpen
  \bibfield  {author} {\bibinfo {author} {\bibfnamefont {J.}~\bibnamefont {Ma}}, \bibinfo {author} {\bibfnamefont {A.~S.}\ \bibnamefont {Nissimagoudar}},\ and\ \bibinfo {author} {\bibfnamefont {W.}~\bibnamefont {Li}},\ }\bibfield  {title} {\bibinfo {title} {{First-principles study of electron and hole mobilities of Si and GaAs}},\ }\href {https://doi.org/10.1103/PhysRevB.97.045201} {\bibfield  {journal} {\bibinfo  {journal} {Phys. Rev. B}\ }\textbf {\bibinfo {volume} {97}},\ \bibinfo {pages} {045201} (\bibinfo {year} {2018})}\BibitemShut {NoStop}%
\bibitem [{\citenamefont {Ponc\'e}\ \emph {et~al.}(2018)\citenamefont {Ponc\'e}, \citenamefont {Margine},\ and\ \citenamefont {Giustino}}]{Ponce2018prb}%
  \BibitemOpen
  \bibfield  {author} {\bibinfo {author} {\bibfnamefont {S.}~\bibnamefont {Ponc\'e}}, \bibinfo {author} {\bibfnamefont {E.~R.}\ \bibnamefont {Margine}},\ and\ \bibinfo {author} {\bibfnamefont {F.}~\bibnamefont {Giustino}},\ }\bibfield  {title} {\bibinfo {title} {Towards predictive many-body calculations of phonon-limited carrier mobilities in semiconductors},\ }\href {https://doi.org/10.1103/PhysRevB.97.121201} {\bibfield  {journal} {\bibinfo  {journal} {Phys. Rev. B}\ }\textbf {\bibinfo {volume} {97}},\ \bibinfo {pages} {121201} (\bibinfo {year} {2018})}\BibitemShut {NoStop}%
\bibitem [{\citenamefont {Ponc\'e}\ \emph {et~al.}(2019)\citenamefont {Ponc\'e}, \citenamefont {Jena},\ and\ \citenamefont {Giustino}}]{Ponce2019prl}%
  \BibitemOpen
  \bibfield  {author} {\bibinfo {author} {\bibfnamefont {S.}~\bibnamefont {Ponc\'e}}, \bibinfo {author} {\bibfnamefont {D.}~\bibnamefont {Jena}},\ and\ \bibinfo {author} {\bibfnamefont {F.}~\bibnamefont {Giustino}},\ }\bibfield  {title} {\bibinfo {title} {Route to high hole mobility in {GaN} via reversal of crystal-field splitting},\ }\href {https://doi.org/10.1103/PhysRevLett.123.096602} {\bibfield  {journal} {\bibinfo  {journal} {Phys. Rev. Lett.}\ }\textbf {\bibinfo {volume} {123}},\ \bibinfo {pages} {096602} (\bibinfo {year} {2019})}\BibitemShut {NoStop}%
\bibitem [{\citenamefont {Park}\ \emph {et~al.}(2020)\citenamefont {Park}, \citenamefont {Zhou}, \citenamefont {Jhalani}, \citenamefont {Dreyer},\ and\ \citenamefont {Bernardi}}]{Park2020prb}%
  \BibitemOpen
  \bibfield  {author} {\bibinfo {author} {\bibfnamefont {J.}~\bibnamefont {Park}}, \bibinfo {author} {\bibfnamefont {J.-J.}\ \bibnamefont {Zhou}}, \bibinfo {author} {\bibfnamefont {V.~A.}\ \bibnamefont {Jhalani}}, \bibinfo {author} {\bibfnamefont {C.~E.}\ \bibnamefont {Dreyer}},\ and\ \bibinfo {author} {\bibfnamefont {M.}~\bibnamefont {Bernardi}},\ }\bibfield  {title} {\bibinfo {title} {Long-range quadrupole electron-phonon interaction from first principles},\ }\href {https://doi.org/10.1103/PhysRevB.102.125203} {\bibfield  {journal} {\bibinfo  {journal} {Phys. Rev. B}\ }\textbf {\bibinfo {volume} {102}},\ \bibinfo {pages} {125203} (\bibinfo {year} {2020})}\BibitemShut {NoStop}%
\bibitem [{\citenamefont {Lee}\ \emph {et~al.}(2020)\citenamefont {Lee}, \citenamefont {Zhou}, \citenamefont {Chen},\ and\ \citenamefont {Bernardi}}]{Lee2020nc}%
  \BibitemOpen
  \bibfield  {author} {\bibinfo {author} {\bibfnamefont {N.-E.}\ \bibnamefont {Lee}}, \bibinfo {author} {\bibfnamefont {J.-J.}\ \bibnamefont {Zhou}}, \bibinfo {author} {\bibfnamefont {H.-Y.}\ \bibnamefont {Chen}},\ and\ \bibinfo {author} {\bibfnamefont {M.}~\bibnamefont {Bernardi}},\ }\bibfield  {title} {\bibinfo {title} {Ab initio electron-two-phonon scattering in gaas from next-to-leading order perturbation theory},\ }\href {https://doi.org/10.1038/s41467-020-15339-0} {\bibfield  {journal} {\bibinfo  {journal} {Nat. Commun.}\ }\textbf {\bibinfo {volume} {11}},\ \bibinfo {pages} {1607} (\bibinfo {year} {2020})}\BibitemShut {NoStop}%
\bibitem [{\citenamefont {Protik}\ and\ \citenamefont {Broido}(2020)}]{Protik2020prb}%
  \BibitemOpen
  \bibfield  {author} {\bibinfo {author} {\bibfnamefont {N.~H.}\ \bibnamefont {Protik}}\ and\ \bibinfo {author} {\bibfnamefont {D.~A.}\ \bibnamefont {Broido}},\ }\bibfield  {title} {\bibinfo {title} {Coupled transport of phonons and carriers in semiconductors: A case study of $n$-doped {GaAs}},\ }\href {https://doi.org/10.1103/PhysRevB.101.075202} {\bibfield  {journal} {\bibinfo  {journal} {Phys. Rev. B}\ }\textbf {\bibinfo {volume} {101}},\ \bibinfo {pages} {075202} (\bibinfo {year} {2020})}\BibitemShut {NoStop}%
\bibitem [{\citenamefont {Brunin}\ \emph {et~al.}(2020)\citenamefont {Brunin}, \citenamefont {Miranda}, \citenamefont {Giantomassi}, \citenamefont {Royo}, \citenamefont {Stengel}, \citenamefont {Verstraete}, \citenamefont {Gonze}, \citenamefont {Rignanese},\ and\ \citenamefont {Hautier}}]{Brunin2020prb}%
  \BibitemOpen
  \bibfield  {author} {\bibinfo {author} {\bibfnamefont {G.}~\bibnamefont {Brunin}}, \bibinfo {author} {\bibfnamefont {H.~P.~C.}\ \bibnamefont {Miranda}}, \bibinfo {author} {\bibfnamefont {M.}~\bibnamefont {Giantomassi}}, \bibinfo {author} {\bibfnamefont {M.}~\bibnamefont {Royo}}, \bibinfo {author} {\bibfnamefont {M.}~\bibnamefont {Stengel}}, \bibinfo {author} {\bibfnamefont {M.~J.}\ \bibnamefont {Verstraete}}, \bibinfo {author} {\bibfnamefont {X.}~\bibnamefont {Gonze}}, \bibinfo {author} {\bibfnamefont {G.-M.}\ \bibnamefont {Rignanese}},\ and\ \bibinfo {author} {\bibfnamefont {G.}~\bibnamefont {Hautier}},\ }\bibfield  {title} {\bibinfo {title} {{Phonon-limited electron mobility in Si, GaAs, and GaP with exact treatment of dynamical quadrupoles}},\ }\href {https://doi.org/10.1103/PhysRevB.102.094308} {\bibfield  {journal} {\bibinfo  {journal} {Phys. Rev. B}\ }\textbf {\bibinfo {volume} {102}},\ \bibinfo {pages} {094308} (\bibinfo {year} {2020})}\BibitemShut {NoStop}%
\bibitem [{\citenamefont {Jhalani}\ \emph {et~al.}(2020)\citenamefont {Jhalani}, \citenamefont {Zhou}, \citenamefont {Park}, \citenamefont {Dreyer},\ and\ \citenamefont {Bernardi}}]{Jhalani2020prl}%
  \BibitemOpen
  \bibfield  {author} {\bibinfo {author} {\bibfnamefont {V.~A.}\ \bibnamefont {Jhalani}}, \bibinfo {author} {\bibfnamefont {J.-J.}\ \bibnamefont {Zhou}}, \bibinfo {author} {\bibfnamefont {J.}~\bibnamefont {Park}}, \bibinfo {author} {\bibfnamefont {C.~E.}\ \bibnamefont {Dreyer}},\ and\ \bibinfo {author} {\bibfnamefont {M.}~\bibnamefont {Bernardi}},\ }\bibfield  {title} {\bibinfo {title} {Piezoelectric electron-phonon interaction from ab initio dynamical quadrupoles: Impact on charge transport in wurtzite {GaN}},\ }\href {https://doi.org/10.1103/PhysRevLett.125.136602} {\bibfield  {journal} {\bibinfo  {journal} {Phys. Rev. Lett.}\ }\textbf {\bibinfo {volume} {125}},\ \bibinfo {pages} {136602} (\bibinfo {year} {2020})}\BibitemShut {NoStop}%
\bibitem [{\citenamefont {Wang}\ \emph {et~al.}(2008)\citenamefont {Wang}, \citenamefont {Zhu}, \citenamefont {Liu}, \citenamefont {Zhang}, \citenamefont {Zheng}, \citenamefont {He}, \citenamefont {Chen},\ and\ \citenamefont {Wen}}]{Wang2008cpl}%
  \BibitemOpen
  \bibfield  {author} {\bibinfo {author} {\bibfnamefont {M.-D.}\ \bibnamefont {Wang}}, \bibinfo {author} {\bibfnamefont {D.-Y.}\ \bibnamefont {Zhu}}, \bibinfo {author} {\bibfnamefont {Y.}~\bibnamefont {Liu}}, \bibinfo {author} {\bibfnamefont {L.}~\bibnamefont {Zhang}}, \bibinfo {author} {\bibfnamefont {C.-X.}\ \bibnamefont {Zheng}}, \bibinfo {author} {\bibfnamefont {Z.-H.}\ \bibnamefont {He}}, \bibinfo {author} {\bibfnamefont {D.-H.}\ \bibnamefont {Chen}},\ and\ \bibinfo {author} {\bibfnamefont {L.-S.}\ \bibnamefont {Wen}},\ }\bibfield  {title} {\bibinfo {title} {Determination of thickness and optical constants of zno thin films prepared by filtered cathode vacuum arc deposition},\ }\href {https://doi.org/10.1088/0256-307X/25/2/106} {\bibfield  {journal} {\bibinfo  {journal} {Chin. Phys. Lett.}\ }\textbf {\bibinfo {volume} {25}},\ \bibinfo {pages} {743} (\bibinfo {year} {2008})}\BibitemShut {NoStop}%
\bibitem [{\citenamefont {Singh}\ \emph {et~al.}(2021)\citenamefont {Singh}, \citenamefont {Tripathi}, \citenamefont {Sulania}, \citenamefont {Kumar},\ and\ \citenamefont {Kumar}}]{Singh2021om}%
  \BibitemOpen
  \bibfield  {author} {\bibinfo {author} {\bibfnamefont {S.}~\bibnamefont {Singh}}, \bibinfo {author} {\bibfnamefont {P.}~\bibnamefont {Tripathi}}, \bibinfo {author} {\bibfnamefont {I.}~\bibnamefont {Sulania}}, \bibinfo {author} {\bibfnamefont {V.~S.}\ \bibnamefont {Kumar}},\ and\ \bibinfo {author} {\bibfnamefont {P.}~\bibnamefont {Kumar}},\ }\bibfield  {title} {\bibinfo {title} {Tuning the optical and electrical properties of magnetron-sputtered cu–zno thin films using low energy ar ion irradiation},\ }\href {https://doi.org/https://doi.org/10.1016/j.optmat.2021.110985} {\bibfield  {journal} {\bibinfo  {journal} {Opt. Mater.}\ }\textbf {\bibinfo {volume} {114}},\ \bibinfo {pages} {110985} (\bibinfo {year} {2021})}\BibitemShut {NoStop}%
\bibitem [{\citenamefont {Sio}\ \emph {et~al.}(2019{\natexlab{a}})\citenamefont {Sio}, \citenamefont {Verdi}, \citenamefont {Ponc\'e},\ and\ \citenamefont {Giustino}}]{Sio2019prb}%
  \BibitemOpen
  \bibfield  {author} {\bibinfo {author} {\bibfnamefont {W.~H.}\ \bibnamefont {Sio}}, \bibinfo {author} {\bibfnamefont {C.}~\bibnamefont {Verdi}}, \bibinfo {author} {\bibfnamefont {S.}~\bibnamefont {Ponc\'e}},\ and\ \bibinfo {author} {\bibfnamefont {F.}~\bibnamefont {Giustino}},\ }\bibfield  {title} {\bibinfo {title} {Ab initio theory of polarons: Formalism and applications},\ }\href {https://doi.org/10.1103/PhysRevB.99.235139} {\bibfield  {journal} {\bibinfo  {journal} {Phys. Rev. B}\ }\textbf {\bibinfo {volume} {99}},\ \bibinfo {pages} {235139} (\bibinfo {year} {2019}{\natexlab{a}})}\BibitemShut {NoStop}%
\bibitem [{\citenamefont {Sio}\ \emph {et~al.}(2019{\natexlab{b}})\citenamefont {Sio}, \citenamefont {Verdi}, \citenamefont {Ponc\'e},\ and\ \citenamefont {Giustino}}]{Sio2019prl}%
  \BibitemOpen
  \bibfield  {author} {\bibinfo {author} {\bibfnamefont {W.~H.}\ \bibnamefont {Sio}}, \bibinfo {author} {\bibfnamefont {C.}~\bibnamefont {Verdi}}, \bibinfo {author} {\bibfnamefont {S.}~\bibnamefont {Ponc\'e}},\ and\ \bibinfo {author} {\bibfnamefont {F.}~\bibnamefont {Giustino}},\ }\bibfield  {title} {\bibinfo {title} {Polarons from first principles, without supercells},\ }\href {https://doi.org/10.1103/PhysRevLett.122.246403} {\bibfield  {journal} {\bibinfo  {journal} {Phys. Rev. Lett.}\ }\textbf {\bibinfo {volume} {122}},\ \bibinfo {pages} {246403} (\bibinfo {year} {2019}{\natexlab{b}})}\BibitemShut {NoStop}%
\bibitem [{\citenamefont {Franchini}\ \emph {et~al.}(2021)\citenamefont {Franchini}, \citenamefont {Reticcioli}, \citenamefont {Setvin},\ and\ \citenamefont {Diebold}}]{Franchini2021nrm}%
  \BibitemOpen
  \bibfield  {author} {\bibinfo {author} {\bibfnamefont {C.}~\bibnamefont {Franchini}}, \bibinfo {author} {\bibfnamefont {M.}~\bibnamefont {Reticcioli}}, \bibinfo {author} {\bibfnamefont {M.}~\bibnamefont {Setvin}},\ and\ \bibinfo {author} {\bibfnamefont {U.}~\bibnamefont {Diebold}},\ }\bibfield  {title} {\bibinfo {title} {Polarons in materials},\ }\href {https://doi.org/10.1038/s41578-021-00289-w} {\bibfield  {journal} {\bibinfo  {journal} {Nat. Rev. Mater.}\ }\textbf {\bibinfo {volume} {6}},\ \bibinfo {pages} {560} (\bibinfo {year} {2021})}\BibitemShut {NoStop}%
\bibitem [{\citenamefont {Dai}\ and\ \citenamefont {Giustino}(2024)}]{Dai2024pnas}%
  \BibitemOpen
  \bibfield  {author} {\bibinfo {author} {\bibfnamefont {Z.}~\bibnamefont {Dai}}\ and\ \bibinfo {author} {\bibfnamefont {F.}~\bibnamefont {Giustino}},\ }\bibfield  {title} {\bibinfo {title} {Identification of large polarons and exciton polarons in rutile and anatase polymorphs of titanium dioxide},\ }\href {https://doi.org/10.1073/pnas.2414203121} {\bibfield  {journal} {\bibinfo  {journal} {Proc. Natl. Acad. Sci. U.S.A.}\ }\textbf {\bibinfo {volume} {121}},\ \bibinfo {pages} {e2414203121} (\bibinfo {year} {2024})}\BibitemShut {NoStop}%
\end{thebibliography}%

\end{document}